\documentclass[12pt,a4paper]{article}
\usepackage{jheppub}  
\usepackage{amssymb} 
\usepackage{amsmath}
\usepackage{mathtools}
\usepackage{amsfonts}    
\usepackage{dsfont}
\usepackage{pdfpages}
\usepackage{verbatim}
\usepackage{tensor}
\usepackage{mathrsfs}
\usepackage[many]{tcolorbox}
\usepackage{tikz}
\usetikzlibrary{patterns,decorations.pathreplacing, decorations.markings,calc,shapes.misc,decorations.pathmorphing}

\newcommand{\ba}{\begin{align}}

\newcommand{\be}{\begin{equation}}
\newcommand{\ee}{\end{equation}}
\def\bd{\begin{tikzpicture}}
\def\ed{\end{tikzpicture}}

\DeclareMathOperator\tr{tr}
\DeclareMathOperator{\lcm}{lcm}

\newcommand\PSL{\text{PSL}}

\newcommand\U{\text{U}}
\newcommand\SL{\text{SL}}
\newcommand\SU{\text{SU}}
\newcommand\OSp{\text{OSp}}
\newcommand\Diff{\text{Diff}}
\newcommand\Map{\text{Map}}

\newcommand\CC{\mathbb{C}}
\newcommand\ZZ{\mathbb{Z}}
\newcommand\RR{\mathbb{R}}
\newcommand\HH{\mathbb{H}}
\newcommand\CP{\mathbb{CP}}

\newcommand\EE{\mathbb{E}}
\newcommand\QQ{\mathbb{Q}}

\newcommand\LL{\mathbb{L}}

\let\H\relax
\DeclareMathOperator\H{H}

\DeclareMathOperator\td{td}
\DeclareMathOperator\ch{ch}

\let\S\relax
\DeclareMathOperator\S{S}
\DeclareMathOperator\AdS{AdS}

\DeclareMathOperator\Hom{Hom}
\DeclareMathOperator\End{End}
\DeclareMathOperator\Aut{Aut}
\DeclareMathOperator\Out{Out}

\DeclareMathOperator\vol{vol}

\let\Im\relax
\DeclareMathOperator\Im{Im}
\DeclareMathOperator*{\Res}{Res}

\newcommand{\Sing}{\nabla}
\newcommand{\Zfake}{\mathtt{Z}}
\newcommand{\bM}{\overline{\mathcal{M}}}
\newcommand{\bC}{\overline{\mathcal{C}}}
\newcommand{\LLleft}{\LL_\circ}
\newcommand{\LLright}{\LL_\bullet}
\newcommand{\psileft}{\psi_\circ}
\newcommand{\psiright}{\psi_\bullet}

\allowdisplaybreaks[1]

\title{Off-shell Partition Functions in 3d Gravity}

\author{Lorenz Eberhardt} 
\affiliation{School of Natural Sciences, Institute for Advanced Study, \\
\hspace*{0.3cm}Einstein Drive 1, Princeton,  NJ 08540, USA}
\emailAdd{elorenz@ias.edu}

\abstract{We explore three-dimensional gravity with negative cosmological constant via canonical quantization. We focus on chiral gravity which is related to a single copy of $\PSL(2,\RR)$ Chern-Simons theory and is simpler to treat in canonical quantization. Its phase space for an initial value surface $\Sigma$ is given by the appropriate moduli space of Riemann surfaces. We use geometric quantization to compute partition functions of chiral gravity on three-manifolds of the form $\Sigma \times \S^1$, where $\Sigma$ can have asymptotic boundaries. Most of these topologies do not admit a classical solution and are thus not amenable to a direct semiclassical path integral computation. We use an index theorem that expresses the partition function as an integral of characteristic classes over phase space. In the presence of $n$ asymptotic boundaries, we use techniques from equivariant cohomology to localize the integral to a finite-dimensional integral over $\bM_{g,n}$, which we evaluate in low genus cases.
Higher genus partition functions quickly become complicated since they depend in an oscillatory way on Newton's constant. There is a precise sense in which one can isolate the non-oscillatory part which we call the fake partition function. We establish that there is a topological recursion that computes the fake partition functions for arbitrary Riemann surfaces $\Sigma$.
There is a scaling limit in which the model reduces to JT gravity and our methods give a novel way to compute JT partition functions via equivariant localization. 
}

\begin{document}

\maketitle

\makeatletter
\g@addto@macro\bfseries{\boldmath}
\makeatother

\section{Introduction}
Three-dimensional gravity with negative cosmological constant is a very interesting arena to explore quantum gravity from the bottom up. It is simple enough to have no dynamical excitations such as gravitational waves, yet complex enough to support black hole solutions \cite{Banados:1992wn}. The theory is essentially topological, but there are boundary graviton excitations in asymptotically AdS spacetimes \cite{Brown:1986nw}. Thus it is one of the simplest arenas for a holographic duality and there have been many different (partially conflicting) proposals about a holographic description of pure three-dimensional gravity, see for example \cite{Coussaert:1995zp, Witten:2007kt, Yin:2007at, Yin:2007gv, Maloney:2007ud, Gaberdiel:2007ve, Giombi:2008vd, Hartman:2014oaa, Keller:2014xba, Benjamin:2019stq, Hartman:2019pcd, Afkhami-Jeddi:2019zci, Benjamin:2020mfz, Cotler:2020ugk, Maxfield:2020ale, Chandra:2022bqq, Schlenker:2022dyo}.

Treating three-dimensional gravity quantum-mechanically  involves in principle the computation of the gravitational path integral, i.e.\ integration over all positive-definite metrics with given boundary conditions. Besides integrating over the metrics in a given topological class of metrics, it also includes a sum over all possible topologies. Such a sum is controllable in a two-dimensional setting such as JT gravity \cite{Jackiw:1984je, Saad:2019lba, Stanford:2019vob} or topological gravity \cite{Witten:1989ig, Witten:1990hr, Dijkgraaf:1990rs, Kontsevich:1992ti}, where this just becomes the genus expansion. In these dilaton gravity theories, there is a dual matrix model which in principle gives a non-perturbative definition of the holographic dual \cite{Douglas:1989ve, Brezin:1990rb, Gross:1989vs}.
Contrary to the two-dimensional case, there is no natural expansion in three dimensions and no good way to control the sum over topologies and as a result the status of three-dimensional quantum gravity is unclear.

Some of the topologies contain a metric representative that solves the Einstein equations. Thus the path integral can be computed semiclassically for such topologies by expanding the metric around the given solution. However, there are many more topologies that do not contain a metric which solves the classical equations of motion, yet they still contribute to the gravitational path integral. We refer to partition functions for such topologies as off-shell partition functions. It is a largely open problem how to compute such off-shell contributions to the gravitational path integral. See however \cite{Cotler:2020ugk} for a direct computation of a two-boundary off-shell partition function.
\medskip

In this paper we explain an efficient method that computes such off-shell partition functions on some geometries exactly. We will use canonical quantization and will thus consider the contribution from manifolds of the form $\Sigma \times \S^1$, where $\Sigma$ is a Riemann surface with possible asymptotic boundaries. In these cases, the gravitational path integral can be computed as a trace over the gravitational Hilbert space associated to the Cauchy surface $\Sigma$. The procedure of canonical quantization is not dependent on the existence of a metric on $\Sigma \times \S^1$ solving the Einstein field equations, In fact, the only saddle geometry of the form $\Sigma \times \S^1$ is thermal $\AdS_3$, which can be obtained by choosing $\Sigma$ to be a hyperbolic disk. Canonical quantization of 3d gravity has of course been considered before, see e.g.\ \cite{DeWitt:1967yk, Kim:2015qoa, Kim:2021nyy} for old and new discussions.

Canonical quantization rests on a good understanding of the relevant phase space of the theory. For three-dimensional gravity, the space space associated to $\Sigma$ is given by $(\mathscr{T}_\Sigma \times \mathscr{T}_\Sigma)/\Map(\Sigma)$ \cite{Moncrief:1989dx, Krasnov:2005dm, Mess:2007, Witten:2007kt, Scarinci:2011np, Kim:2015qoa}. Here $\mathscr{T}_\Sigma$ is the relevant Teichm\"uller space and $\Map(\Sigma)$ is the mapping class group that acts diagonally on $\mathscr{T}_\Sigma \times \mathscr{T}_\Sigma$. It arises because gravity has to be invariant under large diffeomorphisms as well. The symplectic form is given by two copies of the Weil-Petersson symplectic form.
Quantization of Teichm\"uller space is in principle understood \cite{Verlinde:1989ua, Kashaev:1998fc, Chekhov:1999tn}. This means that one can canonically associate a Hilbert space to Teichm\"uller space and the mapping class group defines a projective representation on this Hilbert space. The non-compactness of Teichm\"uller space implies however that the Hilbert space is infinite-dimensional. In particular, the partition function of three-dimensional gravity on $\Sigma \times \S^1$ diverges, because it is just computing the dimension of the Hilbert space associated to $\Sigma$.

For technical convenience, we will mostly consider a simpler version of 3d gravity (from the point of view of canonical quantization), which we call chiral gravity. In the language of $\PSL(2,\RR) \times \PSL(2,\RR)$ Chern-Simons theory, one can choose the levels of the two gauge groups differently. For chiral gravity, one chooses one level to be zero and thus this factor decouples. Correspondingly, the phase space is given by only one copy of Teichm\"uller space divided by the mapping class group $\mathscr{T}_\Sigma/\Map(\Sigma)$, i.e.\ the moduli space of Riemann surfaces $\Sigma$. Thus the phase space admits a canonical compactification by considering the Deligne-Mumford compactification of moduli space. We thus arrive at the nice statement that chiral gravity is equivalent to the quantization of the moduli space of Riemann surfaces. We can hence leverage known mathematical results to study this problem.
Chiral gravity is described by an integer parameter $k$, which is proportional to the dimensionless ratio $\ell_{\AdS}/G_\text{N}$. In particular, chiral gravity would be described holographically by a chiral CFT with left-moving central charge $c=24k$. Such CFTs are extremely constrained, especially if they have a large gap above the vacuum, see \cite{Gaiotto:2007xh, Gaberdiel:2007ve, Gaberdiel:2008xb, Lin:2021bcp}. Some of the ideas about the direct canonical quantization of the theory performed in this paper appeared already earlier in \cite{Maloney:2015ina}. The author treated the theory mainly in the large $k$ limit, where the computations can be approximated by the Weil-Petersson volumes of moduli space.
\medskip

In the bulk of the paper we study the quantization problem on the moduli space of Riemann surfaces. 
We first consider compact surfaces $\Sigma$ in Section~\ref{sec:compact surfaces} where the only quantity of interest is the (finite) dimension of the Hilbert space. We extend this to Riemann surfaces $\Sigma$ with asymptotic boundaries in Section~\ref{sec:adding boundaries}, where one computes partition functions that depends on the modular parameters (or inverse temperatures) of the boundary tori. For example, we compute the following partition functions,
\begin{subequations}
\begin{align} 
Z_0(\beta)&=q^{-k}\prod_{m=2}^\infty \frac{1}{1-q^m}\ , \\
Z_0(\beta_1,\beta_2)&=\prod_{i=1}^2 \prod_{m=1}^\infty \frac{1}{1-q_i^m} \times \frac{1}{1-q_1q_2}\ , \label{eq:intro partition function Z0,2}\\
Z_0(\beta_1,\beta_2,\beta_3)&= \prod_{m=1}^\infty \frac{1}{1-q_i^m} \times \prod_{i=1}^3\frac{1}{1-q_i}\ , \\
Z_0(\beta_1,\beta_2,\beta_3,\beta_4)&= \prod_{m=1}^\infty \frac{1}{1-q_i^m} \times \prod_{i=1}^4\frac{1}{1-q_i}  \left(k+1+\sum_{i=1}^4 \frac{q_i}{1-q_i}\right)\ , \\
Z_1(\beta)&=\prod_{m=1}^\infty \frac{1}{1-q^m} \times  \Bigg(\sum_{\omega=1,\, -1} \frac{\omega^k}{24(1-\omega q)}\left(k+5+\frac{\omega q}{1-\omega q}\right) \nonumber\\
&\qquad\qquad\qquad\quad+\sum_{m=4,\, 6} \sum_{\begin{subarray}{c} \ell=1,\\ 2\ell \ne m \end{subarray}}^{m-1} \frac{\mathrm{e}^{\frac{2\pi i k\ell}{m}}}{m(1-\mathrm{e}^{\frac{4\pi i \ell}{m}})(1-\mathrm{e}^{\frac{2\pi i \ell}{m}} q)}\Bigg)\ , \label{eq:intro partition function Z1,1}\\
Z_2&=\frac{1}{2\pi i} \oint_0 \mathrm{d}t \ \frac{t^{-k-1}}{1-t}\bigg(\frac{1}{(1-t^2)(1-t^3)(1-t^{5})} \nonumber\\
&\qquad\qquad\qquad\qquad\qquad\qquad-\frac{t}{(1-t^4)(1-t^6)(1-t^{12})}\bigg)\ . \label{eq:intro partition function Z2}
\end{align} \label{eq:intro partition functions}%
\end{subequations}
where $q_i=\mathrm{e}^{-\beta_i}$. The notation $Z_g(\beta_1,\dots,\beta_n)$ denotes the partition function on $\Sigma \times \S^1$ where $\Sigma$ is a genus $g$ surface with $n$ asymptotic boundaries with inverse temperatures $\beta_i$.
$Z_0(\beta)$ just corresponds to the thermal $\AdS_3$ partition function and reproduces as expected the vacuum Virasoro character for the correct central charge $c=24k$. Except for this special case, all these partition functions have not appeared before in the literature.
Note that the coefficients of the power series expansion in $q$ of the genus 1 partition function are all integers, even though this is not manifest in the formula. 

Let us explain some general features of the partition functions \eqref{eq:intro partition functions}. First of all, we notice that all of them have $q$-expansions with integer coefficients. Of course this has to be so for any chiral CFT partition function. The partition functions are in general not modular invariant, since the bulk breaks modular invariance explicitly.\footnote{Except in the case $g=0$ and $n=2$. Our formula is not modular invariant and we will discuss the precise reason for this in the Discussion~\ref{sec:discussion}.} Furthermore, the partition functions are of course consistent with the existence of a Virasoro symmetry and they all carry the universal prefactor $\prod_m (1-q^m)^{-1}$, except for the case $g=0$ and $n=1$ which reproduces the vacuum Virasoro character.
All partition functions except for $g=0$, $n=1$ start at $\mathcal{O}(q^0)$. Thus they do not contribute to the boundary partition function below the black hole threshold. This is in agreement with the expectation that there should be no other geometry than thermal AdS contributing to the partition function of the dual CFT \cite{Schlenker:2022dyo}.

The partition functions behave qualitatively different depending on whether $g=0$ or $g \ge 1$. Namely, it will become clear in this paper that the partition function for $g=0$ is always polynomial of degree $n-3$ in $k$ in this case. Things become significantly more complicated for $g \ge 1$, because partition functions generally involve phases of the form $\mathrm{e}^{2\pi i \theta k}$ for rational numbers $\theta$. From a canonical quantization perspective, such terms should be present, since even though the partition function involves more and more complicated fractions such as $\frac{1}{24}$ in \eqref{eq:intro partition function Z1,1}, the $q$-expansion only involves integers. For this reason it is sometimes easier to give a generating function for the partition function as we did in \eqref{eq:intro partition function Z2} for the genus 2 partition function without boundaries. These oscillatory terms arise in the quantization procedure, because for $g \ge 1$ the phase space is no longer a manifold, but an orbifold. In a precise sense, the various terms in eq.~\eqref{eq:intro partition function Z1,1} come from the different possible automorphism groups of genus 1 surfaces with a marked point, i.e.\ the first line comes from the generic $\ZZ_2$ group, while the second line comes from the square torus with $\ZZ_4$ automorphism group and the hexagonal torus with $\ZZ_6$ automorphism group.
Similarly, the higher genus partition functions become quickly very complicated functions of $k$ and exhibit oscillatory behaviour in $k$. 
The oscillatory behaviour is in tension with the usual lore that gravity only give smooth contributions to the CFT partition function or quantities such as the spectral form factor \cite{Cotler:2016fpe}.  
 
 There is a limit where our computations reduce to JT gravity. For this one has to consider the double scaling limit $k \to \infty$, $\beta \to 0$  while keeping
 \be 
 \beta_\text{JT}=\frac{1}{k \beta} \label{eq:intro JT limit}
 \ee
fixed. Physically this makes sense, because we are considering a small thermal circle which effectively dimensionally reduces the theory. Chiral gravity describes aspects of near extremal holes and JT gravity is obtained by considering a further scaling limit \cite{Balasubramanian:2009bg, Nayak:2018qej, Ghosh:2019rcj}. Quantization simplifies in the $k \to \infty$ limit, since this limit is equivalent to an $\hbar \to 0$ limit. For small $\hbar$, every unit volume of phase space is expected to give rise to a quantum state and thus the 3d gravity partition function reduces to the volume of moduli space in the large $k$ limit, which is known to agree with JT gravity.
\medskip

Let us next describe the methods that we use to obtain the partition functions in eq.~\eqref{eq:intro partition functions}. The problem of quantization for a compact genus $g$ Riemann surface can be treated using K\"ahler quantization. It boils down to the determination of the number of holomorphic sections of a line bundle $\mathscr{L}^k$ on moduli space $\bM_{g}$. Such sections can be counted explicitly for low genus where they correspond to modular forms, Jacobi forms and similar objects. In general, the number can be determined from an index theorem.
A major technical complication is that $\bM_{g}$ is not a manifold, but an orbifold. This means that the standard Hirzebruch-Riemann-Roch index theorem that computes the dimension of the space of holomorphic sections of $\mathscr{L}^k$ does not literally apply, because it misses `twisted sectors' from the orbifold structure. We often discuss only the untwisted sector, which following the terminology in the mathematics literature we call fake partition function. For example, in \eqref{eq:intro partition function Z1,1} the fake partition function would correspond to the first term with $\omega=1$, but it misses the oscillatory pieces from the partition function which come from the twisted sectors. For the genus 0 case with asymptotic boundaries the two are equivalent because $\bM_{0,n}$ is actually manifold. Correspondingly, there are no oscillating terms in the genus 0 examples that we listed in \eqref{eq:intro partition functions}.
We solve the problem of computing the fake partition function completely in this paper, but we only compute the actual partition function in some low genus examples, since it quickly becomes very complicated. 

We now explain how to evaluate the fake partition function explicitly, even in the presence of boundaries. The index theorem instructs us to compute the integral
\be 
\int_{\mathcal{M}} \td(\mathcal{M}) \, \mathrm{e}^{k c_1(\mathscr{L})} \label{eq:intro index theorem}
\ee
over the corresponding phase space $\mathcal{M}$.
For compact Riemann surfaces, this integral can be computed using standard techniques from algebraic geometry. Since they might not be very familiar to physicists, we explain the necessary background when needed. We also included a longer Appendix~\ref{app:Mgn algebraic geometry} that reviews the application of the Grothendieck-Riemann-Roch theorem that relates various characteristic classes on moduli space. We fill a small gap in the mathematics literature by computing the characteristic classes of the tangent bundle $\bM_{g,n}$ which has not been previously done in the literature. This is a necessary ingredient to compute the integral \eqref{eq:intro index theorem}.
In the presence of asymptotic boundaries, we explain that one should compute the equivariant version of the integral \eqref{eq:intro index theorem} over the corresponding infinite-dimensional phase spaces. This phase space is probably most familiar in the case of the disk, where it corresponds to $\Diff(\S^1)/\PSL(2,\RR)$. This allows us to use the powerful tool of equivariant localization.  We show that equivariant localization reduces the equivariant integral over the infinite-dimensional phase space to an ordinary integral over $\bM_{g,n}$, where $n$ is the number of asymptotic boundaries.

We then discuss that the appearing integrals over $\bM_{g,n}$ have the correct form to be computed by topological recursion \cite{Eynard:2007kz, Eynard:2011kk, Eynard:2011ga, Dunin-Barkowski:2012kbi}. This allows us to show that there is a dual topological recursion to chiral gravity that computes the fake partition functions. The spectral curve takes the simple form
\be 
\omega_{0,1}(z)=\sin(b z) \sin (b^{-1}z)\, \mathrm{d}z
\ee
where $b$ is related to $k$ via $24k=1+6(b+b^{-1})^2$. However, the connected two-point correlator $\omega_{0,2}$ does take a more complicated form than in random matrix models, see eq.~\eqref{eq:omega02} for the precise form. The topological recursion then gives a complete effective solution to the computation of the fake partition functions.

The existence of the topological recursion allows us to prove an analogue of the dilaton equation of two-dimensional topological gravity. It simply says that
\be 
\Res_{q_{n+1}=\infty} q_{n+1}^{k-2}(1-q_{n+1}) Z^\text{p}_g(q_1,\dots,q_n,q_{n+1})=(2g-2+n) Z^\text{p}_g(q_1,\dots,q_n)
\ee
where $Z^\text{p}_g$ is the partition function of primary states (i.e.\ where the Virasoro character is taken out). By abuse of notation we wrote $Z^\text{p}(q_1,q_2,\dots)$ instead of $Z^\text{p}(\beta_1,\beta_2,\dots)$ since confusion is unlikely.
Our proof is valid for the fake partition function, but we believe that the equation holds true in general, which we check in some low genus cases.

The technique of equivariant localization is very useful and can be also be applied to similar problems in two-dimensional gravity. As we mentioned already above, there is a scaling limit that reduces the theory to JT gravity, see eq.~\eqref{eq:intro JT limit}. For JT gravity, there is a well-established method to compute partition functions with asymptotic boundaries. One first computes the volume of moduli space with geodesic boundaries and then glues `trumpets' to the geodesic boundaries. The method of localization does not need this two-step process and computes the equivariant volume of the corresponding infinite-dimensional moduli spaces in one go. Localization was already explained in \cite{Stanford:2017thb} for the special case of the disk, but there is no obstacle to extending this to higher number of boundaries and genus. As a fun side product, agreement of the two methods of computation proves a famous formula by Mirzakhani on the cohomology class of the Weil-Petersson symplectic form in the presence of geodesic boundaries \cite{Mirzakhani:2006eta}.
\bigskip

This paper is organized as follows. We start in Section~\ref{sec:phase space} by a general discussion of the phase space of three-dimensional gravity. After some general remarks on geometric quantization, we start to quantize gravity on compact Riemann surfaces in Section~\ref{sec:compact surfaces}. We generalize this to the case with asymptotic boundaries in Section~\ref{sec:adding boundaries}, where we explain the localization argument in detail. We also comment more on the scaling limit that reproduces JT gravity. We establish in Section~\ref{sec:topological recursion} the dual topological recursion that computes the fake partition functions of the model. We also discuss the dilaton equation. We end with a discussion of our results in Section~\ref{sec:discussion} and compare it to other literature in the subject. The paper has three appendices. In Appendix~\ref{app:Kawasaki index theorem} we review the orbifold version of the Hirzebruch-Riemann-Roch theorem. 
In Appendix~\ref{app:Mgn algebraic geometry}, we review useful background on the algebraic geometry of $\bM_{g,n}$ and apply it to compute the characteristic classes of the tangent bundle of $\bM_{g,n}$. Finally, we discuss in Appendix~\ref{app:genus 2} the genus 2 partition function in detail using classical invariant theory.

\section{The phase space of gravity and Chern-Simons theory} \label{sec:phase space}
In this section, we discuss the phase space of 3d gravity and its relation to the Chern-Simons description. We first discuss ordinary 3d gravity. Most of the following facts are well-known in the literature, but perhaps not appreciated enough.
\subsection{Gravity and \texorpdfstring{$\PSL(2,\RR) \times \PSL(2,\RR)$}{PSL(2,R) x PSL(2,R)} Chern Simons-theory} \label{subsec:Gravity and Chern Simons}
In first-order formulation, three-dimensional gravity with negative cosmological constant in Lorentzian signature can be described by a dreibein $e_\mu^a$ and a spin-connection $\omega_{\mu\nu}^a$.
 The spin-connection can be dualized to give two $\mathfrak{psl}(2,\RR)$-valued one-forms $e_\mu^a$ and $\omega_\mu^a$. One then defines the combinations
\be 
A^{\pm,a}_\mu=\omega_\mu^a \pm \frac{1}{\ell} e_\mu^a\ , \label{eq:identification PSL(2,R) gauge fields}
\ee
where $\ell$ is the AdS-length related to the cosmological constant as $\Lambda=-\ell^{-2}$. $A^{\pm,a}$ are two $\PSL(2,\RR)$ gauge fields and the Einstein equations are equivalent to the field equations of Chern-Simons theory with levels $k_\text{L}=-k_\text{R}=\frac{\ell}{16G}$ \cite{Witten:1988hc}.\footnote{This differs from the value $\frac{\ell}{4G}$ that is usually stated in the literature by a factor of 4. The reason is that we are working with the gauge group $\PSL(2,\RR)$ instead of $\SL(2,\RR)$, which is more natural in the bosonic context.}

The relation of gravity with Chern-Simons theory is however a bit more subtle even at the classical level. We will now explain the relation between the gravitational phase space $\mathcal{M}_\text{grav}$ and the phase space $\mathcal{M}_{\PSL(2,\RR)} \times \mathcal{M}_{\PSL(2,\RR)}$ of the Chern-Simons theory.
\be 
\mathcal{M}_\text{grav}\overset{?}{\longleftrightarrow} \mathcal{M}_{\PSL(2,\RR)} \times \mathcal{M}_{\PSL(2,\RR)} \ ,
\ee
The phase space of $\PSL(2,\RR)$ Chern-Simons theory consists in general of several disconnected components. Let us assume that the 3-manifold under consideration is of the form $\Sigma \times \RR$, where $\RR$ represents the time direction and $\Sigma$ is a compact Riemann-surface. Then the associated constrained phase space of $\PSL(2,\RR)$ Chern-Simons theory is the moduli space of all flat $\PSL(2,\RR)$-bundles on the Riemann surface $\Sigma$. These in turn are in one-to-one correspondence with the representation variety
\be 
\Hom\left(\pi_1(\Sigma) ,\,  \PSL(2,\RR)\right)/\PSL(2,\RR)\ , \label{eq:representation variety}
\ee
where $\PSL(2,\RR)$ acts by overall conjugation on a homomorphism.\footnote{As it stands, this quotient is not even a Hausdorff space. This can be avoided by taking the GIT-quotient instead, but it will not play a role in this paper.} To each such flat $\PSL(2,\RR)$ bundle, one can associate an integer topological invariant known as the Euler class $e$. Abstractly, it is given by the image of the fundamental class $[\Sigma]$ under the classifying homomorphism $\Sigma \longrightarrow B \PSL(2,\RR)\cong B \S^1 \cong \CP^\infty$. Since $\H^2(\CP^\infty) \cong \ZZ$, this leads to an integer invariant.\footnote{
In more down-to-earth terms, we can choose a canonical basis of paths $\alpha_1,\dots,\alpha_g,\beta_1,\dots,\beta_g$ of $\Sigma$ such that the only relation in $\pi_1(\Sigma)$ is $\alpha_1 \beta_1 \alpha_1^{-1} \beta_1^{-1} \alpha_2 \beta_2 \alpha_2^{-1} \beta_2^{-1} \cdots=1$. Then a homomorphism as in \eqref{eq:representation variety} is an assignment of matrices $\rho(\alpha_i)=A_i$, $\rho(\beta_i)=B_i$ with the same relation. We can then lift each $A_i$ and $B_i$ to the universal cover $\widetilde{\PSL(2,\RR)}$, which we denote by $\widetilde{A}_i$ and $\widetilde{B}_i$. We then have
\be 
Z=\widetilde{A}_1 \widetilde{B}_1 \widetilde{A}_1^{-1} \widetilde{B}_1^{-1}\widetilde{A}_2 \widetilde{B}_2 \widetilde{A}_2^{-1} \widetilde{B}_2^{-1} \cdots\ .
\ee
$Z$ is an element of the center of $\widetilde{\PSL(2,\RR)}$, which can be identified with the integers. Since $Z$ is also independent of the choices made in this construction, it gives an integer invariant of every bundle. }
It is relatively easy see that this invariant cannot take arbitrary values. In fact, we have the basic Milnor-Wood inequality \cite{Milnor, Wood}
\be 
2-2g \le e \le 2g-2\ .
\ee
It was later shown by Goldman \cite{Goldman} that every value satisfying the inequality is in fact realized and the Euler class classifies all connected components of the moduli space $\mathcal{M}_{\PSL(2,\RR)}$ of bundles on a given compact Riemann surface.

There is one special component inside $\mathcal{M}_{\PSL(2,\RR)}$, namely the one with maximal Euler number.\footnote{The dual bundle has opposite Euler number, so the component with minimal Euler number would work as well.} This component is isomorphic to the Teichm\"uller space $\mathscr{T}_\Sigma$ of the surface $\Sigma$. To see this, note that the uniformization theorem states that any Riemann surface with $g\ge 2$ can be endowed with a unique hyperbolic metric. As such it can be written as a quotient of the upper half plane, $\Sigma =\HH/\Gamma$. This means in particular that the cotangent bundle of $\Sigma$ can be realized as a flat $\PSL(2,\RR)$ bundle, since we can write it as $K=(\HH \times \CC)/\Gamma$, where $\Gamma$ acts on $\CC$ via the standard action on one-forms. The Euler class of $K$ coincides with the first Chern class when viewed as a holomorphic bundle and thus $K$ is an element in $\mathcal{M}_{\PSL(2,\RR)}$ with maximal Euler number. This yields an isomorphism to Teichm\"uller space $\mathcal{M}_{\PSL(2,\RR)}^{e=2g-2} \cong \mathscr{T}_\Sigma$.

The phase space of $\PSL(2,\RR) \times \PSL(2,\RR)$ Chern-Simons theory is hence disconnected, but the phase space of gravity is connected. Invertibility of the drei-bein forces us to only consider one of the components of  the Chern-Simons phase space. In fact, it is elementary to see that (up to the action of the mapping class group, to be discussed below)
\be 
\mathcal{M}_\text{grav} \subset \mathscr{T} _\Sigma\times \mathscr{T}_\Sigma\ ,
\ee
but it is non-trivial that the two spaces in fact coincide. This is shown for closed surfaces in \cite{Krasnov:2005dm} and extended to Riemann surfaces with asymptotic boundaries in \cite{Scarinci:2011np}. 

In gravity, we are supposed to gauge all diffeomorphisms, large and small. In particular, the space of dreibeins should be further gauged by the action of the mapping class group, which via the identification \eqref{eq:identification PSL(2,R) gauge fields} is the diagonal mapping class group of $\mathscr{T}_\Sigma \times \mathscr{T}_\Sigma$. This finally means that we identify the gravitational phase space $\mathcal{M}_{\Sigma,\, \text{grav}}$ on $\Sigma \times \RR$ as
\be 
\mathcal{M}_{\Sigma,\, \text{grav}}\cong (\mathscr{T}_\Sigma \times \mathscr{T}_\Sigma)/\Map(\Sigma)\ , \label{eq:gravitational phase space}
\ee
where we recall that $\mathscr{T}_\Sigma$ is the Teichm\"uller space of the surface $\Sigma$ and $\Map(\Sigma)\cong \Out^+(\pi_1(\Sigma))$ is the (orientation-preserving) mapping class group.\footnote{There is also a version of 3d gravity where one considers also non-orientable surfaces and gauges by non-orientable diffeomorphisms. We consider in this paper only orientable gravity in which orientation reversal is ungauged. In the context of JT gravity, all the different possibilities were discussed in \cite{Stanford:2019vob}.} The same conclusion was reached in \cite{Kim:2015qoa}.

We want to emphasize that the gauging of the mapping class group is unnatural on the gauge theory side. In a path integral formalism, we would like to integrate over \emph{all} gauge fields, regardless of whether they are flat or not. However, the mapping class group only acts naturally on flat $\PSL(2,\RR)$ bundles and $\Map(\Sigma)$ is actually not a symmetry of Chern-Simons theory. In order to quantize gravity, we will instead proceed by canonical quantization, where we can directly work with the constrained phase space.

We also mention in passing that there is another description of the gravitational phase space in terms of the cotangent bundle of Teichm\"uller space\cite{Krasnov:2005dm},
\be 
\mathcal{M}_{\Sigma,\, \text{grav}}\cong T^* \mathscr{T}_\Sigma/\Map(\Sigma)\ .
\ee
Both descriptions are valid, but the one in terms of two copies of Teichm\"uller space is much more convenient for the purpose of this paper. The so-called Mess map \cite{Mess:2007} defines an isomorphism between the two descriptions.
\subsection{Symplectic structure, quantization of the level and chiral gravity}
We next discuss the symplectic structure of phase space. This is simple to do in the Chern-Simons description. The symplectic form is the famous Atiyah-Bott symplectic form on the moduli space of flat connections \cite{AtiyahBott},
\be 
\omega=\frac{k}{4\pi} \int_\Sigma \tr \left(\delta A \wedge \delta A\right)\ .
\ee
This formula defines a two-form on the space of \emph{all} $\PSL(2,\RR)$  connections. Since the space of all $\PSL(2,\RR)$ connections is an affine space, $\delta A$ should be thought of as a tangent vector. For a gauge transformation $\delta A=\nabla_A \varepsilon$, $\omega$ vanishes on a flat connection. Thus $\omega$ descends via symplectic reduction to a symplectic form on the moduli space of all flat connections up to gauge transformations. The corresponding moment map is the curvature of $A$. In particular, $\omega$ defines a symplectic form on the Teichm\"uller component of $\mathcal{M}_{\PSL(2,\RR)}$. Both the Weil-Petersson symplectic form and the Atiyah-Bott symplectic form on Teichm\"uller space  are defined in terms of hyperbolic structure on the surface $\Sigma$. Thus they agree essentially by definition, see \cite{Goldman_symplectic} for an extensive discussion.

The Weil-Petersson symplectic form is invariant under the action of the mapping class group action. Thus the symplectic form on the phase space \eqref{eq:gravitational phase space} is given by $\frac{k}{4\pi} \, \omega_\text{WP, L}-\frac{k}{4\pi} \,  \omega_\text{WP, R}$, where L and R denotes the two copies of Teichm\"uller space. 

Let us remark on the normalization of the WP symplectic form since this is not uniform throughout the literature. We will normalize the WP symplectic form such that $\frac{k}{4\pi} \, [\omega_\text{WP}]$ defines an integer cohomology glass in $\mathscr{T}_\Sigma/\Map(\Sigma)$ in the orbifold sense. In particular, the WP symplectic form on the upper half plane in this normalization is
\be 
\frac{\mathrm{d}^2 \tau}{2\, \Im(\tau)^2}\ .
\ee

The level $k$ of $\PSL(2,\RR)$ Chern-Simons theory has to be an integer for consistency. The quantization comes about as follows. In order to define the Chern-Simons action on a 3-manifold $M$ with a $\PSL(2,\RR)$-bundle $V$, one picks a four-manifold $X$ with $\partial X=M$ and an extension of the bundle to all of $X$.\footnote{This is always possible. The obstruction to the existence of such an extension with a $G$-bundle is captured by the oriented cobordism group $\Omega_3(BG) \cong \H_3(G,\ZZ)$, where the isomorphism follows from an application of the Atiyah-Hirzebruch spectral sequence. For $G=\PSL(2,\RR)$, the third homology group vanishes and thus such an extension always exists. See also the discussion in \cite{Dijkgraaf:1989pz}.} However, this does not uniquely define the action because we may pick different four-manifolds $X$ to define the action. For two different four-manifolds, the ambiguity in the action is
\be 
S_X[A]-S_{X'}[A]=\frac{k}{2\pi} \int_{X \cup (-X')} \tr (F \wedge F) \ .
\ee
This is a characteristic class of the bundle on the closed four-manifold $X \cup (-X')$ and as such an integer. Taking care of the various normalization, $S[A]$ is defined up to an element in $2\pi k \, \ZZ$.\footnote{Also every integer can in fact appear, since different choices of four-manifolds are parametrized the oriented cobordism group $\Omega_4(B \PSL(2,\RR)) \cong \ZZ$.} Well-definedness of $\mathrm{e}^{i S[A]}$ imposes then that $k \in \ZZ$.

This argument does however not readily apply to our situation. In fact, the level $k$ in gravity is \emph{not} quantized. The reason for this is that we do not need to define the Chern-Simons action on all $\PSL(2,\RR)$-bundle, but only on those in the Teichm\"uller component. In general, we could consider $\PSL(2,\RR) \times \PSL(2,\RR)$ Chern-Simons theory with levels $(k_\text{L},-k_\text{R})$. The potential ambiguity in the definition of the Chern-Simons theory would hence be
\be 
\delta S=\frac{k_\text{L}}{2\pi} \int_X \tr \left( F_\text{L} \wedge F_\text{L} \right)-\frac{k_\text{R}}{2\pi} \int_X \tr \left( F_\text{R} \wedge F_\text{R} \right)
\ee
for $X$ a closed four-manifold. Since $\PSL(2,\RR)$ is contractible to $\U(1)$, we can actually assume that all gauge fields are $\U(1)$ gauge fields. The fact that we only consider the Teichm\"uller component for both connections means that $\frac{F_\text{L}}{2\pi}=-\frac{F_\text{R}}{2\pi}$ are negatives of each other as cohomology classes. This means that
\be 
\delta S=\frac{1}{2\pi} (k_\text{L}-k_\text{R}) \int_X F_\text{L} \wedge F_\text{L}\ .
\ee
This is integer provided that 
\be 
k_\text{L}-k_\text{R} \in \ZZ\ , \label{eq:kL kR quantization}
\ee
but there is no condition on $k_\text{L}$ and $k_\text{R}$ separately. 

This is of course what we would expect from the point of view of gravity. However it also means that there is a generalization of three-dimensional gravity where we can choose $(k_\text{L},k_\text{R})$ independently, while keeping their difference as an integer. The symplectic form on the phase space \eqref{eq:gravitational phase space} clearly reads
\be 
\omega=\frac{k_\text{L}}{4\pi} \, \omega_\text{WP, L}-\frac{k_\text{R}}{4\pi} \, \omega_\text{WP, R}\ .
\ee
One can also see the quantization condition on the levels from the perspective of canonical quantization. The Bohr-Sommerfeld quantization condition states that $\int_C \omega \in \ZZ$ for every two-cycles in the phase space, i.e.\ $\omega$ should define an integer cohomology class. Since Teichm\"uller space is contractible, the phase space contracts on the diagonal $\mathscr{T}_\Sigma/\Map(\Sigma) $ with symplectic form $\frac{k_\text{L}-k_\text{R}}{4\pi} \omega_\text{WP}$. The normalization of the Weil-Petersson symplectic form is such that this is an integer cohomology class, provided that \eqref{eq:kL kR quantization} is satisfied.

For generic choices of $k_\text{L}$ and $k_\text{R}$, the phase space is non-compact. Thus we would find a divergent result if we would try to compute the gravity partition function on a 3-manifold of the form $\Sigma \times \S^1$ for $\Sigma$ a compact surface. This is because in canonical quantization this partition function is given by the dimension of the Hilbert space on $\Sigma$. Three-dimensional gravity is topological (except for boundary excitations that are absent for compact 3-manifolds) and thus the partition function is just the dimension of the Hilbert space associated to $\Sigma$. But since the phase space is non-compact, the Hilbert space is infinite-dimensional. This fact is problematic for a proper understanding of 3d gravity. The divergence also happens also for many partition functions with asymptotic boundary, see the Discussion~\ref{sec:discussion}. There is a special case where the problem does not appear and the theory simplifies. This is the case with $k_\text{R}=0$, but $k_\text{L} \in \ZZ_{\ge 0}$.\footnote{This chiral gravity is different from topologically massive gravity that was considered e.g.\ in \cite{Deser:1981wh, Deser:1982vy, Li:2008dq}.}  
In this special case, the theory can be described in terms of one copy of $\PSL(2,\RR)$ Chern-Simons theory and the phase space becomes just $\mathscr{T}_\Sigma/\Map(\Sigma)$, i.e.\ the moduli space of Riemann surfaces.

Famously, the moduli space of Riemann surfaces admits a natural compactification -- the so-called Deligne-Mumford compactification, where one allows also nodal surfaces, i.e.\ Riemann surfaces consisting of several components touching at their connecting nodes. This compactification is natural to consider in the present context because it appears directly as the compactification of the moduli space of hyperbolic surfaces. However the following results crucially depend on the chosen compactification and this dependence essentially reflects the type of singularities one allows in the spatial manifolds. We will comment more on this point in the Discussion~\ref{sec:discussion}.
The symplectic form is now given simply by $\frac{k}{4\pi} \, \omega_\text{WP}$, where we write $k \equiv k_\text{L}$. The Weil-Petersson form extends smoothly to the boundary divisors and hence defines a symplectic form on the compactified phase space $\bM_\Sigma =\bM_g$ for $\Sigma$ a genus $g$ Riemann surface. Thus quantizing chiral gravity amounts to quantizing the moduli space of Riemann surfaces. Contrary to the generic situation in 3d gravity, this phase space is compact and we can expect to obtain finite partition functions on compact 3-manifolds such as $\Sigma \times \S^1$.

\subsection{Adding boundaries and the universal phase space}
So far, we mostly discussed the case where $\Sigma$ is a compact Riemann surface. It is however physically far more interesting to allow asymptotic boundaries for $\Sigma$. In this case, it is well-known that 3d gravity with AdS boundary conditions has boundary excitations \cite{Brown:1986nw}. The gravitational Hilbert space carries a representation of two copies of the Virasoro algebra (or one copy in the case of chiral gravity).

In principle, $\Sigma$ could have any number of asymptotic AdS boundaries and for each boundary there is an action of the corresponding Virasoro algebra. In a holographic setting, we are interested in computing the partition function given by the following trace over the Hilbert space
\be 
Z(\beta_1,\dots,\beta_n)=\tr_{\mathcal{H}} \left(\mathrm{e}^{-\sum_i \beta_i H_i} \right)\ , \label{eq:connected partition function}
\ee
where $H_i$ is the Hamiltonian associated to the $i$-th boundary of $\Sigma$. For a single boundary, this would compute a part of a putative boundary CFT partition function. There would be further contributions from three-dimensional manifolds with asymptotic AdS boundary conditions, but which are not of the form $\Sigma \times \S^1$.
The original AdS/CFT correspondence \cite{Maldacena:1997re} conjectures a single CFT dual to various string background, but in recent years it has become clearer that AdS/CFT dualities based on pure gravity presumably do not admit such a clean duality. Rather it was found that the boundary description of gravity often involves \emph{ensembles} of CFTs. This is clearest in two-dimensional JT gravity where the duality can essentially be fully understood and proven perturbatively in the genus expansion \cite{Saad:2019lba}. There are also various hints that similar statements might also hold for three-dimensional gravity, but it is much less clear whether the theory actually exists at the quantum level. In this paper we corroborate these claims for chiral gravity, where we are actually able to compute many of the partition functions involving several boundaries.

Our strategy for computing \eqref{eq:connected partition function} is somewhat indirect, but quite general (and we expected that the techniques applied in this paper can be generalized further). The phase space for chiral gravity on a non-compact surface $\Sigma$ was analyzed in \cite{Scarinci:2011np} and found to be given by the \emph{universal} Teichm\"uller space divided by the mapping class group, which we may call universal moduli space. This space is well-known when $\Sigma$ is the hyperbolic disk, where it is given by $\Diff(\S^1)/\PSL(2,\RR)$ (the mapping class group is trivial in this instance), but in principle an analogous space exists when $\Sigma$ has any number of boundaries and genus. This phase space carries additionally $n$ distinct $\U(1)$-actions corresponding to the actions of the $n$ Hamiltonians. Mathematically, \eqref{eq:connected partition function} is then simply an equivariant version of the dimension. As we shall explain in detail, it can be computed using an equivariant index theorem, which in turn can be simplified using equivariant localization.

\section{Quantization for compact surfaces} \label{sec:compact surfaces}
In the last section we argued that chiral gravity can be understood in a Hamiltonian formalism by quantizing the constrained phase space $\bM_g$ with symplectic structure given by the Weil-Petersson symplectic form $\frac{k}{4\pi}\omega_{\text{WP}}$. We explain now in more detail how the theory is quantized on a compact Riemann surface $\Sigma$. We consider the case of a Riemann surface with asymptotic boundaries in the next Section.

\subsection{The quantization problem}
In principle, the quantization problem is straightforward, since there are no local operators and no symmetries in the theory that we would like to preserve at the quantum level. Since the phase space $\bM_g$ has a K\"ahler structure, we will be able to proceed by K\"ahler quantization. This associates a canonical Hilbert space to the quantization and hence solves the quantization problem completely. Since the phase space is compact, this dimension turns out to be finite. Hence our only task will be to determine the dimension of the associated Hilbert, which depends on the genus $g$ of the surface and $k$. 

Before explaining the specific details, we want to recall some notions from geometric quantization that transform our problem into a question of algebraic geometry. We will first discuss these notions for a generic smooth phase space $\mathcal{M}$ which we assume to be K\"ahler. and discuss the generalization to phase spaces with orbifold singularities below.
 In geometric quantization, one starts by picking a line bundle $\mathscr{L}$ (the so-called prequantum line bundle) over the phase space $\mathcal{M}$ such that $c_1(\mathscr{L})=\omega$. If $\mathcal{M}$ is K\"ahler such a line bundle always exists; this is the content of Lefschetz' $(1,1)$-theorem.
However, such a line bundle is not necessarily unique and different choices are parametrized by $\mathrm{H}^1(\mathcal{M},\mathrm{U}(1))$.
 The Hilbert space of the problem are the polarized sections of this line bundle. Thus to proceed one has to pick a polarization that tells us which coordinates on phase space are the `momenta' and which ones are the `positions' of the problem, since the wavefunction should only depend on either positions or momenta. In case $\mathcal{M}$ is a K\"ahler manifold, we can pick the K\"ahler polarization, so that the Hilbert space is formed by all \emph{holomorphic} sections. 
  In other words, $\mathcal{H}=\H^0(\mathcal{M},\mathscr{L})$.\footnote{
 There is another step that is sometimes done in the literature, which is the `metaplectic correction'. If we would follow the recipe as stated so far, one sometimes runs into problems with the ground state energies. For example, one finds that the energy levels of the harmonic oscillator are $n \in \ZZ_{\ge 0}$ (in units where $\hbar \omega=1$) instead of the usual $n+\frac{1}{2}$. To correct for this in geometric quantization, one changes the line bundle under consideration to $\mathscr{L}$ tensored with a choice of square root of the canonical line bundle of phase space. Such a square root may or may not exist depending on the specific situation. In the case of the moduli space of Riemann surfaces, the canonical bundle does generically not admit a square root. Thus we will not include a metaplectic correction in our analysis, since it would lead to inconsistencies. \label{footnote:metaplectic correction} } In our context, the line bundle belonging to the symplectic form $\frac{k}{4\pi} \omega_\text{WP}$ is by construction the $k$-th power of the basic line bundle with curvature form $\frac{1}{4\pi} \omega_\text{WP}$. Thus we will write in the following $\mathscr{L}^k$ for the prequantum line bundle.

So far, we have explained that in K\"ahler quantization, states of the Hilbert space correspond to holomorphic sections of the line bundle $\mathscr{L}^k$. We are only interested in the dimension of this Hilbert space, which corresponds to the cohomology group $\mathcal{H}_k=\H^0(\mathcal{M},\mathscr{L}^k)$, which as we promised is a problem purely in the realm of algebraic geometry.\footnote{We could refine the quantization problem by keeping track of other operators in the theory, such as geodesic length operators \cite{Chekhov:1999tn, Chekhov:2000tw, Teschner:2005bz}.} 

There is a further simplification that helps in computing this cohomology group. Assume that the symplectic form comes in families, so that $\omega^{(k)}=k \omega$ with $k \in \ZZ_{\ge 1}$. Then $k$ plays the role of $\hbar^{-1}$. The corresponding line bundle associated to $\omega^{(k)}$ is simply the $k$-th power of the line bundle for $k=1$. Furthermore, the curvature form of line bundle $\mathscr{L}$ coincides by construction with the K\"ahler form on phase space, which in turn is related to a (positive definite) metric. Such a line bundle is called positive in algebraic geometry.
The Kodaira vanishing theorem now states with these assumptions that for $k$ large enough, higher cohomology groups vanish,
\be 
\H^n(\mathcal{M},\mathscr{L}^k)=0
\ee
for $n \ge 1$ and $k \gg 0$. Thus, the dimension of the Hilbert space of geometric quantization can be computed as the holomorphic Euler characteristic of the line bundle $\mathscr{L}^k$. Thus, we have
\be 
\dim \mathcal{H}_k = \dim \H^0(\mathcal{M},\mathscr{L}^k) 
= \sum_{n \ge 0} (-1)^n \dim \H^n(\mathcal{M},\mathscr{L}^k)
= \chi (\mathcal{M},\mathscr{L}^k) 
\ee
for $k$ large enough.
For smooth $\mathcal{M}$, the latter quantity can be computed from the Riemann-Roch-Hirzebruch index theorem,
\begin{align}
\dim \mathcal{H}_k&= \int_{\mathcal{M}} \td (\mathcal{M})\,    \mathrm{e}^{k c_1(\mathscr{L})}\ . \label{eq:index theorem}
\end{align}
Here $\td(\mathcal{M})$ is the Todd class of the tangent bundle of $\mathcal{M}$. It can be defined as the cohomology class of
\be 
\det \left(\frac{R}{1-\mathrm{e}^{-R}}\right)\ ,
\ee
where $R$ is the $\End(T\mathcal{M})$-valued Ricci 2-form of the tangent bundle with respect to any choice of connection. This expression should be viewed as a formal power series that terminates since $R$ is a two-form and $\mathcal{M}$ is finite dimensional. We can then take a determinant in the space $\End(T\mathcal{M})$ to obtain an ordinary differential form (of mixed degree) and hence a cohomology class.

Recalling that $k c_1(\mathscr{L})=\omega^{(k)}$, this shows the semiclassical intuition that the number of states is roughly given by the volume of phase space. Indeed, for large $k$, we can neglect the Todd genus on the right, since the dominant contribution comes from the top power of the the exponential $\mathrm{e}^{k c_1(\mathscr{L})}$, and so
\be 
\dim \mathcal{H}_k \sim \frac{k^{\dim(\mathcal{M})}}{\dim(\mathcal{M})!} \int_{\mathcal{M}} c_1(\mathscr{L})^{\dim(\mathcal{M})}=\left(\frac{k}{4\pi}\right)^{\dim(\mathcal{M})} \vol(\mathcal{M})\ ,
\ee
where $\dim(\mathcal{M})$ is the complex dimension of $\mathcal{M}$. This connection was exploited in \cite{Witten:1991we} to determine the volume of the moduli space of flat bundles on a Riemann surface for compact gauge groups.

Finally, we explain the modifications to this procedure in the case when $\mathcal{M}$ is an orbifold. The geometric quantization procedure is completely analogous as long as one is careful about the notion of a line bundle on orbifolds. However, for the purposes of this work, it is important to mention that the index theorem \eqref{eq:index theorem} is in general wrong for orbifold phase spaces. In that case, there is a refinement called the Kawasaki index theorem. We describe it in Appendix~\ref{app:Kawasaki index theorem}.  The integral \eqref{eq:index theorem} evaluates to a rational number that is sometimes called the `fake' Euler characteristic of the line bundle. The orbifold index theorem expresses the Euler characteristic of the line bundle instead as an integral over the so-called inertia stack of the phase space. For $\bM_{g,n}$, the inertia stack consists of several components that essentially capture curves together with their automorphism groups. The original moduli space appears as one component, namely of curves with trivial automorphism groups. So the `fake Euler characteristic' is the first in a number of terms. As we discuss in more detail in Section~\ref{subsec:corrections}, these corrections as suppressed and oscillatory in $k$.  The inertia stack of $\bM_{g,n}$ has been studied, see e.g.~\cite{Cornalba_automorphisms} and it seems possible to improve upon the results in this paper by incorporating also the other sectors of the inertia stack.

We will use the following notation for the partition function. Let $Z_g$ denote the genus $g$ partition function (that equals $\dim \mathcal{H}_k$). We will later extend this notation to $Z_g(\beta_1,\dots,\beta_n)$ in the presence of asymptotic boundaries. Since we will often work with the naive index theorem, we will denote the fake partition function as $\Zfake_g$ and $\Zfake_g(\beta_1,\dots,\beta_n)$ in the case of boundaries. For $g=0$, the moduli space does not have orbifold singularities and consequently $Z_0(\beta_1,\dots,\beta_n)=\Zfake_0(\beta_1,\dots,\beta_n)$. 

\subsection{\texorpdfstring{$\bM_{1,1}$}{M1,1} and modular forms} \label{subsec:M11 modular forms}
The preceding discussion was rather abstract and we want to exemplify it in the simple example of $\bM_{1,1}$, where everything can be discussed very explicitly.
 Recall that $\bM_{1,1}$ is also the space of two-dimensional lattices $\Lambda \subset \CC$, which can always be brought into the form $\ZZ \oplus \ZZ \tau$ with $\Im \tau>0$ and $\tau \sim \tau+1 \sim -\frac{1}{\tau}$. A  generic lattice described by $\tau$ has only a $\ZZ_2$ automorphism given by inversion at the origin, whereas the lattice with $\tau=i$ has a $\ZZ_4$ automorphism given by rotation by 90 degrees and the lattice with $\tau=\mathrm{e}^{\frac{\pi i}{3}}$ has a $\ZZ_6$ automorphism given by 60 degree rotation.

There is a fundamental line bundle $\mathscr{L}^k$ on $\bM_{1,1}$ (in the sense that it generates the group of line bundles), whose sections are functions $F$ of the lattice $\Lambda$ with the homogeneity property
\be 
F(c \Lambda)=c^k F(\Lambda)\ .
\ee
The first Chern class of the bundle is indeed given by the Weil-Petersson symplectic form $\frac{1}{4\pi} \omega_\text{WP}$.
This is the same as a level $k$ modular forms since we can write the same property also as
\be 
f\left(\frac{a \tau+b}{c \tau+d} \right)=(c \tau+d)^k f(\tau)\ , \qquad \begin{pmatrix}
a & b \\ c & d
\end{pmatrix}\in \SL(2,\ZZ)\ .
\ee
Hence
\be 
\mathcal{H}_k=\mathbb{M}_{k}(\mathrm{SL}(2,\ZZ))\ ,
\ee
the space of modular forms of level $k$. It is well-known that the ring of modular forms is freely generated by the Eisenstein series $E_4$ and $E_6$. This immediately gives
\begin{align} 
\dim \mathcal{H}_k &=[t^{k}] \frac{1}{(1-t^4)(1-t^6)} \\
&=\begin{cases}
0 & k \quad \text{odd}\ , \\
\lfloor\frac{k}{12}\rfloor +1& k\equiv 0,\, 4,\, 6,\, 8,\, 10 \bmod 12 \ , \\
\lfloor\frac{k}{12}\rfloor & k\equiv 2 \bmod 12 \ .
\end{cases} \label{eq:dimensions modular forms}
\end{align}
This is clearly a somewhat complicated and highly oscillatory function and this will not be reproduced by the naive index theorem (which would predict a polynomial answer in $k$). Let us proceed anyway and compute the right hand side of the index theorem \eqref{eq:index theorem}. To do so, we first have to determine the first Chern class of $\mathscr{L}$. This can be done by noting that $E_4$ is a section of $\mathscr{L}$ and it has a simple zero at the point of enhanced symmetry $\mathrm{e}^{\frac{2\pi i}{3}}$ with an automorphism group of size 6. Consequently,\footnote{In the presence of orbifold singularities, any divisor with enhanced symmetries is always counted with the additional factor $|\Aut|^{-1}$, where $\Aut$ is the automorphism group.}
\be 
\int_{\bM_{1,1}} c_1(\mathscr{L})=\frac{1}{24}\ .
\ee
The same conclusion can be reached by noting $E_6$ that has a simple zero at $\tau=i$.

The canonical bundle of $\mathcal{M}_{1,1}$ is generated by the forms $f(\tau)\mathrm{d}\tau$. This combination has to be invariant under modular transformations, which tells us that $f$ is a modular form of weight $2$. Hence the canonical bundle corresponds to modular forms of weight $2$ and so $\mathscr{K}_{\mathcal{M}_{1,1}}=\mathscr{L}^2$. However, the canonical bundle of $\bM_{1,1}$ behaves differently from the canonical bundle of $\mathcal{M}_{1,1}$ at the boundary divisor for $\tau \to i\infty$. Setting $q=\mathrm{e}^{2\pi i \tau}$, we have $\mathrm{d}\tau=\frac{\mathrm{d}q}{2\pi i q}$ and hence regularity at the cusp means that $f(\tau)$ is in fact a cusp from of weight 2. Thus the canonical bundle of $\bM_{1,1}$ is in fact the weight 2 modular forms twisted by the boundary divisor $\mathscr{D}$, which corresponds to the added point $\tau=i\infty$ in the compactification. This means that sections are constrained to have first order zeros on the boundary divisor $\mathscr{D}$, 
\be 
\mathscr{K}_{\bM_{1,1}} \cong \mathscr{L}^2(-\mathscr{D})\ .
\ee
One can simplify this further by noting that the modular discriminant $\Delta(\tau)=\eta(\tau)^{24}$ is a holomorphic section of $\mathscr{L}^{12}$. It has a simple zero at $\mathscr{D}$, which tells us that $\mathscr{L}^{12} \cong \mathscr{O}(\mathscr{D})$. Here $\mathscr{O}$ denotes the trivial line bundle on $\bM_{1,1}$ and $\mathscr{O}(\mathscr{D})$ is the trivial line bundle twisted by the divisor $\mathscr{D}$ as above, i.e.\ sections of $\mathscr{O}(\mathscr{D})$ are holomorphic functions with a simple pole at $\mathscr{D}$.
Thus\footnote{We can confirm this by checking that the Gauss-Bonnet theorem works out, since the top Chern class of the tangent bundle gives the orbifold Euler characteristic of $\bM_{1,1}$. This works, since the holomorphic tangent bundle is $\mathscr{L}^{10}$ and so
\be 
\int_{\bM_{1,1}} c_1(\mathscr{L}^{10})=\frac{5}{12}=\chi(\bM_{1,1})=\chi(\mathcal{M}_{1,1})+\frac{1}{2}\chi(\mathcal{M}_{0,3})\ ,
\ee
since the boundary divisor of $\bM_{1,1}$ is isomorphic to $\mathcal{M}_{0,3}/\ZZ_2$. The $\ZZ_2$ interchanges the two nodes in the nodal sphere.}
\be 
\mathscr{K}_{\bM_{1,1}} \cong \mathscr{L}^{-10}\ .
\ee

We can now compute the right hand side of the naive index theorem \eqref{eq:index theorem}
\be 
\int_{\bM_{1,1}} \td(\bM_{1,1}) \, \mathrm{e}^{k c_1(\mathscr{L})}=\int_{\bM_{1,1}} \left(k c_1(\mathscr{L})-\frac{1}{2} c_1( \mathscr{K}_{\bM_{1,1}} )\right)=\frac{k+5}{24}\ .
\ee
This clearly does not reproduce the actual dimension count given by eq.~\eqref{eq:dimensions modular forms}. However, it gives in a sense an averaged measure of the number of sections. Indeed, since there are roughly $\frac{k}{12}$ sections for $k$ odd and none for $k$ even, there are about $\frac{k}{24}$ on average. One can be more precise and even understand the constant term. It comes about because there are 5 cases in \eqref{eq:dimensions modular forms} which involve the correction $+1$. We will explain it in more detail in Appendix~\ref{app:Kawasaki index theorem}, where we discuss the modified index theorem for orbifolds.

In fact, since the result of the naive index theorem reproduces many properties that one would expect from the Euler characteristic (of the cohomology taking values in the line bundle), it is sometimes called `fake' Euler characteristic (e.g.\ in \cite{Givental:2011ds}), thus motivating our terminology of fake partition function.

\subsection{Bundles and classes on moduli space} \label{subsec:bundles on Mgn}
Let us now explain various vector bundles and line bundles on moduli space that are of interest to us. We will discuss the moduli space $\bM_{g,n}$ with punctures, since they will be needed once we add boundaries to the Riemann surfaces. The following is all very standard in algebraic geometry. A physicist-friendly introductory account can for example be found in \cite{Zvonkine_intro}. 

There are $n$ line bundles $\LL_i$ on $\bM_{g,n}$, whose fiber at a punctured surface consists of the cotangent space at the $i$-th marked point. The different font is meant to avoid confusions with the prequantum line bundle $\mathscr{L}^k$.\footnote{There is some care required to specify how these line bundles are defined for singular curves. A more rigorous definition can be given as follows. Let $\pi: \bC_{g,n} \to \bM_{g,n}$ be the universal curve. Let $\omega_\pi$ be the line bundle on $\bC_{g,n}$ consisting of holomorphic differentials on the fiber. Sections are allowed to have simple poles at the nodes as long as the residues on the two branches of the node are opposite. This extends the definition of $\omega_\pi$ to the boundary of $\bC_{g,n}$, which technically is called the relative dualizing sheaf since it satisfies Serre-duality for the fiber. Let now $\sigma_i: \bM_{g,n} \to \bC_{g,n}$ be the sections that take a curve to itself together with the $i$-th marked point. One then defines $\LL_i=\sigma_i^*(\omega_\pi)$. This description is explained and used further in Appendix~\ref{app:Mgn algebraic geometry}. \label{footnote:definition Li}} 
 The corresponding first Chern class is denoted by $\psi_i=c_1(\LL_i)$. Intersection numbers of psi-classes 
 \be 
\int_{\bM_{g,n}} \psi_1^{d_1} \psi_2^{d_2} \cdots \psi_n^{d_n} 
 \ee
 are determined by Witten's conjecture (Kontsevich's theorem) \cite{Witten:1990hr, Kontsevich:1992ti}.
 
 We also have natural vectorbundles on moduli space. The Hodge bundle $\EE$ is a $g$-dimensional vector bundle whose fibers consists of the space of holomorphic differentials $\H^0(\Sigma,K)$ on the curve.\footnote{As in the previous footnote \ref{footnote:definition Li}, some care has to be taken to properly define the Hodge bundle near the boundary divisors. By definition the fiber of $\EE$ is given by sections of $\omega_\pi$, which are allowed to have simple poles at the nodes.} More generally, we can consider the push-forward of any line bundle $\mathscr{L}$ on the Riemann surface to moduli space. The fiber at a point in moduli space described by the surface $\Sigma$ is given by the formal linear combination
\be 
\H^0(\Sigma,\mathscr{L}) - \H^1(\Sigma,\mathscr{L}) \ ,
\ee
which should be interpreted as an element of K-theory. This is just a fancy way to allow oneself to consider formal linear combinations of vector bundles, where addition can be identified as taking a direct sum. More practically, we consider this always for situations where $\H^1(\Sigma,\mathscr{L})$ is either trivial or vanishing, in which case $\H^0(\Sigma,\mathscr{L})$ alone is a well-defined vector bundle. In general one needs to consider the formal difference of the two cohomologies, because cohomology can jump, but the Riemann-Roch theorem guarantees that jumps cancel out of this combination. 
 Finally, every vector bundle gives rise to the so-called determinant line bundle by taking the top exterior power. In particular, the formal linear combination above gives an honest line bundle of the form
\be 
\det\H^0(\Sigma,\mathscr{L}) \otimes (\det \H^1(\Sigma,\mathscr{L}))^{*}\ .
\ee
An easy computation shows that the first Chern class of the determinant line bundle equals the first Chern class of the underlying vector bundle.

Recall also that there is a forgetful map 
\be 
\pi: \bM_{g,n+1} \longrightarrow \bM_{g,n}
\ee
that forgets one of the marked points (say $z_{n+1}$) and stabilizes the curve afterwards if necessary.\footnote{Since we forget a marked point, the component with the marked point of the nodal curve might become unstable through this procedure. It is then necessary to contract components until the curve becomes stable again, which is called stabilization.} There are other natural gluing morphisms that embed products of lower-dimensional moduli spaces into $\bM_{g,n}$ as boundary divisors. 

We can use these maps to push forward and pull back various cohomology classes. In particular, the important $\kappa$-classes (Morita-Mumford-Miller classes) are defined as
\be 
\kappa_m=\pi_*(\psi_{n+1}^{m+1})\ ,
\ee
where the pushforward $\pi_*$ is integration over the fiber of the forgetful map. Their importance for us lies in the fact that $\kappa_1$ describes the cohomology class of the Weil Petersson form \cite{Wolpert:1983, Wolpert:1986}
\be 
\kappa_1=\frac{1}{4\pi} [\omega_\text{WP}]\ .
\ee
Let us now summarize all the natural classes in $\H^\bullet(\bM_{g,n},\QQ)$ that we have mentioned
\begin{enumerate}
\item The $\psi$-classes $\psi_1$, \dots, $\psi_n$.
\item The $\kappa$-classes $\kappa_m$.
\item The boundary class  that is Poincar\'e dual to the boundary divisor in $\bM_{g,n}$, which we will denote by  $\Delta_1$.\footnote{It is often denoted by $\delta$ in the literature.} The subscript will become clear below where we consider a generalization of this class.
Since the boundary divisor has several components, one can further write $\Delta_1$ as a sum of these boundary components. If we pinch a cycle, then we can either get two surfaces of genus $g_1+g_2=g$ (called separating divisor) or one surface with genus $g-1$ (called non-separating divisor). In the former case, some punctures go on one side of the pinched cycle and thus the boundaries are labelled by two sets $\mathcal{I} \sqcup \mathcal{J}=\{z_1,\dots,z_n\}$ and genera $g_1+g_2=g$. Stability imposes some conditions, e.g. if $g_1=0$, then $|\mathcal{I}| \ge 2$ etc. Finally, there is a single boundary corresponding to the pinching of a non-separating cycle. Following the literature on the subject, we denote the separating divisor classes by $\delta_{g_1,\mathcal{I}}$ with $\mathcal{I} \subset \{1,2,\dots,n\}$ and the non-separating divisor by $\delta_\text{irr}$. We also denote the corresponding divisors by $\mathscr{D}_{g_1,\mathcal{I}}$ and $\mathscr{D}_\text{irr}$. In particular, the boundary divisor that we called $\mathscr{D}$ in Section~\ref{subsec:M11 modular forms} for $\bM_{1,1}$ is actually $\mathscr{D}_\text{irr}$. Hence
\be 
\Delta_1=\frac{1}{2} \sum_{g_1=0}^g \sum_{\begin{subarray}{c} \mathcal{I} \subset \{1,\dots,n\} \\ \text{stable} \end{subarray}} \delta_{g_1,\mathcal{I}}+\frac{1}{2} \delta_\text{irr}\ . \label{eq:definition Delta1}
\ee
The factor of $\frac{1}{2}$ is present because the first sum overcounts the separating boundary classes by a factor of 2. The non-separating divisor has generically a $\ZZ_2$ automorphism that exchanges the two nodes and hence should be counted with a factor $\frac{1}{2}$.
\item We need a slight generalization of the previous boundary class that we denote by $\Delta_\ell \in \H^4(\bM_{g,n},\QQ)$. For a boundary divisor $\mathscr{D}_{h,\mathcal{I}}$ or $\mathscr{D}_\text{irr}$, the corresponding nodal surface has two nodes and we have two $\psi$-classes associated to it that we will call $\psileft$ and $\psiright$. Since the two nodes are not labelled, only symmetric combinations of $\psileft$ and $\psiright$ will be well-defined. We then define $\Delta_\ell$ as the pushforward of the class $(\psileft+\psiright)^{\ell-1}$ from the boundary. In more details, letting $\xi_{h,\mathcal{I}}$ and $\xi_\text{irr}$ denote the inclusion maps of the corresponding boundary divisors, we define
\be 
\Delta_\ell=\frac{1}{2} \sum_{g_1=0}^g \sum_{\begin{subarray}{c} \mathcal{I} \subset \{1,\dots,n\} \\ \text{stable} \end{subarray}} (\xi_{g_1,\mathcal{I}})_*\left((\psileft+\psiright)^{\ell-1}\right)+\frac{1}{2} (\xi_\text{irr})_* \left((\psileft+\psiright)^{\ell-1}\right)\ .
\label{eq:definition Delta}
\ee
Since $\mathscr{D}_{h,\mathcal{I}}$ is the image of $\xi_{h,\mathcal{I}}$ and $\mathcal{D}_\text{irr}$ is the image of $\xi_\text{irr}$, this is consistent with our previous definition of $\Delta_1$.
\item The characteristic classes of various natural vector bundles over $\bM_{g,n}$. The Chern classes of the Hodge line bundle are usually denoted by $\lambda_j\equiv c_j(\EE_g)$ and hence the first Chern class of the Hodge line bundle is denoted by $\lambda_1$. Another important vector bundle in our analysis is the bundle of quadratic differentials over moduli space, which we will denote it by $\EE^{(2)}$. A more careful definition taking into account the behaviour at the degenerations is given in Appendix~\ref{app:Mgn algebraic geometry}. The final important vector bundle is the (co)tangent bundle on $\bM_{g,n}$. One might think that the cotangent bundle $T^* \bM_{g,n}$ is isomorphic to $\EE^{(2)}$, but this is incorrect because they have a different behaviour near the boundary divisors of $\bM_{g,n}$. We discuss the precise difference in Appendix~\ref{app:Mgn algebraic geometry}. We denote the canonical bundle on $\bM_{g,n}$ by $\mathscr{K}$. It is the determinant line bundle of the cotangent bundle, $\mathscr{K} \cong \det T^* \bM_{g,n}$. 
\end{enumerate}
In the Appendix~\ref{app:Mgn algebraic geometry}, we review and extend the standard computation \cite{Mumford1983} that expresses the Chern classes of various line bundles in terms of the other set of classes. We have in particular for the first Chern classes
\begin{subequations} \label{eq:first Chern classes line bundles}
\begin{align}
\lambda_1&=\frac{1}{12}\left(\kappa_1-\sum_i \psi_i+\Delta_1\right)\ , \label{eq:Mumford formula}\\
c_1(\EE^{(2)})&=\frac{1}{12}\left(13\kappa_1-\sum_i \psi_i+\Delta_1\right) \\
c_1(\mathscr{K})&=\frac{1}{12}\left(13 \kappa_1-\sum_i \psi_i-11 \Delta_1\right)\ . \label{eq:first Chern class canonical bundle}
\end{align}
\end{subequations}
The first is the celebrated formula by Mumford \cite{Mumford1983}, while the first Chern class of the canonical line bundle was derived in \cite{Harris_Kodaira}.
This shows in particular how to construct a line bundle with first Chern class given by $\kappa_1$. We can choose for the prequantum line bundle
\be 
\mathscr{L}\equiv \det \EE^{(2)}\otimes \det (\EE)^{-1}\ .
\ee
We already mentioned above that different choices of the prequantum line bundle are parametrized by the cohomology group $\H^1(\bM_{g,n},\mathrm{U}(1))$. In the case of the moduli space, this group is trivial, since the moduli space is simply connected (see e.g.~\cite{boggi_pikaart_2000}). Thus, the choice of prequantum line bundle is unique. 

In fact, much more is known for the moduli space of curves. A theorem due to Arbarello and Cornalba states that the Picard group (i.e.\ the group of all line bundles) is freely generated (for genera $g \ge 3$) by the classes $\lambda_1$, $\psi_i$ for $i=1,\dots,n$ and the boundary divisors of moduli space \cite{Harer_Picard, Arbarello_Picard}.\footnote{
We should note that this theorem shows that the canonical bundle $\mathscr{K}$ on moduli space does not possess a square root (at least for genera $g \ge 3$), since its Chern class cannot be written as an integer linear combination of $\lambda$, $\psi$ and $\delta$ classes. Thus there is no consistent quantization scheme that employs a metaplectic correction.}

\subsection{Vanishing theorems} \label{subsec:vanishing theorems}
As we have mentioned before, the Kodaira vanishing theorem guarantees that the cohomology groups $\H^n(\bM_{g,n},\mathscr{L}^k)$ vanish for $n>0$ and sufficiently high $k$. We would like to make this more quantitative now. The Kodaira vanishing theorem actually ensures that this vanishing is true if $\mathscr{L}^k \otimes \mathscr{K}^{-1}$ is a positive line bundle. 

The set of positive line bundles on $\bM_{g}$ is known, but it is only conjecturally known on $\bM_{g,n}$. Fulton's conjecture is a  characterization for them on $\bM_{g,n}$ \cite{Cornalba_ample, Gibney}. It states that a line bundle on $\bM_{g,n}$ is positive if and only if its first Chern class has positive intersection with all dimension 1 strata (which is a necessary condition for positivity). We will content ourselves here with the simpler discussion on $\bM_{g}$, where the full conjecture is not needed and the set of positive line bundles is known. We have on $\bM_{g}$ by combining eqs.~\eqref{eq:first Chern classes line bundles}
\be 
c_1(\mathscr{L}^k \otimes \mathscr{K}^{-1})=(12k-13) \lambda_1-(k-2)\Delta_1\ . \label{eq:first Chern class Lk K inverse}
\ee
Cornalba and Harris' theorem \cite{Cornalba_ample} says that a line bundle with first Chern class $a \lambda_1-b \Delta_1$ with $a>0$ and $b>0$ is positive when $a>11b$. In our case, this condition is satisfied when $k>2$. Thus we will in the following restrict $k$ to this range. Assuming the validity of Fulton's conjecture, one can also check that the line bundle $\mathscr{L}^k$ is positive on $\bM_{g,n}$ provided that $k>2$.

This condition of $k > 2$ is sufficient to ensure vanishing of higher cohomology groups.
The formulas that we will derive seem completely regular even for $k = 2$ and thus we suspect that this restriction could be relaxed to include $k=2$ as well.
\subsection{The naive index theorem} \label{subsec:naive index theorem}
In the following, we will apply the naive index theorem to compute $\dim \mathcal{H}_k$. As we have explained, the index theorem does not compute the actual dimension of the Hilbert space, but rather an `averaged dimension', which we will call the fake partition function. In particular, the index theorem will not yield integer dimensions.

We start by computing the Todd class of the tangent bundle. This is straightforward with the help of the formulas of the Chern characters of the tangent bundle given in eq.~\eqref{eq:tangent bundle Chern characters}. For the purpose of relating characteristic classes, one can invoke the splitting principle which basically says that we can assume that a given vectorbundle is a direct sum of line bundles. Let $x_i$ denote the first Chern classes of the individual line bundles (called the Chern roots). Then it is an exercise in symmetric function theory to relate different characteristic classes. 
For the Chern character of the tangent bundle we have in particular
\be 
\ch(\bM_{g,n})=\sum_i \mathrm{e}^{x_i} =\sum_{m=0}^\infty \sum_{i=1}^{3g-3+n} \frac{x_i^m}{m!}
\ee
The Todd class is defined in terms of the Chern roots as follows:
\begin{align} 
\td(\bM_{g,n})&=\prod_{i=1}^{3g-3+n} \frac{x_i}{1-\mathrm{e}^{-x_i}} \\
&=\exp\left(\sum_{i=1}^{3g-3+n} \log\left( \frac{x_i}{1-\mathrm{e}^{-x_i}}\right)\right)\ .
\end{align}
The Taylor series of the appearing function can be computed as follows. Notice that
\begin{align} 
\frac{\mathrm{d}}{\mathrm{d}x} \log\left(\frac{x}{1-\mathrm{e}^{-x}}\right)&=\frac{1}{x}-\frac{1}{\mathrm{e}^x-1}=\frac{1}{2}-\sum_{m=1}^\infty \frac{B_{2m}}{(2m)!} x^{2m-1}\ , \label{eq:Taylor expansion log Bernoulli}
\end{align}
where the last equality follows from the definition of the Bernoulli numbers. Integrating once leads to
\begin{align}
\td(\bM_{g,n})&=\exp\left(\sum_{i=1}^{3g-3+n} \left(\frac{x_i}{2}- \sum_{m=1}^\infty \frac{B_{2m}}{(2m)(2m)!} x_i^{2m}\right)\right) \\
&=\exp\left(\frac{1}{2}\ch_1(T\bM_{g,n})-\sum_{m=1}^\infty \frac{B_{2m}}{2m} \ch_{2m}(T\bM_{g,n}) \right) \\
&=\exp\left(-\frac{1}{2} c_1(\mathscr{K})-\sum_{m=1}^\infty \frac{B_{2m}}{(2m)(2m)!} \left(\kappa_{2m}-\Delta_{2m}\right) \right) \ , \label{eq:todd class tangent bundle}
\end{align}
where in the last equality we used the formula \eqref{eq:even Chern characters tangent bundle Mgnbar} for the Chern characters of the tangent bundle. Here the boundary class $\Delta_{2m}$ defined in \eqref{eq:definition Delta} appears.
For the first Chern class of the canonical bundle, we can use eq.~\eqref{eq:first Chern class canonical bundle}.

We can now assemble the index theorem. Since the first Chern class of the prequantum line bundle is by definition $c_1(\mathscr{L}^k)=k \kappa_1$, we learn that the fake partition function equals
\begin{multline}
\Zfake_{g,n}=\int_{\bM_{g,n}} \exp\Bigg[\left(k-\frac{13}{24}\right) \kappa_1+\frac{1}{24}\sum_i \psi_i+\frac{11}{24}\Delta_1 \\
- \sum_{m=1}^\infty \frac{B_{2m}}{(2m)(2m)!}\left(\kappa_{2m}-\Delta_{2m}\right) \Bigg]\ , \label{eq:integral index theorem}
\end{multline}
where we added a subscript $n$ since we are computing the integral on $\bM_{g,n}$.

For future reference, let us record the value of this integral for the without punctures in the following cases
\begin{subequations}
\begin{align}
\Zfake_2&= \frac{1}{34560}(86k^3+369k^2+391k-186)\ ,\label{eq:index theorem 20} \\
\Zfake_3&= \frac{176557 k^6}{77414400}+\frac{116377 k^5}{25804800}-\frac{3367 k^4}{2211840}-\frac{24071 k^3}{3096576}\nonumber\\
&\qquad-\frac{691 k^2}{691200}+\frac{75011
   k}{19353600}-\frac{1651}{2903040} \ .\label{eq:index theorem 30}
\end{align}
\end{subequations}
We obtained these results by using the Sage program \texttt{admcycles} \cite{admcycles}.

\subsection{The case of genus 2}
We want to illustrate the treatment so far with the case of $\bM_{2}$. Our task was the determination of the number of sections of $\mathscr{L}^k$. In the case of genus 2 surfaces, this can be done explicitly and the number of sections is most conveniently packed into a Hilbert series:
\begin{align}
P(t)&=\sum_{k=0}^\infty t^k \dim \H^0(\bM_{2},\mathscr{L}^k) \\
&=\frac{1}{1-t}\left(\frac{1}{(1-t^2)(1-t^3)(1-t^{5})}-\frac{t}{(1-t^4)(1-t^6)(1-t^{12})}\right)\ . \label{eq:genus 2 Hilbert series}
\end{align}
This result is derived in Appendix~\ref{app:genus 2} and we copied the main result here.

Let us discuss the features of this formula. We can extract a specific dimension $\dim \H^0(\bM_{2},\mathscr{L}^k)$ from this formula by computing the corresponding contour integral $\oint_0 \mathrm{d}t\ t^{-k-1} P(t)$. The contour integral can then be rewritten as a contour integral around various roots of unity. Thus, we get
\be 
\dim \H^0(\bM_{2},\mathscr{L}^k)=-\sum_{\text{roots of unity }t_*} \Res_{t=t_*} \ t^{-k-1} P(t)\ . \label{eq:dimension roots of unity}
\ee
It is now important to note that roots of unity other than 1 give oscillating contributions to the answer. We also note that a pole of order $p$ gives a contribution of order $\mathcal{O}(k^{p-1})$ for large $k$. Thus the most dominant contribution in the large $k$ limit comes from the pole at $t=1$, since this is the only fourth order pole of $P(t)$. This gives the explicit asymptotic formula
\be 
\dim \mathcal{H}_k \sim \left(\frac{1}{180}-\frac{1}{1728}\right)k^3= \frac{43k^3}{8640}\ .
\ee
On the other hand, the index theorem predicts \eqref{eq:index theorem 20}.
Thus the index theorem predicts asymptotically for $k \to \infty$ exactly half as many sections as there are in reality. This is actually expected in this case, since any genus 2 surface has a $\ZZ_2$ automorphism.\footnote{This classical fact follows directly from the fact that $f(z)=\frac{\omega_1(z)}{\omega_2(z)}$ with $\omega_i(z)$ the two holomorphic differentials gives a degree 2 map to the Riemann sphere. A surface of this form is called hyperelliptic and the $\ZZ_2$ automorphism interchanges the two sheets of the degree 2 map.} Because a genus 2 surface has a non-trivial automorphism group, there is a correction of order $\mathcal{O}(k^3)$ to the index theorem. See Appendix~\ref{app:Kawasaki index theorem} for a precise explanation for this. The correction is actually the same as the original term, up to a possible phase $(-1)^k$. In this case, the phase is absent and we get a result twice as high than predicted from the naive index theorem. The fact that the phase is unity is explained in footnote \ref{footnote:hyperelliptic correction phase} below.
In the case of modular forms on $\bM_{1,1}$, we have the same generic $\ZZ_2$ automorphism. There the factor $(-1)^k$ is present, which leads to the asymptotic dimension formula $\frac{k}{24}(1+(-1)^k)$, whereas the naive index theorem would predict the asymptotic dimension $\frac{k}{24}$.

Let us also remark that when we compute the dimension from the Hilbert series as in \eqref{eq:genus 2 Hilbert series}, we get various oscillating terms. They have a common period that is given by the least common multiple of all the roots of unity that appear. In this case, the period is
\be 
\lcm(2,3,5,4,6,12)=60\ . \label{eq:genus 2 period}
\ee
Thus we could for example write a polynomial formula for $\dim \mathcal{H}_k$ if $k$ is a multiple of 60 given by
\be 
Z_2=\dim \mathcal{H}_k =\frac{43 k^3+642 k^2+2808 k+8640}{8640}\quad\text{for}\quad k \equiv 0 \bmod 60\ .
\ee
\subsection{Corrections to the naive index theorem} \label{subsec:corrections}
Let us mention a couple of observations on the true behaviour of $Z_g=\dim \mathcal{H}_k$ compared to the fake dimension $\Zfake_g$ that is computed by the naive index theorem. First of all, it might seem up to this point that the use of the index theorem is essentially unnecessary, since it only gives us partial information about the number of sections of $\mathscr{L}^k$. However, the naive index theorem without accounting the corrections from Kawasaki's orbifold index theorem is still useful. 

First of all, we should underline that all stable curves of genus 0 have trivial automorphism group. Thus $\bM_{0,n}$ does not have orbifold singularities and the naive index theorem gives the correct answer for $g=0$.

It also follows from the form of Kawasaki's index theorem that corrections to the naive index generically have a phase factor of the form $\mathrm{e}^{2\pi i r k}$ for $r$ a rational number. This is because corrections are associated to non-trivial automorphism groups of surfaces and are roughly expressed as integrals over the locus in $\bM_{g,n}$ with a prescribed automorphism group.
This phase is the eigenvalue is the action of the automorphism group on sections of $\mathscr{L}^k$ on the locus with given automorphism group. We should mention that $r$ can be zero and there can be non-trivial non-oscillatory corrections to the naive index theorem. For example, the locus of hyperelliptic surfaces in $\bM_g$ gives rise to a non-oscillatory correction.\footnote{To see this, notice that the normal bundle is $g-2$-dimensional. Hence the $\ZZ_2$-automorphism acts as $-1$ on $g-2$ of the $3g-3$ quadratic differentials. The $g$ holomorphic differentials are all odd under the hyperelliptic involution, since they can be written as $\omega_j(z)=\frac{z^{j-1}\, \mathrm{d}z}{y}$, where $y$ is defined as in eq.~\eqref{eq:hyperelliptic surface}. Under the hyperelliptic involution $y \mapsto -y$ and $z$ stays invariant. Thus all the holomorphic differentials are odd. In total, this means that the action on $\mathscr{L}^k=(\det \EE^{(2)})^k \otimes (\det \EE)^{-k}$ is $(-1)^{k(g-2-g)}=1$ for any $k$ and thus the contribution is not oscillatory. \label{footnote:hyperelliptic correction phase}}

It also follows from simple dimension considerations that a locus of enhanced automorphism symmetries of dimension $N$ leads to a contribution of order $\mathcal{O}(k^N)$ to the index theorem. In $\bM_{g,n}$, the locus of smallest codimension with non-trivial automorphism is given by the divisor $\mathscr{D}_{1,\emptyset}$ as defined in Section~\ref{subsec:bundles on Mgn}, i.e.\ the locus where a genus 1 surface without punctures is connected to the rest of the surface. We can evaluate the leading correction to the index theorem coming from this locus. The details are explained in Appendix~\ref{app:Kawasaki index theorem} and the result is
\be 
(-1)^k \int_{\bM_{g-1,n+1} \times \bM_{1,1}} \frac{\td(\bM_{g-1,n+1} \times \bM_{1,1}) \, \mathrm{e}^{k \kappa_1}}{1+\mathrm{e}^{\psileft+\psiright}}\ , \label{eq:leading correction index theorem}
\ee
where $\psileft$ and $\psiright$ are the two $\psi$-classes associated to the node.

Let us finally comment on the general behaviour of the `recurrence time' for large genus, i.e.\ the smallest common period for all the oscillatory contributions in $k$. As we have mentioned above, this `recurrence time' is 60 for $\bM_{2}$, whereas it is well-known to be $12$ in the case of $\bM_{1,1}$.
Hurwitz' theorem on automorphism groups of Riemann surfaces states that the order of the automorphism group is bounded by $84(g-1)$. A very rough upper bound on the behaviour of the recurrence time is thus
\be 
T(g)\le \lcm(1,2,\dots,84(g-1))\ .
\ee
One could of course greatly improve this, for example the values in the interval $[40(g-1)+1,84(g-1)-1]$ can never be attained. On the other hand, one can easily see that the values $\{1,2,\dots,2g+2\}$ are always attained, since for the hyperelliptic surface
\be 
y^2=\prod_{i=1}^{2g+2} (z-\lambda_i)
\ee
we can choose $\lambda_j=\mathrm{e}^{\frac{2\pi i j}{n}}$ for $j=1,\dots,n$ and send the remaining $\lambda_j$ to 0 at the same speed. The resulting surface is in general nodal and has a cyclic automorphism group of order $n$.\footnote{Actually for $n$ odd the automorphism group is or order $2n$ since the rotation around the origin has to be combined with the hyperelliptic involution.}
Thus there are in any case always of the order $g$ many automorphisms.
It is a direct consequence of the prime number theorem that
\be 
\log \lcm(1,2,\dots,84(g-1)) \sim 84g
\ee
asymptotically. Thus we expect that asymptotically, the recurrence time behaves as
\be 
\log T(g) \sim Ag
\ee
for an exponent $2 \le A \le 84$. Thus we conclude that the recurrence time will grow exponentially as $g \to \infty$.
\section{Adding boundaries} \label{sec:adding boundaries}
In this section, we explain how to extend the results of the previous Section to the case where the Riemann surface $\Sigma$ is allowed to have asymptotic boundaries. In this case, we want to repeat the story in an equivariant setting that keeps track of the $\U(1)$-actions associated to the boundaries of moduli space. The simplest case is given when $\Sigma$ is the hyperbolic disk, where the relevant moduli space is $\Diff(\S^1)/\PSL(2,\RR)$.

\subsection{Equivariant index theorem and localization}
Since the phase spaces are infinite-dimensional, there is no reasonable way to evaluate the integral of the index theorem. Instead, the right thing to do in this circumstance was discussed in \cite{Stanford:2017thb}. One notices $\Diff(\S^1)/\PSL(2,\RR)$ and the corresponding generalizations to arbitrary boundaries and punctures carry a $\U(1)$ action (or actually a $\U(1)$ action for each boundary) that rotate the asymptotic boundary. It is then natural to consider the equivariant version of the index theorem.

Physically, it is also sensible that we do not just want to compute the number of states in this setting. We expect that the 3d gravity theory under consideration will be dual to a 2d CFT (when suitably interpreted) and this computation should compute the contribution of the background $\Sigma \times \S^1$ to the boundary partition function. The boundary partition function can be written as $\tr (\mathrm{e}^{-\sum_i \beta_i H_i})$, where $H_i$ is the boundary Hamiltonian on the $i$-th boundary.
In the setting of equivariant cohomology, this is precisely the definition of the equivariant Euler characteristic, see e.g.~\cite{AtiyahSegalII}. For this, one should note that the boundary dynamics is entirely right-moving so that the rotation operator is in fact equivalent to the boundary Hamiltonian.
$\beta$ is then identified with the equivariant parameter.

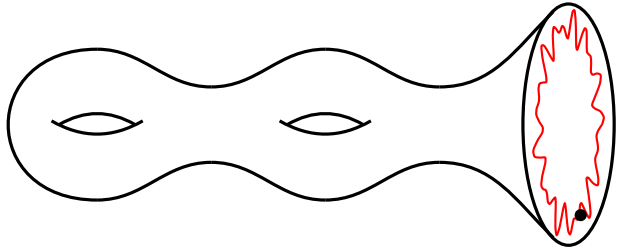
\begin{figure}
\begin{center}
\begin{tikzpicture}
\draw[very thick,out=0, in=180] (-1.5,.5) to (0,1);
\draw[very thick,out=0, in=180] (0,1) to (1.5,.5);
\draw[very thick,in=180,out=0] (-3,1) to (-1.5,.5);
\draw[very thick,in=180,out=0] (-3,-1) to (-1.5,-.5);
\draw[very thick,in=180,out=180, looseness=2] (-3,-1) to (-3,1);
\draw[very thick,out=0,in=225] (1.5,.5) to (3,1.5);
\draw[very thick,out=0, in=180] (-1.5,-.5) to (0,-1);
\draw[very thick,out=0, in=180] (0,-1) to (1.5,-.5);
\draw[very thick,out=0,in=-225] (1.5,-.5) to (3,-1.5);
\draw[very thick, bend right=30] (-.6,.05) to (.6,.05); 
\draw[very thick, bend left=30] (-.5,0) to (.5,0); 
\draw[very thick, bend right=30] (-3.6,.05) to (-2.4,.05); 
\draw[very thick, bend left=30] (-3.5,0) to (-2.5,0); 
\draw[very thick] (3.2,0) circle (.6 and 1.6);
\draw[smooth, thick, samples=100,domain=0:360, red] plot({3.2+0.4*cos(\x)*(1+0.1*sin(19*\x)+0.1*cos(13*\x))}, {1.3*sin(\x)*(1+0.1*sin(19*\x)+0.1*cos(13*\x))});
\fill (3.36,-1.2) circle (.08);
\end{tikzpicture}
\end{center}
\caption{A typical surface in $\mathscr{M}_2^{(1)}$. The red wiggly surface is schematically the cutoff surface and the black denotes the marked point on it.} \label{fig:cutoff surface}
\end{figure}

To fix notation, let us denote the relevant moduli spaces by $\overline{\mathscr{M}}^{(n)}_{g,m}$, where the superscript $(n)$ stands for the number of asymptotic boundaries and we add $m$ additional punctures. If $m=0$, we write $\overline{\mathscr{M}}^{(n)}_{g}$. These moduli spaces consist of all nodal stable Riemann surfaces with $n$ asymptotic boundaries.\footnote{Contrary to what the notation might suggest, these moduli spaces are actually not compact (which is fairly clear in the simplest example $\mathscr{M}_0^{(1)}=\overline{\mathscr{M}}_0^{(1)}=\Diff(\S^1)/\PSL(2,\RR)$. However, we will see that equivariant localization reduces everything to an integral over $\bM_{g,n}$.}  The simplest case to keep in mind is $\overline{\mathscr{M}}_0^{(1)}=\Diff(\S^1)/\PSL(2,\RR)$. 
For an asymptotic boundary, we imagine that the surface locally looks like the disk which is described by the infinite-dimensional moduli space $\Diff(\S^1)/\PSL(2,\RR)$. One should think of these infinitely many moduli as a gluing of the surface to its boundary. One useful way to think about this is inspired from the treatments of JT gravity \cite{Maldacena:2016upp, Stanford:2017thb} as follows. We can give the surface a hyperbolic metric so that it looks asymptotically like a hyperbolic cylinder. One then cuts off the boundary very far out. This cutoff is specified by a function $\Diff(\S^1)$, as described in detail in \cite{Maldacena:2016upp} for the case of the disk. Thus we think of a surface pictorially as surface whose asymptotic boundary is cut off in some irregular way, as illustrated in Figure~\ref{fig:cutoff surface}. To make this pictorial way of thinking about the moduli space correct, we also need to mark one point on the cutoff surface, i.e.\ we have to remember which point on the boundary corresponds to the origin. Indeed, this is part of the data of specifying a diffeomorphism $\S^1 \to \S^1$.

The equivariant index theorem says now that the equivariant index can be computed as an integral
\be 
\int_{\overline{\mathscr{M}}^{(n)}_{g}} \td(\overline{\mathscr{M}}^{(n)}_{g}) \, \mathrm{e}^{k \kappa_{1}}\ , \label{eq:equivariant integral}
\ee
where all the ingredients are interpreted equivariantly. This means in particular that the Todd class entering here is really the equivariant Todd class and $\kappa_1$ receives an equivariant completion, $\kappa_1=(\kappa_1)_1+(\kappa_1)_0$, where $(\kappa_1)_0$ is a 0-form that we discuss below. 
 The reader does not need to know the details of equivariant cohomology to follow the discussion. One only needs to know that an equivariantly closed $\alpha$ form has admixtures of various different degrees less than the original degree of $\alpha$, that we denote by $(\alpha)_n$ or $\alpha_n$ if there is no risk of confusion. This is the Cartan model of equivariant cohomology.
 The same caveat as in the case without boundaries applies. Since $\overline{\mathscr{M}}_g^{(n)}$ is actually an orbifold for $g \ge 1$, we should use an equivariant version of Kawasaki's index theorem described in Appendix~\ref{app:Kawasaki index theorem}. The equivariant integral \eqref{eq:equivariant integral} computes the fake partition function $\Zfake_g(\beta_1,\dots,\beta_n)$. We will discuss in Section~\ref{subsec:alternative derivation} how to in principle compute the exact partition function $Z_g(\beta_1,\dots,\beta_n)$. This is much more complicated and we will do so only in simple examples.

We can apply localization of equivariant cohomology to compute equivariant integrals over $\mathscr{M}^{(n)}_{g,m}$, which are then interpreted as the partition functions of chiral 3d gravity on the corresponding surface with asymptotic boundary. Equivariant localization reduces the integral to an ordinary integral over the fixed-point locus of the $\U(1)^n$ action.

To understand the nature of the fixed point set, let's first consider the simple case of the disk partition function, where the moduli space is $\mathscr{M}_0^{(1)}=\overline{\mathscr{M}}_0^{(1)}=\Diff(\S^1)/\PSL(2,\RR)$. As described above, we think of such surfaces as a disk with the asymptotic boundary removed. The $\U(1)$ action simply acts via rotations of this cutoff disk. So clearly the only fixed point of the $\U(1)$ action is the disk with a round cutoff. Thus the equivariant integral localizes to a single point. We treat this simple case in Section~\ref{subsec:disk partition function}. 

Let us next discuss the case of $\overline{\mathscr{M}}_{0,m}^{(1)}$, since it can still be easily described. This moduli space described the same disk with an asymptotic wiggly cutoff as above, but with the choice of $n$ additional points on the disk. The $\U(1)$-action again acts by a rotation of the disk. Thus it is clear that the only fixed surfaces are surfaces where all punctures are at the center of the disk and we take a round cut off. Since we are considering the compactification of $\overline{\mathscr{M}}_{0,m}^{(1)}$, the corresponding surface is thus actually a nodal surface, where instead of allowing all marked points to coincide at the origin, we blow this region up and get a sphere attached to the disk. The fixed point set itself is isomorphic to $\bM_{0,m+1}$, where one of the punctures is the node.

\begin{figure}
\begin{center}
\begin{tikzpicture}
\draw[very thick, out=0, in=240] (-3,0) to (1.5,2.6);
\draw[very thick, out=0, in=120] (-3,0) to (1.5,-2.6);
\draw[very thick, out=240, in=120] (1.5,2.6) to (1.5,-2.6);
\draw[very thick, bend right=30] (-.7,.05) to (.5,.05); 
\draw[very thick, bend left=30] (-.6,0) to (.4,0); 
\draw[very thick, out=310, in=180] (-5.5,1.5) to (-3,0);
\draw[very thick, out=50, in=180] (-5.5,-1.5) to (-3,0);
\draw[very thick] (-5.6,0) circle (.5 and 1.5);
\draw[thick, red] (-5.6,0) circle (.4 and 1.2);
\draw[very thick, bend left=40,<-] (-6.1,1.8) to node[above] {$\U(1)_1$} (-5.1,1.8);
\draw[very thick, out=310, in=180, rotate=120] (-5.5,1.5) to (-3,0);
\draw[very thick, out=50, in=180, rotate=120] (-5.5,-1.5) to (-3,0);
\draw[very thick, rotate=120] (-5.6,0) circle (.5 and 1.5);
\draw[thick, red, rotate=120] (-5.6,0) circle (.4 and 1.2);
\draw[very thick, out=310, in=180, rotate=240] (-5.5,1.5) to (-3,0);
\draw[very thick, out=50, in=180, rotate=240] (-5.5,-1.5) to (-3,0);
\draw[very thick, rotate=240] (-5.6,0) circle (.5 and 1.5);
\draw[thick, red, rotate=240] (-5.6,0) circle (.4 and 1.2);
\draw[very thick, bend left=40,<-] (4.4,4.7) to node[right] {$\U(1)_2$} (4.4,3.7);
\draw[very thick, bend right=40,->] (4.4,-4.7) to node[right] {$\U(1)_3$} (4.4,-3.7);
\draw  (-1.1,-0.1) node[cross out, draw=black, very thick, minimum size=5pt, inner sep=0pt, outer sep=0pt] {};
\draw  (.3,.6) node[cross out, draw=black, very thick, minimum size=5pt, inner sep=0pt, outer sep=0pt] {};
\end{tikzpicture}
\end{center}
\caption{The fixed point set in $\overline{\mathscr{M}}_{g,m}^{(n)}$ is given by $\bM_{g,n+m}$. Here we drew a $g=1$ surface with three asymptotic boundaries and two marked points. In order for the surface to be invariant under the rotations of the three boundaries, it has to be nodal and the boundary cut offs (that we drew in red) are round. Thus the fixed point set is naturally isomorphic to $\bM_{1,5}$ where two of the punctures are the punctures that we denoted by a cross in the figure and three punctures correspond to the nodes in the drawing.} \label{fig:fixed point set Mgn}
\end{figure}
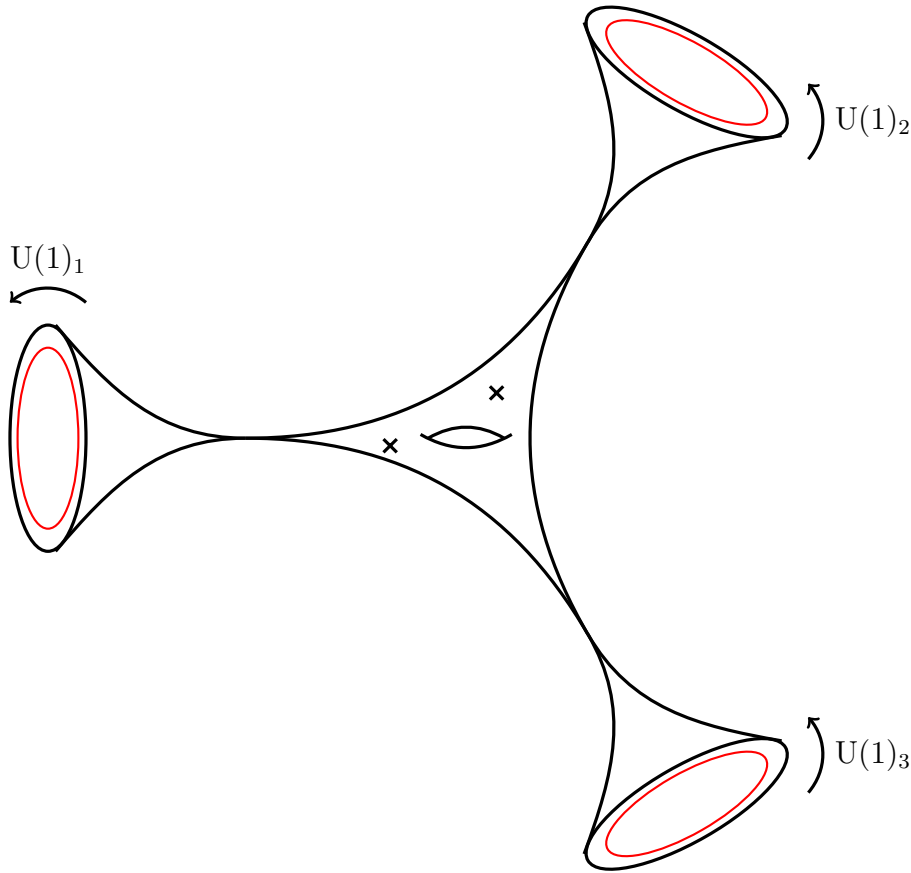

The same reasoning carries over to arbitrary $\overline{\mathscr{M}}_{g,m}^{(n)}$. The fixed point set consists of surfaces with $n+m$ punctures and to the $n$ punctures disks are attached at a single node. The disk themselves have a round asymptotic cut off. In particular, the fixed point set is isomorphic to $\bM_{g,m+n}$. We drew an example of a fixed surface in Figure~\ref{fig:fixed point set Mgn}. All surfaces in the fixed point set are of this form.

This means that the equivariant localization theorem reduces equivariant integrals over the infinite-dimensional moduli spaces $\overline{\mathscr{M}}^{(n)}_{g}$ to finite-dimensional integrals over the usual moduli spaces $\bM_{g,n}$, i.e.\
\be 
\int_{\overline{\mathscr{M}}^{(n)}_{g}} \td(\overline{\mathscr{M}}^{(n)}_{g})\, \mathrm{e}^{k \kappa_1}=\int_{\bM_{g,n}} \alpha \label{eq:equivariant localization 1}
\ee
for some equivariant form $\alpha$ (depending on the equivariant parameters) on moduli space. Our main task in this section is to figure out the form $\alpha$. This reduces the problem of computing partition functions with boundaries back to the integrals that we studied in the previous section. It will turn out that one can naturally express the equivariant form $\alpha$ in terms of ordinary forms. The final result for $\alpha$ is given in eq.~\eqref{eq:multiboundary correlator}.
Abstractly, $\alpha$ is given by
\be 
\alpha = \frac{\mathrm{e}^{k \kappa_{1}} \td(T\overline{\mathscr{M}}^{(n)}_{g,n})}{e(\mathcal{N})}\ ,  \label{eq:equivariant localization 2}
\ee
where $\mathcal{N}$ is the normal bundle to the fixed point set and $e$ is the equivariant Euler class. The division can always formally be performed, since $e(\mathcal{N})_0 \ne 0$ on a fixed point of the action.

\subsection{Disk partition function} \label{subsec:disk partition function}
We start by applying the localization formula \eqref{eq:equivariant localization 1} and \eqref{eq:equivariant localization 2} to the disk partition function, where everything is very explicit. This computation is very similar to the computation explained in \cite{Stanford:2017thb}. Since $\overline{\mathscr{M}}_0^{(1)}$ is a Virasoro coadjoint orbit, our discussion is equivalent to their quantization \cite{Witten:1987ty}.

As already explained above, the fixed point set here is just a single point. In this case the equivariant Euler class has a simple description. Let $\Sigma_0$ be the fixed point surface. Then there is a well-defined $\U(1)$-action on the normal space $\mathcal{N}=T_{\Sigma_0} \overline{\mathscr{M}}_0^{(1)}$. The action on $\mathcal{N}$ decomposes into irreducible representations of charge $\bigoplus_n (m_n) \oplus (-m_n)$. This defines the $m_n$ up to sign. The overall sign $\prod_n m_n$ naturally follows from the orientation of the tangent space. Then $\prod_n (\beta m_n)$ is the restriction of the equivariant Euler class of the normal bundle to $\Sigma_0$.

For computations, we identify $\S^1 \cong [0,2\pi)$. A diffeomorphism $\varphi:\S^1 \to \S^1$ gives then rise to an element in $\Diff(\S^1)/\PSL(2,\RR)$ where $\PSL(2,\RR)$ acts by 
\be 
f=\tan\left(\frac{\varphi}{2}\right) \longmapsto \frac{a f+b}{c f+d}\ .
\ee
By definition, diffeomorphisms $\varphi(\tau)$ in $\Diff(\S^1)$ satisfy $\varphi'(\tau)>0$ everywhere.
The $\U(1)$ acts by `time' translations $\varphi(\tau) \mapsto \varphi(\tau-\theta)$.
A convenient way to fix the gauge invariance is then to require that $\varphi$ satisfies
\be 
\varphi(0)=0\ , \qquad \varphi'(0)=1\ , \qquad \varphi''(0)=0\ .
\ee
The unique fixed point in the $\U(1)$ action on $\Diff(\S^1)/\PSL(2,\RR)$ is the equivalence class of the identity diffeomorphism. Geometrically this corresponds to the disk with round cutoff that we discussed above.

Let us work out the indices $m_n$ that are necessary for the determination of the equivariant Euler class. We can consider small perturbations of $\varphi$ of the form $\delta_n^{(1)} \varphi(\tau)=\cos(n \tau)$ and $
\delta_n^{(2)} \varphi(\tau)=\sin(n \tau)$. By applying a $\PSL(2,\RR)$ gauge transformation, one can achieve that these preserve the gauging $\varphi(0)=0$, $\varphi'(0)=1$ and $\varphi''(0)=0$, which gives the following corrected deformations
\begin{align}
\delta_n^{(1)} \varphi(\tau)&=\cos(n \tau)-1-n^2(\cos(\tau)-1)\ , \\
\delta_n^{(2)} \varphi(\tau)&=\sin(n \tau)-n \sin(\tau)\ .
\end{align}
So clearly, the deformation is only non-trivial for $n \ge 2$, since for $n=1$ the deformation is part of the $\PSL(2,\RR)$ gauge freedom. We continue with the non-gauge fixed forms of the deformations. Under $\tau \to \tau+\theta$, $\delta_n^{(1)} \varphi$ and $\delta_n^{(2)} \varphi$ transform in the $\U(1)$ representation of charge $(n) \oplus (-n)$. We thus conclude that the indices $m_n$ take the form $m_n=n$ for $n \ge 2$.

We can similarly work out the restriction of the equivariant Todd class.  By definition of the equivariant Euler class, the combinations $\beta m_n$ correspond in fact to the the restrictions of the equivariant Chern roots to the fixed point. Thus we have from the definition of the Todd class,
\be 
\td(\overline{\mathscr{M}}_0^{(1)})_0=\prod_{n=2}^\infty \frac{\beta m_n}{1-\mathrm{e}^{-\beta m_n}}=\prod_{n=2}^\infty \frac{n \beta}{1-\mathrm{e}^{-n \beta}}\ .
\ee

It remains to work out the restriction of the equivariant class $\kappa_1$. For this, we are going to use its definition in terms of the pushforward of the equivariant $\psi$-class on the moduli space of a disk together with one marked point, which is given by $\overline{\mathscr{M}}^{(1)}_1 \equiv \Diff(\S^1)/\S^1$. By definition, the $\psi$-class is the first Chern class of the cotangent bundle at the marked point. The $\U(1)$-action on the tangent space has index $1$, since this corresponds precisely to the missing index $m_1=1$ above. This means that the index on the cotangent space is $-1$ and consequently $\psi$ restricts to $-\beta$. 
We thus have
\be 
\int_{\overline{\mathscr{M}}_0^{(1)}} \kappa_1= \int_{\overline{\mathscr{M}}_{0,1}^{(1)}}  \psi^2 =\frac{(-\beta)^2}{\prod_{n=1}^\infty (n \beta)}=\frac{\beta}{\prod_{n=2}^\infty (n \beta)}\ ,
\ee
which means that the restriction of $\kappa_1$ to the fixed point is $\beta$. Here we used again that the Euler class of the normal bundle on $\Diff(\S^1)/\S^1$ includes the missing index $m_1$.

We can finally assemble the localization formula \eqref{eq:equivariant localization 1} and \eqref{eq:equivariant localization 2} and get
\be 
Z_0(\beta)=\Zfake_0(\beta)= \int_{\overline{\mathscr{M}}_0^{(1)}}  \td(\overline{\mathscr{M}}_0^{(1)}) \, \mathrm{e}^{k \kappa_{1}} =\mathrm{e}^{k \beta}\prod_{n=2}^\infty \frac{1}{1-\mathrm{e}^{-n \beta}}\ .
\ee
We see that the Euler class cancels partially with the Todd class and the infinite product is absolutely converging for $\beta>0$. In this case, the fake and true partition functions agree because of the absence of orbifold singularities.
\subsection{The vacuum Virasoro character} \label{subsec:vacuum Virasoro character}
We have obtained the disk partition function
\be 
Z_0(\beta)=\Zfake_0(\beta)= \mathrm{e}^{k \beta}\prod_{n=2}^\infty \frac{1}{1-\mathrm{e}^{-n \beta}}\ ,
\ee
which agrees with the Virasoro vacuum character of central charge
\be 
c=24k\ .
\ee
Physically, this result is not surprising. We considered a chiral version of gravity that is expected to be dual to a chiral two-dimensional CFT (or perhaps an ensemble thereof). The boundary of a disk times a circle is a torus and thus our computation is expected to give the vacuum contribution to the torus partition function, which has to take the universal form we computed. Of course, this result has been obtained previously using different formalisms, see e.g.\ \cite{Yin:2007gv, Giombi:2008vd, Chen:2015uga, Cotler:2018zff}. 

The relation $c=24k$ is also not surprising. The chiral gravity theory under consideration is only consistent for $k \in \ZZ$. Correspondingly, chiral CFTs suffer from a gravitational anomaly unless $c \in 24\ZZ$. Absence of anomalies on both sides thus essentially force this relation. Considering the relation $k=\frac{\ell}{16\pi G }$ explained in Section~\ref{subsec:Gravity and Chern Simons} of $k$ with the AdS radius $\ell$ and Newton's constant, this relation is nothing else than the expected Brown-Henneaux central charge \cite{Brown:1986nw}.

We should note that for pure gravity, it is often claimed that there is a one-loop correction of $+13$ to the central charge, see e.g.\ \cite{Cotler:2018zff}. Such a one-loop correction is clearly inconsistent in the holomorphic setting. We can see where such a correction would come from the present formalism. We would obtain it by including the metaplectic correction that is mentioned in footnote~\ref{footnote:metaplectic correction} in our quantization scheme. With metaplectic correction, we would need to look at sections of the line bundle  $\mathscr{L}^k \otimes \sqrt{\mathscr{K}}$, where $\sqrt{\mathscr{K}}$ is a choice of square root of the canonical bundle. As we have mentioned already in footnote~\ref{footnote:metaplectic correction}, such a square root generically does not exist on moduli space, which signals inconsistency of this quantization scheme. However, in the case of the universal moduli space $\Diff(\S^1)/\mathrm{PSL}(2,\RR)$, it does exist and we would need to compute the integral
\be 
\int_{\overline{\mathscr{M}}^{(1)}_0}  \td(\overline{\mathscr{M}}^{(1)}_0) \, \mathrm{e}^{k \kappa_1+\frac{1}{2}c_1(\mathscr{K})} =\mathrm{e}^{(k+\frac{13}{24}) \beta}\prod_{n=2}^\infty \frac{1}{1-\mathrm{e}^{-n \beta}}\ ,
\ee
where we used that eq.~\eqref{eq:first Chern class canonical bundle} simplifies to $c_1(\mathscr{K})=\frac{13}{12}\kappa_1$ in this case, since there are neither $\psi$-classes nor boundary classes for $\overline{\mathscr{M}}^{(1)}_0$.

\subsection{The general case}\label{subsec:general case}
After having evaluated the equivariant localization in the case of the disk, we now explain the general case. To simplify the discussion, we assume in the following that there is only one asymptotic boundary.

$e(\mathcal{N})$ is the equivariant Euler class of the normal bundle of the fixed point set. All forms appearing here are understood to be equivariant. The main work involved is to obtain a good understanding of the normal bundle. To achieve this, we proceed in two steps. 

As a first step, we recall some standard facts about the normal bundle at a generic boundary divisor $\mathscr{D}$ in $\bM_{g,n}$ which directly generalize to $\overline{\mathscr{M}}^{(n)}_{g,m}$. Let us denote the normal bundle by $\mathcal{N}_1$ in order to distinguish it from the normal bundle appearing in the localization formula. 
Near the degeneration, the surface can locally be described as
\be 
(x-a)(y-b)=q \label{eq:degeneration}
\ee
for $q>0$. For $q=0$, the surface degenerates and $x=a$ and $y=b$ describes the two branches of the nodal surfaces. Thus $q$ is the modulus that describes the normal direction in $\bM_{g,n}$ to the divisor $\mathscr{D}$. All other moduli do not play a role in the following discussion. Since the discussion is entirely local, we should imagine that all of $x-a$, $y-b$ and $q$ are very small in this discussion. We can thus to first order replace $x\to \mathrm{d}x$ (at $a$), $y \to\mathrm{d}y$ (at $b$) and $q\to \mathrm{d}q$. We thus see that $\mathrm{d}q$ behaves like $\mathrm{d}x \otimes \mathrm{d}y$. Since
$\mathrm{d}q$ is locally a section of the conormal bundle $\mathcal{N}_1^*$, we learn that
\be 
\mathcal{N}_1^* \cong \LLleft \otimes \LLright\ , \label{eq:conormal bundle divisor}
\ee
where $\LLleft$ and $\LLright$ are the standard cotangent bundles at the punctures $a$ and $b$. $\mathscr{D}$ is always isomorphic to a finite quotient of products of moduli spaces, and our notation of $\LLleft$ and $\LLright$ is consistent with the earlier notation $\LL_i$ for these line bundles.

We now determine in a second step the normal bundle $\mathcal{N}_2$ of the fixed point set $\bM_{g,n+1}$ inside the boundary divisor $\mathscr{D}$ of $\mathscr{M}_{g,n}^{(1)}$. $\mathscr{D}$ is the boundary divisor where the asymptotic boundary is pinched off as in Figure~\ref{fig:fixed point set Mgn}. This is almost trivial. Since 
\be 
\mathscr{D} \cong \bM_{g,n+1} \times \mathscr{M}_{0,1}^{(1)} \equiv \bM_{g,n+1} \times \Diff(\S^1)/\S^1
\ee
naturally factorizes into a finite-dimensional part and the moduli space containing only the one node, we have for the normal bundle of the fixed point set inside $\mathscr{D}$
\be 
\mathcal{N}_2=T_{\Sigma_0} \mathscr{M}_{0,1}^{(1)}\ ,
\ee
where $\Sigma_0$ is the disk with a marked point at the center and a round cutoff. For the full normal bundle of the fixed point set inside $\mathscr{M}_{g,n}^{(1)}$, 
this implies in particular that
\begin{subequations}
\begin{align} 
e(\mathcal{N})&=e(\mathcal{N}_1) e(\mathcal{N}_2)\ , \\
\td(T \overline{\mathscr{M}}_{g,n}^{(1)}) &=\td(T \bM_{g,n+1}) \td(\mathcal{N}_1) \td(\mathcal{N}_2)\ .
\end{align}
\end{subequations}
We can now work on each factor in turn.

Since the divisor $\mathscr{D}$ is a cartesian product, all characteristic classes of $\mathcal{N}_2$ reduce to 0-forms when restricted to the fixed point set and are thus universal functions of $\beta$. We will now see that  $\mathcal{N}_2$ supplies the necessary Virasoro characters to the partition function. This function is computed as in the case of the disk and is given again by the infinite product over all the indices $m_n$. The only difference is that the gauge invariance is only given by $\U(1)$ instead of $\PSL(2,\RR)$. This has the effect that the index $m_1=1$ survives now.  Thus as in Section~\ref{subsec:disk partition function}, we have
\be 
\frac{\td(\mathcal{N}_2)}{e(\mathcal{N}_2)} =\prod_{n=1}^\infty \frac{m_n \beta}{m_n \beta(1-\mathrm{e}^{-m_n \beta})}=\prod_{n=1}^\infty \frac{1}{1-\mathrm{e}^{-n \beta}}
\ee
From the discussion so far, $\alpha$ in \eqref{eq:equivariant localization 1} is hence given by
\be 
\alpha=\frac{\mathrm{e}^{k \kappa_{1}}}{\prod_{n=1}^\infty (1-\mathrm{e}^{-n \beta}) } \,    \td(T\bM_{g,n+1})\,  \frac{\td(\LLleft^{-1} \otimes \LLright^{-1})}{e(\LLleft^{-1} \otimes \LLright^{-1})}\ ,
\ee
where all forms are still understood equivariantly.

Let's discuss next the last factor. Define $\psileft=c_1(\LLleft)$ and $\psiright=c_1(\LLright)$ as the equivariant $\psi$-classes of the two nodes connecting the bulk of the surface and the attached asymptotic disks. We assume that $\psileft$ corresponds to the node on $\bM_{g,n+1}$, whereas $\psiright$ corresponds to the node on the attached disk. Then from the definition of the Euler and the Todd class, we have
\be 
\frac{\td(\LLleft^{-1} \otimes \LLright^{-1})}{e(\LLleft^{-1} \otimes \LLright^{-1})}=\frac{1}{1-\mathrm{e}^{\psileft+\psiright}}\ . \label{eq:equivariant todd class Euler class quotient}
\ee
It remains to figure out what $\psileft$ and $\psiright$  restrict to. Clearly, $\psiright$ does not have 
 support on the integration over $\bM_{g,n+1}$ and has to restrict to a number proportional to $\beta$. We actually worked this out already above when we discussed the disk partition function and found that it restricts to $-\beta$.
 
Finally, the restriction of $\psileft$ is very easy to see. The $\U(1)$-action is trivial on the cotangent bundle at $a$ and thus the equivariant cohomology coincides with the ordinary one. Thus $\psileft$ actually becomes the ordinary $\psi$-class on $\bM_{g,n+1}$ when restricted to the fixed point set. Hence on the fixed point set
 \be 
 \frac{\td(\LLleft^{-1} \otimes \LLright^{-1})}{e(\LLleft^{-1} \otimes \LLright^{-1})} = \frac{1}{1-\mathrm{e}^{-\beta+\psileft}}\ .
 \ee
Next, we need to work out the restriction of the equivariant $\kappa_1$ to the fixed point locus. $\kappa_1$ just restricts to the ordinary $\kappa_1$ on $\bM_{g,n+1}$ for essentially the same reason that the equivariant $\psi$-class restricted to the ordinary $\psi$-class $\psileft$ above. By definition $\kappa_1=\pi_*(\psi^2_{n+2})$, where $\pi:\bM_{g,n+2} \to \bM_{g,n+1}$ is the forgetful morphism. But since the equivariant $\psi_{n+2}$-class restricts to the ordinary $\psi_{n+2}$-class, the same follows also for $\kappa_1$. 
Finally, we need  the restriction of the Todd class of $\bM_{g,n+1}$. But this is again essentially trivial, since the $\U(1)$-action on $\bM_{g,n+1}$ is trivial. In this case, we can just forget about the fact that the Todd class was equivariant.

Thus we have
\begin{align} 
\Zfake_{g}(\beta)=\int_{\overline{\mathscr{M}}^{(1)}_{g}} \td(\overline{\mathscr{M}}_{g}^{(1)}) \mathrm{e}^{k \kappa_1} =\frac{1}{\prod_{m=1}^\infty (1-\mathrm{e}^{-m \beta}) } \int_{\bM_{g,1}} \frac{\mathrm{e}^{k \kappa_1} \, \td(\bM_{g,1})}{1-\mathrm{e}^{-\beta+\psileft}} \ .
\end{align}

It is now straightforward to generalize this to the case with multiple boundaries. The analysis of the equivariant localization can be carried out analogously. We now get several factors of the form \eqref{eq:equivariant todd class Euler class quotient}. The final result then becomes the simple formula 
\vspace{.2cm}
\begin{tcolorbox}[ams align]
Z_g(\beta_1,\dots,\beta_n)=\prod_{i=1}^n \frac{1}{\prod_{m=1}^\infty (1-\mathrm{e}^{-m \beta_i})}  \int_{\bM_{g,n}} \frac{\mathrm{e}^{k \kappa_1} \td(\bM_{g,n})}{\prod_{i=1}^n (1-\mathrm{e}^{-\beta_i+\psi_i})}  \ . \label{eq:multiboundary correlator}
\end{tcolorbox}
\noindent We renamed $\psileft$ as $\psi_i$, in accordance with the notation in Section~\ref{subsec:bundles on Mgn}.
We also note that we can go through the same steps as in Section~\ref{subsec:naive index theorem} to reexpress the Todd class in terms of other classes:
\begin{multline} 
\int_{\bM_{g,n}} \frac{\mathrm{e}^{k \kappa_1} \td(\bM_{g,n})}{\prod_{i=1}^n (1-\mathrm{e}^{-\beta_i+\psi_i})}
=\int_{\bM_{g,n}}  \prod_{i=1}^n \frac{\mathrm{e}^{\frac{1}{24} \psi_i}}{1-\mathrm{e}^{-\beta_i+\psi_i}}  \\
\times \exp\left(\left(k-\frac{13}{24}\right) \kappa_1+\frac{11}{24}\, \Delta_1- \sum_{m=1}^\infty \frac{B_{2m}}{(2m)(2m)!}\left(\kappa_{2m}-\Delta_{2m}\right) \right) \ . \label{eq:equivariant partition function}
\end{multline}
The formula \eqref{eq:multiboundary correlator} covers almost all the cases of interest, except for $g=0$ and $n=1$ and $n=2$. The case of $n=1$ is the disk partition function that we discussed above. The case $n=2$ is also easy enough to deal with. In this case, the fixed point consists of a single point, namely two disks attached at a node. One computes the partition function using equivariant localization just as before. The only small change is the Euler and Todd class of $\LLleft^{-1} \otimes \LLright^{-1}$. Before the equivariant class $\psiright$ restricted $-\beta$ and $\psileft$ restricted to the ordinary $\psi$-class. In this case however, $\psiright$ restricts to $-\beta_2$ and $\psileft$ restricts to $-\beta_1$.  Thus one finds for the two-boundary wormhole
\be 
Z_0(\beta_1,\beta_2)=\frac{1}{(1-\mathrm{e}^{-\beta_1-\beta_2})\prod_{m=1}^\infty (1-\mathrm{e}^{-m \beta_1}) (1-\mathrm{e}^{-m \beta_2}) }\ .
\ee

\subsection{Alternative derivation} \label{subsec:alternative derivation}
We now present an alternative derivation of the same result. It is more elementary than the argument using equivariant localization and actually gives a formula for the full partition function including oscillatory terms in terms of quantities on $\bM_{g,n}$. However, we also wanted to present the argument using equivariant localization since it readily generalizes to other integrals of this type (such as the JT gravity partition function). Additionally, the argument we will present now is not directly applicable to the disk partition function.

One can already guess the result we will derive from \eqref{eq:multiboundary correlator}. Setting $q_i=\mathrm{e}^{-\beta_i}$, we can expand the result in $q_i$. The fake partition function of primary states (i.e.\ factoring out the infinite product prefactors) reads then
\be 
\Zfake_g^\text{p}(\beta_1,\dots,\beta_n)= \sum_{\ell_1,\dots,\ell_n=0}^\infty \prod_{i=1}^n q_i^{\ell_i} \int_{\bM_{g,n}} \mathrm{e}^{k \kappa_1+\sum_i \ell_i \psi_i} \, \td(\bM_{g,n})
\ee
The appearing integral in turn is the `naive' index theorem for the number of sections of the line bundle $\mathscr{L}^k  \otimes \bigotimes_i \LL_i^{\ell_i}$, where we recall that $\LL_i$ was defined to be the line bundle whose fiber consists of the cotangent space at the $i$-th marked point. One may hence suspect that
\be 
Z_g^\text{p}(\beta_1,\dots,\beta_n)=\sum_{\ell_1,\dots,\ell_n=0}^\infty \prod_{i=1}^n q_i^{\ell_i} \ \dim \H^0\left(\bM_{g,n},\mathscr{L}^k \otimes  \bigotimes_i \LL_i^{\ell_i}\right) \label{eq:primary partition function sections}
\ee
is true \emph{exactly}. In particular, the right hand side is a power series in $q_i$ whose coefficients are all integer. This is of course required for a consistent Hilbert space interpretation of the theory. This is what we shall now show.

We do this in two steps. First we reduce the problem to the computation of the partition function of primary states. This is achieved by integrating out the `boundary wiggles' as follows. 
We admit that there still is a way to compute the partition function as an integral over $\overline{\mathscr{M}}_g^{(n)}$. Formally, the integrand is given by the integrand of the equivariant Kawasaki index theorem described in Appendix~\ref{app:Kawasaki index theorem}. This integral is actually an integral over the so-called inertia stack $\overline{\mathscr{M}}_g^{(n)}$, which is roughly a disconnected union of the moduli space $\overline{\mathscr{M}}_g^{(n)}$ and the set of divisors of surfaces with enhanced symmetry. But by inserting the corresponding Poincar\'e dual classes, we can assume that we are simply computing an integral over $\overline{\mathscr{M}}_g^{(n)}$. 
We can then replicate a similar step that we performed in the previous analysis in Section~\ref{subsec:general case}. Instead of completely localizing the integral, we only localize it to a bigger submanifold whose normal bundle we denoted by $\mathcal{N}_2$ in the previous derivation. This submanifold will be identified with the total space of the direct sum of line bundles $\bigoplus_i \LL_i$ over $\bM_{g,n}$, so let us give it the name $\bM^{\LL}_{g,n}$ for now.

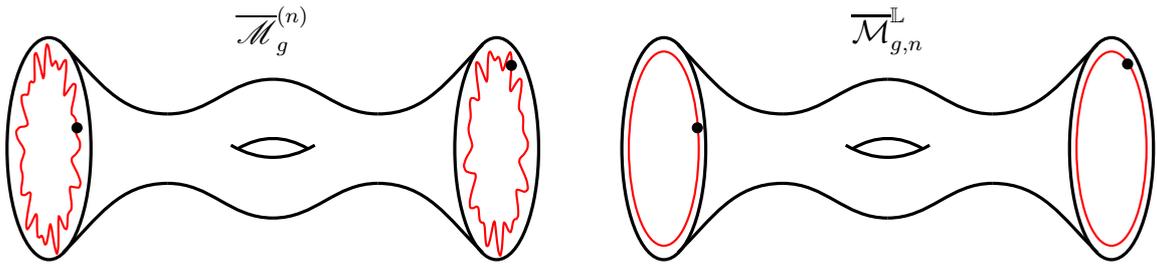
\begin{figure}
\begin{center}
\begin{tikzpicture}[scale=.92]
\draw[very thick,out=0, in=180] (-1.5,.5) to (0,1);
\draw[very thick,out=0, in=180] (0,1) to (1.5,.5);
\draw[very thick,in=180,out=-45] (-3,1.5) to (-1.5,.5);
\draw[very thick,out=0,in=225] (1.5,.5) to (3,1.5);
\draw[very thick,out=0, in=180] (-1.5,-.5) to (0,-1);
\draw[very thick,out=0, in=180] (0,-1) to (1.5,-.5);
\draw[very thick,in=180,out=45] (-3,-1.5) to (-1.5,-.5);
\draw[very thick,out=0,in=-225] (1.5,-.5) to (3,-1.5);
\draw[very thick, bend right=30] (-.6,.05) to (.6,.05); 
\draw[very thick, bend left=30] (-.5,0) to (.5,0); 
\draw[very thick] (3.2,0) circle (.6 and 1.6);
\draw[very thick] (-3.2,0) circle (.6 and 1.6);
\draw[smooth, thick, samples=100,domain=0:360, red] plot({3.2+0.4*cos(\x)*(1+0.1*sin(12*\x)+0.1*cos(17*\x))}, {1.3*sin(\x)*(1+0.1*sin(12*\x)+0.1*cos(17*\x))});
\draw[smooth, thick, samples=100,domain=0:360, red] plot({-3.2+0.4*cos(\x)*(1+0.1*sin(13*\x)+0.1*cos(19*\x))}, {1.3*sin(\x)*(1+0.1*sin(13*\x)+0.1*cos(19*\x))});
\fill (3.41,1.2) circle (.08);
\fill (-2.8,.3) circle (.08);
\node at (0,1.7) {$\overline{\mathscr{M}}_g^{(n)}$};
\end{tikzpicture}
\hspace{.8cm}
\begin{tikzpicture}[scale=.92]
\draw[very thick,out=0, in=180] (-1.5,.5) to (0,1);
\draw[very thick,out=0, in=180] (0,1) to (1.5,.5);
\draw[very thick,in=180,out=-45] (-3,1.5) to (-1.5,.5);
\draw[very thick,out=0,in=225] (1.5,.5) to (3,1.5);
\draw[very thick,out=0, in=180] (-1.5,-.5) to (0,-1);
\draw[very thick,out=0, in=180] (0,-1) to (1.5,-.5);
\draw[very thick,in=180,out=45] (-3,-1.5) to (-1.5,-.5);
\draw[very thick,out=0,in=-225] (1.5,-.5) to (3,-1.5);
\draw[very thick, bend right=30] (-.6,.05) to (.6,.05); 
\draw[very thick, bend left=30] (-.5,0) to (.5,0); 
\draw[very thick] (3.2,0) circle (.6 and 1.6);
\draw[very thick] (-3.2,0) circle (.6 and 1.6);
\draw[thick,  red] (-3.2,0) circle (.5 and 1.4);
\draw[thick,  red] (3.2,0) circle (.5 and 1.4);
\fill (3.43,1.22) circle (.08);
\fill (-2.72,.3) circle (.08);
\node at (0,1.7) {$\bM_{g,n}^{\LL}$};
\end{tikzpicture}
\end{center}
\caption{This picture illustrates the reduction from the full partition function to the primary partition function and the reduction of the associated moduli spaces. Here we drew the case $g=1$ and $n=2$.} \label{fig:moduli spaces with asymptotic boundaries}
\end{figure}

Geometrically, this localization corresponds to restricting ourselves to surfaces with an asymptotic round cutoff instead of an arbitrary wiggly cutoff. As in $\overline{\mathscr{M}}_g^{(n)}$, the moduli space comes also with the additional data of a marked point on the asymptotic boundary. We illustrated the two relevant moduli spaces $\overline{\mathscr{M}}_g^{(n)}$ and $\bM_{g,n}^{\LL}$ in Figure~\ref{fig:moduli spaces with asymptotic boundaries}. As before in Section~\ref{subsec:general case}, the normal space to $\bM_{g,n}^{\LL}$ inside $\overline{\mathscr{M}}_g^{(n)}$ can be identified with $n$ copies of $T \overline{\mathscr{M}}_{0,1}^{(1)}$. The Euler class and the Todd class conspire as before to give the Virasoro character. 

After reducing to $\bM_{g,n}^\LL$, we now have to compute the number of sections of the line bundle $\mathscr{L}^k$ (that is obtained by pulling back the line bundle $\mathscr{L}^k$ from all of $\overline{\mathscr{M}}_g^{(n)}$ to $\bM_{g,n}^\LL$). The $\U(1)^n$ action is the obvious one; it simply moves around the marked points on the asymptotic boundaries. We next show that $\bM_{g,n}^\LL$ is in fact the total space of the direct sum of line bundles $\bigoplus_i \LL_i$ on $\bM_{g,n}$.

The reason for this was essentially also mentioned already earlier in Section~\ref{subsec:general case} where we discussed the normal bundle of the divisor $\mathscr{D}$ of fixed points of the $\U(1)^n$-action. The surfaces described by the moduli space $\bM_{g,n}^{\LL}$ can be obtained by gluing $n$ disks to a genus $g$ surface with $n$ punctures. Each of these gluings is locally of the same form as eq.~\eqref{eq:degeneration} and hence
requires one additional complex gluing parameter. In the hyperbolic world, we would identify the absolute value of this gluing parameter by the geodesic length $b$ of the `neck' in the surface and the phase $\theta \in [0,2\pi]$ with the location of the marked point on the asymptotic boundary (i.e.\ the twist of the hyperbolic gluing). The gluing parameters do not have any non-trivial topology, since we could set
\be 
z=b \mathrm{e}^{i \theta} \in \CC\ .
\ee
The Deligne-Mumford compactification of moduli space allows us to also consider the $b \to 0$ limit.
Thus $\bM_{g,n}^{\LL}$ has the topology of a sum of line bundles over $\bM_{g,n}$, where the gluing parameter describes the point in the fiber. As we have explained before, the gluing parameter is in fact a section of the line bundle $\LL_i$. In this case, there is no contribution from the corresponding line bundle over the disk moduli space to \eqref{eq:conormal bundle divisor} since that moduli space is trivial. This shows that 
\be 
\bM_{g,n}^\LL \equiv \bigoplus_i \LL_i\ ,
\ee
where the right-hand side should be understood as the total space. 

A section of $\mathscr{L}^k$ over $\bM_{g,n}^{\LL}$ can in local coordinates thus be written as a function 
\be 
f(\boldsymbol{m},\mathrm{d}z_1,\dots,\mathrm{d}z_n)\ ,
\ee
where $\boldsymbol{m}$ stands for all the moduli in $\bM_{g,n}$ and $z_1,\dots,z_n$ are the $n$ marked points on the surface. The $\U(1)^n$ action acts by rotation of the differentials. In other words, a section with charges $(\ell_1,\dots,\ell_n)$ is of the form
\be 
f(\boldsymbol{m}) \prod_{i=1}^n (\mathrm{d}z_i)^{\ell_i}\ ,
\ee
i.e.\ a section of $\mathscr{L}^k \otimes \bigotimes_i \LL_i^{\ell_i}$. This establishes the claim \eqref{eq:primary partition function sections}.

\subsection{The \texorpdfstring{$g=1$, $n=1$ and $n=2$}{g=1, n=1 and n=2} partition functions} \label{subsec:g=1 partition functions}
We can use the formula \eqref{eq:primary partition function sections} to compute the $g=1$, $n=1$ partition function. We have $\mathscr{L}^k \otimes \LL_1^{\ell}=\mathscr{L}^{k+\ell}$, since $\kappa_1$ is identified with $\psi_1$ on $\bM_{1,1}$. As discussed in Section~\ref{subsec:M11 modular forms}, the number of sections of $\mathscr{L}^{k+\ell}$ is given by
\be 
\frac{1}{2\pi i} \oint_0 \mathrm{d}x \  \frac{x^{-k-\ell-1}}{(1-x^4)(1-x^6)}\ ,
\ee
since the ring of modular forms is generated by $E_4$ and $E_6$. Thus we get
\begin{align}
Z_1^\text{p}(\beta)&=\frac{1}{2\pi i} \sum_{\ell=0}^\infty q^\ell \oint_0 \mathrm{d}x \  \frac{x^{-k-\ell-1}}{(1-x^4)(1-x^6)} \ .
\end{align}
We can evaluate the integral explicitly by taking instead residue at all poles different from the pole at $x=0$. This gives
\begin{align}
Z_1^\text{p}(\beta)&=-\sum_{\ell=0}^\infty q^\ell \sum_{\omega=1,\, -1,\, i,\, -i,\, \mathrm{e}^{\frac{\pi i}{3}},\, \mathrm{e}^{\frac{2\pi i}{3}},\, \mathrm{e}^{\frac{4\pi i}{3}},\, \mathrm{e}^{\frac{5\pi i}{3}}}\Res_{x=\omega} \frac{x^{-k-\ell-1}}{(1-x^4)(1-x^6)} \\
&=\sum_{\ell=0}^\infty q^\ell \Bigg(\frac{(1+(-1)^{k+\ell})( k+\ell+5)}{24}+ \sum_{\omega=i,\, -i} \frac{\omega^{k+\ell}}{8} \nonumber \\
&\qquad\qquad\qquad\qquad+\sum_{\omega=\mathrm{e}^{\frac{\pi i}{3}},\, \mathrm{e}^{\frac{2\pi i}{3}},\, \mathrm{e}^{\frac{4\pi i}{3}},\, \mathrm{e}^{\frac{5\pi i}{3}}}  \frac{\omega^{k+\ell}}{6(1-\omega^2)}\Bigg) \\
&=\sum_{\omega=1,\, -1} \frac{\omega^k}{24(1-\omega q)}\left(k+5+\frac{\omega q}{1-\omega q}\right)+\sum_{\omega=i,\, -i} \frac{\omega^k}{8(1-\omega q)} \nonumber\\
&\qquad\qquad\qquad\qquad+\sum_{\omega=\mathrm{e}^{\frac{\pi i}{3}},\, \mathrm{e}^{\frac{2\pi i}{3}},\, \mathrm{e}^{\frac{4\pi i}{3}},\, \mathrm{e}^{\frac{5\pi i}{3}}}  \frac{\omega^{k}}{6(1-\omega^2)(1-\omega q)}\ .
\end{align}
The naive index theorem and hence the fake partition function capture the first term with $\omega=1$.

We can repeat the same computation in the case of $g=1$ and $n=2$ using that sections of line bundles on $\bM_{1,2}$ can be realized as weak Jacobi forms. We will use standard facts about weak Jacobi forms, see e.g.\ \cite{EichlerZagier}. On $\bM_{1,2}$, we have the isomorphism $\LL_1 \cong \LL_2$. A section of $\mathscr{L}^k \otimes \LL_1^{\ell_1} \otimes \LL_2^{\ell_2}$ is in fact a weak Jacobi form of weight $k$ and index $m=\frac{1}{2}(k+\ell_1+\ell_2)$. The ring of weak Jacobi forms with half-integral index is generated by the Eisenstein series $E_4$ and $E_6$ as well as the forms 
\begin{align} 
\phi_{-1,\frac{1}{2}}(z,\tau)&=\frac{\vartheta_1(z,\tau)}{\eta(\tau)^3}\ , \\
\phi_{0,1}(z,\tau)&=4\sum_{i=2,3,4} \frac{\vartheta_i(z,\tau)^2}{\vartheta_i(\tau)^2}\ , \\
\phi_{0,\frac{3}{2}}(z,\tau)&=\frac{\vartheta_1(2z,\tau)}{\vartheta_1(z,\tau)}\ .
\end{align}
These generators satisfy a single relation, namely
\be 
\phi_{0,\frac{3}{2}}^2=\frac{1}{432}\left(\phi_{0,1}^3+2\phi_{-1,\frac{1}{2}}^6 E_6-3 \phi_{0,1} \phi_{-1,\frac{1}{2}}^4 E_4\right)\ .
\ee
We thus have
\begin{align} 
\dim \H^0(\bM_{1,2},\mathscr{L}^k \otimes \LL_1^{\ell_1} \otimes  \LL_2^{\ell_2}) &=\Res_{x=0}  \Res_{y=0}  \frac{x^{-k-1}y^{-k-\ell_1-\ell_2-1}(1+y^3)}{(1-x^4)(1-x^6)(1-y^2)(1-x^{-1}y)} \\
&= \Res_{x=0} \frac{(1-x+x^2)x^{-2k-\ell_1-\ell_2-1}-x^{-k}}{(1-x)(1-x^4)(1-x^6)} \ .
\end{align}
One can rewrite this as a sum over residues of various roots of unity and sum over $\ell_1$ and $\ell_2$. This gives
\begin{align}
Z_1(\beta_1,\beta_2)&= \sum_{\ell_1,\, \ell_2=0}^\infty \dim \H^0(\bM_{1,2},\mathscr{L}^k \otimes \LL_1^{\ell_1} \otimes  \LL_2^{\ell_2})\  q_1^{\ell_1} q_2^{\ell_2} \\
&= -\sum_{\omega \text{ roots of unity}} \Res_{x=\omega} \frac{1}{(1-x)(1-x^4)(1-x^6)}\nonumber\\
&\qquad\qquad\times\left(\frac{(1-x+x^2)x^{-2k-1}}{(1-x^{-1}q_1)(1-x^{-1} q_2)}-\frac{x^{-k}}{(1-q_1)(1-q_2)}\right)\ .
\end{align}
Of course we could compute the residues and sum them up which would lead to the exact partition function, but let us refrain from doing so. We only mention that taking the residue at $x=1$ leads to the dominant contribution for large $k$ and reproduces the fake partition function that is computed by the naive index theorem above.

\subsection{Comparison with JT gravity}
Let us now explain the limit in which our computations reduce to JT gravity \cite{Jackiw:1984je}.
In the computations of the dimension of the Hilbert space of a compact surface, the JT answer could be recovered in the limit $k \to \infty$. Indeed, the relevant moduli space is finite-dimensional and so the leading term of \eqref{eq:integral index theorem} in a large $k$ limit is simply proportional to the Weil-Petersson volume of moduli space,
\be 
\int_{\bM_g} \mathrm{e}^{k \kappa_1} = k^{3g-3} \vol_\text{WP}(\bM_g)\ ,
\ee
which is what JT gravity is computing.

 This is clearly no longer the case for the disk partition function. However, physically JT gravity is obtained from the three-dimensional gravity model by compactifying the thermal circle, i.e.\ considering the limit $\beta \to 0$. It will turn out that the correct identification is in fact
\be 
\beta^\text{JT}=\frac{1}{k \beta}\ . \label{eq:betaJT identification}
\ee
Using this scaling limit, we find for the disk partition function
\be 
Z_0(\beta)= \frac{\mathrm{e}^{(k-\frac{1}{24}) \beta} (1-\mathrm{e}^{-\beta})}{\eta\left(\frac{i\beta}{2\pi}\right)}  
= \frac{\sqrt{\beta}\mathrm{e}^{(k-\frac{1}{24}) \beta} (1-\mathrm{e}^{-\beta})}{\sqrt{2\pi}\,   \eta(\frac{2\pi i}{\beta}) } 
\sim \frac{\beta^{\frac{3}{2}}\mathrm{e}^{k\beta+\frac{\pi^2}{6 \beta}}}{\sqrt{2\pi}} 
\ee
With the identification \eqref{eq:betaJT identification}, we hence get
\be 
Z_0(\beta)\sim \frac{\mathrm{e}^{\frac{1}{\beta^\text{JT}}+\frac{\pi^2}{6} k \beta^\text{JT}}}{\sqrt{2\pi} k^{\frac{3}{2}}(\beta^\text{JT})^{\frac{3}{2}}}\ ,
\ee
which agrees with the standard expression for the disk partition function of JT gravity \cite{Saad:2019lba} up to convention-dependent normalizations.
Note that the fact that the temperatures are inverse to each other is physically expected, because in our case $\beta$ measures the radius of the thermal circle in the three-dimensional geometry, whereas it measured the circumference of the disk in the JT gravity interpretation. The two are related by a modular transformation in the putative boundary CFT. The term $\mathrm{e}^{\frac{\pi^2}{6} k \beta^\text{JT}}$ can be viewed as an infinite normalization constant.\footnote{It would be absent in the usual treatment of JT gravity, where one zeta-function regularizes the infinite product $\prod_{n=2}^\infty  \frac{n}{\beta^\text{JT}}$ that appears in the disk partition function. Here we instead regularized it by interpreting it as a limit of the Dedekind eta-function. This is a divergent term in the regularization and as usual one should discard it and keep only the finite contribution.}

Let us also connect our discussion of the localization in the multi-boundary case to the standard formulas of JT gravity. In this case, we could run the same discussion. The only modification would be the omission of the Todd class. We can alternatively directly take the scaling limit on $k$ and $\beta$ of \eqref{eq:multiboundary correlator}. 
This leads to
\begin{align}
\lim_{\begin{subarray}{c} k \to \infty, \\ \beta \to 0 \end{subarray}} Z_g(\beta_1,\dots,\beta_n)&=\prod_{i=1}^n \frac{\mathrm{e}^{\frac{\pi^2}{6} k \beta^\text{JT}}}{\sqrt{2\pi k \beta^\text{JT}}} \int_{\bM_{g,n}} \frac{\mathrm{e}^{k \kappa_1}}{\prod_{i=1}^n \big(1-\mathrm{e}^{-\frac{1}{k\beta_i^\text{JT} }+\psi_i} \big)} \ .
\end{align}
Here we used again the finite-dimensionality of the remaining moduli space integral, which shows that the Todd class cannot contribute to the leading large $k$ behaviour. We can then rescale all characteristic classes by $k$ which gives a factor of $k^{3g-3+n}$. Finally, we expand the denominator:
\begin{align}
\lim_{\begin{subarray}{l} k \to \infty, \\ \beta \to 0 \end{subarray}} Z_g(\beta_1,\dots,\beta_n)&=k^{3g-3+n} \prod_{i=1}^n \frac{\mathrm{e}^{\frac{\pi^2}{6} k \beta^\text{JT}}}{\sqrt{2\pi k \beta^\text{JT}}} \int_{\bM_{g,n}} \frac{\mathrm{e}^{\kappa_1}}{\prod_{i=1}^n \big(1-\big(1-\frac{1}{k\beta_i^\text{JT} }+\frac{\psi_i}{k}\big) \big)} \\
&=k^{3g-3+\frac{3n}{2}} \prod_{i=1}^n \sqrt{\frac{\beta^\text{JT}}{2\pi }} \mathrm{e}^{\frac{\pi^2}{6} k \beta^\text{JT}} \int_{\bM_{g,n}} \frac{\mathrm{e}^{\kappa_1}}{\prod_{i=1}^n (1-\beta_i^\text{JT} \psi_i)} \ .
\end{align}
After discarding again the infinite regularization constants $\mathrm{e}^{\frac{\pi^2}{6} k \beta^\text{JT}}$, this is indeed the correct formula for the multi-boundary correlators in JT gravity \cite{Saad:2019lba, Okuyama:2020ncd}. Usually boundary correlators in JT gravity are computed by gluing the trumpet contributions to the volumes of moduli spaces with geodesic boundaries. Thus our discussion constitutes an alternative derivation of this result in the complex world (as opposed to the hyperbolic point of view that is more naturally taken in JT gravity).

 Note also that the prefactor $k^{3g-3+\frac{3n}{2}}$ can be written as $k^{-\frac{3}{2} \chi(\Sigma)}$, where $\chi(\Sigma)$ is the Euler characteristic of the surface. Thus $k^{\frac{3}{2}}$ plays the role of $\mathrm{e}^{-S_0}$, where $S_0$ is the usual genus counting parameter of JT gravity. Since $k$ is chosen to be large in this model, higher genera are not suppressed and this leads actually to strongly-coupled JT gravity.

As promised in the Introduction, we can compare our computation using localization to the usual computation of JT partition functions using hyperbolic geometry \cite{Saad:2019lba}. In the latter one uses the volume of the moduli space of hyperbolic surfaces with geodesic boundaries. A famous formula due to Mirzakhani says that the cohomology class of the Weil-Petersson form in the presence of $n$ geodesic boundaries of length $b_1,\dots,b_n$ is \cite{Mirzakhani:2006eta}
\be 
\omega^{b_1,\dots,b_n}_\text{WP}=\omega_\text{WP}+\frac{1}{\pi} \sum_i b_i^2 \psi_i\ . \label{eq:Mirzakhani relation}
\ee
This formula can then be used to compute the volume the moduli space of Riemann surfaces with geodesic boundaries. One then accounts for the contribution from the boundary wiggles by gluing the bulk of the surface to the so-called `trumpet' partition functions
\be 
Z_g^\text{JT}(\beta_1,\dots,\beta_n)=\int \prod_{i=1}^n  \mathrm{d} b_i\, b_i \ \prod_{i=1}^n Z_\text{trumpet}(b_i,\beta_i) \int_{\bM_{g,n}} \mathrm{e}^{\frac{1}{4\pi}\omega^{b_1,\dots,b_n}_\text{WP}} 
\ee
Comparison of the two methods readily proves \eqref{eq:Mirzakhani relation}. This is perhaps not surprising since the identity was derived in \cite{Mirzakhani:2006eta} by using symplectic reduction and the Duistermaat-Heckman theorem, which is in essence equivalent to the equivariant localization we have employed here.

\section{Topological recursion for the fake partition functions} \label{sec:topological recursion}
Since we phrased everything in terms of standard intersection theory on moduli space, one can apply the machinery of topological recursion to find a spectral curve that computes the fake partition functions of chiral 3d gravity given by eq.~\eqref{eq:omega01}. 
\subsection{A short review of topological recursion} \label{subsec:review topological recursion}
Let us recall the basic ingredients of topological recursion. Topological recursion was defined in \cite{Eynard:2007kz}. We will not need the most general case and correspondingly only explain the restricted case. Technically, the spectral curve that we will determine (given in eq.~\eqref{eq:omega01}) has a single branch point at $z=0$ of order 2, which simplifies the discussion significantly. We will choose coordinates such that the local Galois inversion (i.e.\ the map exchanging the two sheets) just takes the form $z \mapsto
 -z$. Topological recursion is an abstraction of the loop equation of matrix models, see e.g.\ \cite{DiFrancesco:1993cyw, Eynard:2015aea}. However, the topological recursion that will appear is slightly more general than the one coming from matrix models, since the two-point function $\omega_{0,2}(z)$ does not take the universal form typical for matrix models. 

Topological recursion computes a family of meromorphic multi-differentials that we denote by $\omega_{g,n}(z_1,\dots,z_n)$. They are symmetric and one-forms in each entry,
 starting from the two basic cases of $\omega_{0,1}(z)$ and $\omega_{0,2}(z_1,z_2)$ that are considered to be the initial data. We can actually treat all differentials as formal power series around the branch point $z_i=0$, although it will turn out that in the case of interest all differentials are in fact meromorphic.

One defines the recursion kernel from $\omega_{0,2}(z_1,z_2)$  as follows,
\be 
K(z_0,z)\equiv \frac{\int^z_{-z} \omega_{0,2}(z_0,-)}{4\, \omega_{0,1}(z)}\ .
\ee
Higher differentials $\omega_{g,n}(z_1,\dots,z_n)$ are then constructed as follows:
\begin{align}
\omega_{g,n}(z_1,\dots,z_n)&\equiv\Res_{\zeta=0} K(z_1,\zeta)\Bigg( \omega_{g-1,n+1}(\zeta,-\zeta,z_2,\dots,z_n)\nonumber\\
&\qquad+\sum_{h=0}^g \sum_{\begin{subarray}{c} \mathcal{I} \sqcup \mathcal{J} =\{z_2,\dots,z_n\} \\
(h,\mathcal{I}) \ne (0,\varnothing) \\
(h,\mathcal{J}) \ne (g,\varnothing) 
\end{subarray} } \omega_{h,|\mathcal{I}|+1}(\zeta,\mathcal{I})\omega_{h,|\mathcal{J}|+1}(-\zeta,\mathcal{J})\Bigg)\ , \label{eq:topological recursion}\\
\omega_g&\equiv\frac{1}{2g-2}\Res_{z=0} \Bigg(\omega_{g,1}(z) \int^z \mathrm{d} \zeta\  \omega_{0,1}(\zeta)\Bigg)\ .
\end{align}
These equations are equivalent to the loop equations of random matrix theory. In particular for $2g-2+n>0$, $\omega_{g,n}(z_1,\dots,z_n)$ translates to the following resolvent in the matrix model
\be 
\omega_{g,n}(z_1,\dots,z_n)=\left \langle \prod_{i=1}^n \tr \left(\frac{2 z_i\, \mathrm{d}z_i}{z_i^2+H} \right) \right \rangle_g\ ,
\ee
where $H$ is the Hamiltonian of the matrix model.

Our main interest in topological recursion comes from the fact that these differentials can be expressed in terms of a sum of intersection numbers on moduli space \cite{Eynard:2011kk, Dunin-Barkowski:2012kbi, Eynard:2011ga} as follows
\begin{multline}
\omega_{g,n}(z_1,\dots,z_n)=2^{3g-3+n} \sum_{\Gamma \in \mathcal{G}_{g,n}} \frac{1}{|\Aut(\Gamma)|} \\
\times \int_{\bM_\Gamma} \prod_{v \in \mathcal{V}_\Gamma} \mathrm{e}^{\sum_{m \ge 0} \hat{t}_m \kappa_m} \prod_{e\in \mathcal{E}_\Gamma} \left(\sum_{\ell,m \ge 0} B_{2\ell,2m} \psileft^\ell \, \psiright^{m} \right) \prod_{i=1}^n \left(\sum_{\ell \ge 0} \psi_i^{\ell} \, \mathrm{d} \eta_\ell(z_i) \right)\ . \label{eq:topological recursion intersection number}
\end{multline}
There is various technology and notation that goes into this equation which we explain in turn.
\begin{enumerate}
\item The sum runs over all stable graphs of type $(g,n)$, whose set is denoted by $\mathcal{G}_{g,n}$. A stable graph is a convenient way to parametrize boundary strata in $\bM_{g,n}$. It encodes the topological type of a nodal surface. In a stable graph, every vertex labels a smooth component of the curve. A vertex has a number of half-edges attached to it, some of them corresponding to nodes of the smooth component of the surface and some corresponding to the external punctures. We connect two half-edges corresponding to nodes if the nodal surface is glued along these two nodes. Finally, every vertex is labeled by the genus of the corresponding component. For stability of the graph, we require that the Euler character of every component is negative. The genus of the nodal surface is given by the sum of the genera of all the vertices and the first Betti number of the graph, $g=\sum_v g_v+b_1(\Gamma)$. The codimension of the stratum parametrized by $\Gamma$ is given by the number of internal edges of the stable graph, since these correspond to pinched cycles in the surface. An example of a stable graph and the corresponding nodal surface is depicted in Figure~\ref{fig:stable graph example}.
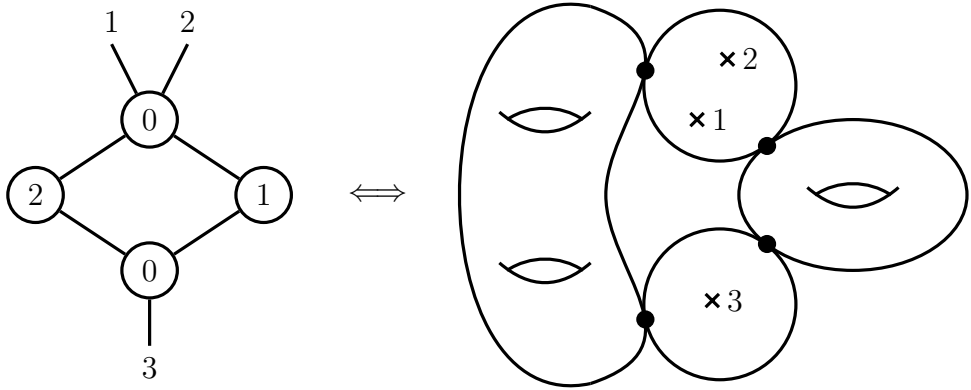
\begin{figure}
\begin{center}
\begin{tikzpicture}
\begin{scope}
\node[shape=circle,draw=black, very thick] (A) at (-1.5,0) {2};
\node[shape=circle,draw=black, very thick] (B) at (0,1) {0};
\node[shape=circle,draw=black, very thick] (C) at (0,-1) {0};
\node[shape=circle,draw=black, very thick] (D) at (1.5,0) {1};
\draw[very thick] (A) to (B);
\draw[very thick] (A) to (C);
\draw[very thick] (B) to (D);
\draw[very thick] (C) to (D);
\draw[very thick] (B) to (-.5,2) node[above] {1};
\draw[very thick] (B) to (.5,2) node[above] {2};
\draw[very thick] (C) to (0,-2) node[below] {3};
\node at (3,0) {$\Longleftrightarrow$};
\end{scope}

\begin{scope}[shift={(8.5,0)}]
\draw[very thick] (.75,0) circle (1.5 and 1);
\draw[smooth, very thick] (.15,.1) .. controls (.55,-.25) and (.95,-.25) .. (1.35,.1);
\draw[smooth, very thick] (.25,0) .. controls (.55,.2) and (.95,.2) .. (1.25,0);
\draw[very thick] (-1,1.45) circle (1);
\draw[very thick] (-1,-1.45) circle (1);
\draw[very thick] (-2.5,0) ..controls (-2.5,1) and (-1.2,2) .. (-2.7,2.5);
\draw[very thick] (-2.5,0) ..controls (-2.5,-1) and (-1.2,-2) .. (-2.7,-2.5);
\draw[very thick] (-2.7,-2.5) ..controls (-5,-3) and (-5,3) .. (-2.7,2.5);
\draw[smooth, very thick] (-3.9,1.1) .. controls (-3.5,0.75) and (-3.1,0.75) .. (-2.7,1.1);
\draw[smooth, very thick] (-3.8,1) .. controls (-3.5,1.2) and (-3.1,1.2) .. (-2.8,1);
\draw[smooth, very thick] (-3.9,-.9) .. controls (-3.5,-1.25) and (-3.1,-1.25) .. (-2.7,-.9);
\draw[smooth, very thick] (-3.8,-1) .. controls (-3.5,-.8) and (-3.1,-.8) .. (-2.8,-1);
\fill (-1.98,1.65) circle (.12);
\fill (-1.98,-1.65) circle (.12);
\fill (-.38,.65) circle (.12);
\fill (-.38,-.65) circle (.12);
\draw  (-1.3,1) node[cross out, draw=black, very thick, minimum size=5pt, inner sep=0pt, outer sep=0pt] {};
\draw  (-.9,1.8) node[cross out, draw=black, very thick, minimum size=5pt, inner sep=0pt, outer sep=0pt] {};
\draw  (-1.1,-1.4) node[cross out, draw=black, very thick, minimum size=5pt, inner sep=0pt, outer sep=0pt] {};
\node at (-1,1) {1};
\node at (-.6,1.8) {2};
\node at (-.8,-1.4) {3};
\end{scope}
\end{tikzpicture}
\end{center}
\caption{An example of a stable graph describing a boundary stratum in $\bM_{4,3}$ of codimension 4. This stable graph has no symmetries and the moduli space of the stratum is naturally given by $\bM_{\Gamma} \cong \bM_{2,2} \times \bM_{0,4} \times \bM_{0,3} \times \bM_{1,2}$.} \label{fig:stable graph example}
\end{figure}
In the formula \eqref{eq:topological recursion intersection number}, $v$ runs over all the vertices $\mathcal{V}_\Gamma$ of the stable graph. $g_v$ denotes the corresponding genus and $n_v$ the number of half-edges at each vertex. The moduli space describing a particular stable graph is given by $\bM_\Gamma/\Aut(\Gamma)$, where
\be 
\bM_\Gamma \cong \prod_{v \in \mathcal{V}_\Gamma} \bM_{g_v,n_v}\ .
\ee
We lift the integral to the product of moduli spaces $\bM_\Gamma$, which leads to the factor $|\Aut(\Gamma)|^{-1}$ in eq.~\eqref{eq:topological recursion intersection number}.
Finally, $e$ runs over all internal edges $\mathcal{E}_\Gamma$ in \eqref{eq:topological recursion intersection number} and $i$ runs over all external punctures. The $\psi$-classes that appear in the product over the edges are the $\psi$-classes of the two nodes corresponding to the edge. (We should perhaps call them $(\psileft)_e$ and $(\psiright)_e$, but we do not want to clutter the notation.) Since the expressions are all symmetric in $\psileft$ and $\psiright$, we do not have to specify which $\psi$ belongs to which branch of the surface at the node. $\psi_i$ in the last factor always denotes the $\psi$-classes of the external punctures.
\item The quantities $B_{2\ell,2m}$ are related to the Taylor coefficients of $\omega_{0,2}(z_1,z_2)$:
\be 
\omega_{0,2}(z_1,z_2)=\left(\frac{1}{(z_1-z_2)^2}+2\pi \sum_{\ell,m \ge 0} \frac{B_{2\ell,2m}}{\Gamma(\ell+\frac{1}{2})\Gamma(m+\frac{1}{2})} \, z_1^{2\ell} z_2^{2m}\right) \mathrm{d}z_1\, \mathrm{d}z_2\ . \label{eq:definition omega02}
\ee
$\omega_{0,2}(z_1,z_2)$ could have also odd powers in the series expansion, but these do not contribute to the topological recursion.
\item The forms $\mathrm{d}\eta_\ell(z)$ are defined as
\begin{align}
\mathrm{d}\eta_\ell(z)&=\frac{(2\ell+1)!!}{2^\ell z^{2\ell+2}}+\sum_{m\ge 0} \frac{B_{2\ell,2m} 2^{m+1}}{(2m-1)!!} \, z^{2m} \mathrm{d}z \label{eq:definition etal powerseries}\\
&=\frac{\Gamma(\ell+\frac{1}{2})}{\Gamma(\frac{1}{2})} \, \Res_{\zeta=0} \zeta^{-2\ell-1} \omega_{0,2}(z,\zeta)\ . \label{eq:definition etal residue}
\end{align}
\item Finally, the numbers $\hat{t}_m$ are defined as follows. First we expand
\be 
\omega_{0,1}(z)=\sum_{m=0}^\infty \frac{\Gamma(\frac{1}{2}) t_m }{(2m+1) \Gamma(m+\frac{1}{2})} \, z^{2m+2} \mathrm{d}z\ .
\ee
The hatted coefficients are then defined through the equality of the following two power series
\be 
\sum_{m\ge 0} t_m u^m =\exp\left(-\sum_{m \ge 0} \hat{t}_m u^m \right) \label{eq:Kontsevich times definition}
\ee
around $u=0$.
\end{enumerate}
An important special case of this construction is the case where $B_{2\ell,2m}=0$ for all $\ell$, $m \ge 0$. In this case, the only term that survives in the sum over stable graphs is the trivial graph of codimension 0 that represents the bulk of moduli space, since any pinching gives rise to an edge, which is accompanied with factors of $B_{2\ell,2m}$. This is equivalent to requiring that
\be 
\omega_{0,2}(z_1,z_2)=\frac{\mathrm{d}z_1 \, \mathrm{d}z_2}{(z_1-z_2)^2}\ ,
\ee
which is the standard form of random matrix theory. 
In our case, we will however need the more general version, since the formula \eqref{eq:equivariant partition function} involves pushforwards of integrals from the boundary strata of moduli space.

\subsection{Reducing the integral to a sum of stable graphs} \label{subsec:reducing to stable graphs}
We are interested in computing the integral \eqref{eq:integral index theorem} and its generalization \eqref{eq:equivariant partition function} to the case with asymptotic boundaries.
We can first consider the simpler case without asymptotic boundaries. The following discussion is completely unchanged if we also include the additional factors in \eqref{eq:equivariant partition function} due to the presence of asymptotic boundaries. Most of this formula is already in the correct form for the topological recursion. In particular, we will have from \eqref{eq:equivariant partition function}
\begin{subequations} \label{eq:hat t}
\begin{align} 
\hat{t}_1&=k-\frac{13}{24}\ , \\
\hat{t}_{2m}&=-\frac{B_{2m}}{(2m)(2m)!}\ ,
\end{align}
\end{subequations}
while all other odd $\hat{t}_{2m+1}$ vanish for $m \ge 1$.
The more difficult part is to deal with the terms involving the pushforward of boundary classes. The result will be that \eqref{eq:equivariant partition function} can be written as \eqref{eq:multiboundary correlator sum over stable graphs} whose form is suitable for topological recursion. Let's recall the integral of interest \eqref{eq:integral index theorem}
\begin{align}
\Zfake_{g,n} = \int_{\bM_{g,n}} \exp\left(\left(k-\frac{13}{24}\right) \kappa_1+\frac{11}{24}\Delta_1- \sum_{m=1}^\infty \frac{B_{2m}}{(2m)(2m)!}\left(\kappa_{2m}-\Delta_{2m}\right) \right)\ , 
\end{align}
It follows from the general results of \cite{Dunin-Barkowski:2012kbi} that an integral of the form \eqref{eq:integral index theorem} can always be written as a sum over stable graphs (and from reading the following paragraphs it should become clear why this is so). It thus suffices to look at one particular stable graph to read off the value of $B_{2\ell,2m}$.  Recall from Section~\ref{subsec:bundles on Mgn} eq.~\eqref{eq:definition Delta} that $\Delta_\ell$ is the pushforward of a boundary class from all codimension 1 boundary strata, which we can write as 
\be 
\Delta_\ell=\sum_{\begin{subarray}{c} \Gamma \in \mathcal{G}_{g} \\
\dim \bM_\Gamma=\dim \bM_g -1 \end{subarray}}\frac{1}{|\Aut(\Gamma)|} \,  (\xi_\Gamma)_*\left((\psileft+\psiright)^{\ell-1} \right)\ .
\ee
Here, $\xi_\Gamma:\bM_\Gamma \longrightarrow \bM_g$ is the inclusion of the stratum. As we shall now explain, one can compute intersection numbers involving the $\Delta_\ell$'s by repeatedly pulling back to the boundary divisors. 

Let us consider an intersection of the form
\be 
\int_{\bM_g} \Delta_\ell \, \Psi \ , \label{eq:pullback bMg Psi}
\ee
where $\Psi$ is any other class. Then we have by general properties of the pushforward and pullback
\be 
\int_{\bM_g} \Delta_\ell \, \Psi =\sum_{\begin{subarray}{c} \Gamma \in \mathcal{G}_{g} \\
\dim \bM_\Gamma=\dim \bM_g -1 \end{subarray}}\frac{1}{|\Aut(\Gamma)|} \int_{\bM_\Gamma} (\psileft+\psiright)^{\ell-1} \, \xi_\Gamma^*(\Psi)\ .
\ee
Since $\Psi$ will be given as a product of various other classes, we need to discuss the pullback of the classes that enter in \eqref{eq:integral index theorem}.
We have essentially by definition
\be 
\xi_\Gamma^*(\psi_i)=\psi_i\ , \qquad \xi_\Gamma^*(\kappa_m)=\kappa_m\ , \qquad \xi_\Gamma^*(\lambda_i)=\lambda_i\ ,
\ee
where $\kappa_m$ on a boundary divisor of the form $\bM_\Gamma$ means the sum of the $\kappa$-classes on the various factors associated to the vertices, with the same comment applying to $\lambda_1$. 
The only slightly non-trivial part  is the pullback of the $\lambda$-classes. On the separating degeneration, the Hodge bundle $\EE_g$ is naturally isomorphic to the outer product $\EE_{g_1} \oplus \EE_{g_2}$ on $\bM_{g_1,n_1+1} \times \bM_{g_2,n_2+1}$ which gives the claimed formula for $\lambda_i$. For the non-separating degeneration, the $\EE_g$ is naturally isomorphic to $\EE_{g-1} \oplus \mathscr{O}$, where $\mathscr{O}$ is the trivial line bundle. The map from $\EE_g$ to $\mathscr{O}$ is given by taking the residue of the differential at one of the nodes. Since $\mathscr{O}$ does not contribute to the Chern classes, we conclude again that $\lambda_i$ pulls back to itself.\footnote{The residue is only defined up to a sign since it involves choosing a branch of the surface. This means that the claimed isomorphism is true up to 2-torsion. However Chern-classes do not detect 2-torsion.}

With these pullbacks in place, we can work out the pullback of $\Delta_1$ to a codimension 1 boundary stratum. For this we use Mumford's formula \eqref{eq:Mumford formula} twice
\begin{align}
\xi_\Gamma^*(\Delta_1)&=\xi_\Gamma^*\left(12\lambda_1-\kappa_1+\sum_i \psi_i\right) \\
&=12\lambda_1-\kappa_1+\sum_i \psi_i \\
&=\Delta_1-\psileft-\psiright\ ,
\end{align} 
where $\psileft$ and $\psiright$ are the two new $\psi$-classes at the nodes. Thus perhaps surprisingly, $\Delta_1$ does not pull back to itself. The correction comes essentially from the self-intersection of $\Delta_1$. A similar formula applies to $\Delta_\ell$. We can work it out by considering the intersection
\be 
\int_{\bM_g} \Delta_1^m \Delta_\ell \ .
\ee
On the one hand, taking $\Delta_\ell$ and $\Psi=\Delta_1^m$ in \eqref{eq:pullback bMg Psi}, this is equal to
\be 
\int_{\bM_g} \Delta_\ell \, \Psi =\sum_{\begin{subarray}{c} \Gamma \in \mathcal{G}_{g} \\
\dim \bM_\Gamma=\dim \bM_g -1 \end{subarray}}\frac{1}{|\Aut(\Gamma)|} \int_{\bM_\Gamma} (\psileft+\psiright)^{\ell-1}  (\Delta_1-\psileft-\psiright)^m  \ .
\ee
On the other hand, we can treat one of the $\Delta_1$'s as the special $\Delta$ and $\Psi=\Delta_\ell \Delta_1^{m-1}$ in \eqref{eq:pullback bMg Psi}, which gives
\be 
\int_{\bM_g} \Delta_\ell \, \Psi =\sum_{\begin{subarray}{c} \Gamma \in \mathcal{G}_{g} \\
\dim \bM_\Gamma=\dim \bM_g -1 \end{subarray}}\frac{1}{|\Aut(\Gamma)|} \int_{\bM_\Gamma}(\Delta_1-\psileft-\psiright)^{m-1} \xi_\Gamma^* (\Delta_\ell )
\ee
Thus we learn that
\be 
\xi_\Gamma^*(\Delta_\ell)=\Delta_\ell- (\psileft+\psiright)^{\ell} \ . \label{eq:Deltal pullback}
\ee
We can apply this trick recursively to rewrite the integral under consideration as a sum of integrals involving only $\psi$-classes of the nodes and the $\kappa$-classes. In fact, since $\xi_\Gamma^* (\kappa_m)=\kappa_m$ they are spectators to the recursive pullback that we have discussed here.
Assuming that the integral after recursively eliminating all the boundary classes has the correct form of \eqref{eq:topological recursion intersection number}, it is now easy to work out what the coefficients $B_{2\ell,2m}$ should be by considering a particular stable graph. We will take the graph corresponding to the non-separating degeneration
\be 
\Gamma= \ \begin{tikzpicture}[baseline={([yshift=-.5ex]current bounding box.center)}]
\node[shape=circle,draw=black, very thick] (A) at (0,0) {$g-1$};
\draw[very thick] (A) .. controls (1.7,-1.2) and (1.7,1.2) ..  (A);
\end{tikzpicture}
\ee
such that $\bM_\Gamma \cong \bM_{g-1,2}$.
For this graph,  we can work out the contribution to
\begin{align}
\int_{\bM_{g,n}} \exp\left(\frac{11}{24}\Delta_1+\sum_{m=1}^\infty \frac{B_{2m}}{(2m)(2m)!}\Delta_{2m} \right) \label{eq:boundary class integral}
\end{align} 
as follows. First note that the expression will obviously only depend on $\psileft+\psiright$, since this is true for the pullback \eqref{eq:Deltal pullback}. Thus we will clearly have
\be 
\sum_{\ell,m \ge 0} B_{2\ell,2m} \psileft^{\ell}\,  \psiright^{m}=f(\psileft+\psiright) \label{eq:B in terms of f}
\ee
for some function $f$. The contribution of $\psi$'s of the boundary divisor of the non-separating degeneration to \eqref{eq:boundary class integral} is
\begin{align}
\int_{\bM_{g,n}} \prod_{i=1}^m \Delta_{\ell_i}&=\frac{1}{2}\int_{\bM_\Gamma \cong  \bM_{g-1,2}} (\psileft+\psiright)^{\ell_1-1} \prod_{i = 2}^m \left(-(\psileft+\psiright)^{\ell_i}\right) \nonumber\\
&\qquad\qquad+ \text{further boundary classes} \\
&=\frac{(-1)^{m+1}}{2}\int_{\bM_\Gamma } (\psileft+\psiright)^{\sum_i \ell_i-1} + \text{further boundary classes}\ .
\end{align} 
So to obtain the function $f(\psileft+\psiright)$ in \eqref{eq:B in terms of f}, we simply have to replace $\Delta_\ell$ with $-(\psileft+\psiright)^\ell$ in the integrand \eqref{eq:boundary class integral} and multiply the result with $(-\psileft-\psiright)^{-1}$. We should also subtract the constant term because it can never contribute to any intersection number. Thus we have
\begin{align}
f(x)&=- x^{-1} \left(\exp\left(-\frac{11}{24}x-\sum_{m=1}^\infty \frac{B_{2m}}{(2m)(2m)!}x^{2m} \right)-1\right) \\
&=\frac{1}{x}+\frac{\mathrm{e}^{\frac{x}{24}}}{1-\mathrm{e}^x}\ .
\end{align}
Here we used the same Taylor expansion as in \eqref{eq:Taylor expansion log Bernoulli}. Let us now recall the definition of the Bernoulli polynomials in terms of the generating series,
\be 
\frac{x\, \mathrm{e}^{x t}}{\mathrm{e}^x-1}=\sum_n B_n(t) \frac{x^n}{n!} \label{eq:generating function Bernoulli polynomials}
\ee
We can hence write 
\be 
f(x)=-\sum_{n=1}^\infty B_{n+1}\left(\frac{1}{24}\right) \frac{x^{n}}{(n+1)!}
\ee
and thus obtain the explicit formula
\be 
B_{2\ell,2m} =-\frac{B_{\ell+m+1}\left(\frac{1}{24}\right)}{(\ell+m+1)!} \binom{\ell+m}{\ell} \ .
\ee
It should now also be clear that we can also repeatedly pullback these classes to higher codimension strata. Since the only effect of pulling back is the modification of $\Delta_\ell$ according to \eqref{eq:Deltal pullback}, it follows that upon further pulling back the integrand has the structure as in eq.~\eqref{eq:topological recursion intersection number}. 

The same discussion applies in the presence of multi-boundary correlators, since the additional $\psi$-classes in \eqref{eq:equivariant partition function} just pullback to themselves under the procedure that we have described. We thus conclude that we can rewrite the multi-boundary fake partition function \eqref{eq:equivariant partition function} as follows:
\begin{multline}
\Zfake_g^\text{p}(\beta_1,\dots,\beta_n)= \sum_{\Gamma \in \mathcal{G}_{g,n}} \frac{1}{|\Aut(\Gamma)|} \int_{\bM_\Gamma} \prod_{e \in \mathcal{E}_\Gamma} \left(\frac{1}{\psileft+\psiright}+\frac{\mathrm{e}^{\frac{\psileft+\psiright}{24}}}{1-\mathrm{e}^{\psileft+\psiright}}\right)  \\
\times  \prod_{v \in \mathcal{V}_\Gamma} \exp\left(\left(k-\frac{13}{24}\right) \kappa_1- \sum_{m=1}^\infty \frac{B_{2m}}{(2m)(2m)!}\kappa_{2m}\right)\prod_{i=1}^n \frac{\mathrm{e}^{\frac{\psi_i}{24}}}{1-\mathrm{e}^{-\beta_i+\psi_i}}  \ . \label{eq:multiboundary correlator sum over stable graphs}
\end{multline}
Here we included a superscript p to indicate that this the partition function counting Virasoro primary states. In other words, we have factored out the infinite product $\prod_{i=1}^n \prod_{m=1}^\infty (1-\mathrm{e}^{-\beta_i m})^{-1}$. 
The terms $\frac{1}{\psileft+\psiright}$ cancel out once one expands the term depending on $\psileft$ and $\psiright$ as a power series around $\psileft=\psiright=0$ and hence everything is well-defined.
It is interesting to see that essentially the same function appears in the $\psi$-classes for both the nodes and the external punctures. In our derivation they arose in completely different ways.

\subsection{Assembling the topological recursion}
We can now deduce the initial data for the topological recursion for our model. For $\omega_{0,2}(z_1,z_2)$, we have to compute the following sum appearing in eq.~\eqref{eq:definition omega02}
\begin{multline}
2\pi \sum_{\ell,m \ge 0} \frac{B_{2\ell,2m}}{\Gamma(\ell+\frac{1}{2})\Gamma(m+\frac{1}{2})} z_1^{2\ell} z_2^{2m} \\
= -2\pi \sum_{s=0}^\infty \sum_{\ell=0}^s  \frac{B_{s+1}\left(\frac{1}{24}\right)}{(s+1) \, \ell! \, \Gamma(\ell+\frac{1}{2})(s-\ell)!\, \Gamma(s-\ell+\frac{1}{2})}  z_1^{2\ell}z_2^{2s-2\ell} \ .
\end{multline}
We can use the Legendre duplication formula to write
\be 
\Gamma\left(\ell+\frac{1}{2}\right)\ell!=4^{-\ell}\sqrt{\pi}\, (2\ell)!\ .
\ee
We thus get
\begin{align}
= -2\sum_{s=0}^\infty \sum_{\ell=0}^s  \frac{4^sB_{s+1}\left(\frac{1}{24}\right)}{(s+1) \, (2\ell)!(2s-2\ell)!}  z_1^{2\ell}z_2^{2s-2\ell}
\end{align}
We next extend the sum over $\ell$ to half-integers and include a factor $(-1)^{2\ell}$. This is allowed since odd powers of $z_1$ and $z_2$ never enter the topological recursion. We then get
\be 
= -2\sum_{s=0}^\infty \frac{4^sB_{s+1}\left(\frac{1}{24}\right)}{(2s)!\, (s+1)}  \sum_{r=0 }^{2s}  \binom{2s}{r} (-1)^rz_1^{r}z_2^{2s-r}= -2\sum_{s=0}^\infty \frac{4^sB_{s+1}\left(\frac{1}{24}\right)}{(2s)!\, (s+1)} (z_1-z_2)^{2s}\ .
\ee
There does not seem to be an elementary way to evaluate this infinite power series. However it is absolutely convergent for any choice of $z_1$ and $z_2$.
Thus
\vspace{.2cm}
\begin{tcolorbox}[ams align]
\omega_{0,2}(z_1,z_2)=\left(\frac{1}{(z_1-z_2)^2}-2\sum_{s=0}^\infty \frac{4^sB_{s+1}\left(\frac{1}{24}\right)}{(2s)!\, (s+1)} (z_1-z_2)^{2s}\right) \mathrm{d}z_1\, \mathrm{d}z_2\ . \label{eq:omega02}
\end{tcolorbox}
\noindent
It is nice that $\omega_{0,2}(z_1,z_2)$ only depends on the difference of $z_1$ and $z_2$ and is hence `translation invariant'.

To compute $\omega_{0,1}(z)$, we first need to compute the `times' $t_m$ entering in \eqref{eq:Kontsevich times definition}:
\begin{align}
\sum_{m \ge 0} t_m u^m&= \exp\left(-\left(k-\frac{13}{24}\right)u+\sum_{m=1}^\infty \frac{B_{2m}}{(2m)(2m)!} u^{2m} \right) \\
&=\frac{1}{u} \exp\left(\left(\frac{25}{24}-k\right)u\right)-\frac{1}{u} \exp\left(\left(\frac{1}{24}-k\right)u\right)\ ,
\end{align}
where we used the times $\hat{t}_m$ given in \eqref{eq:hat t}.
Hence we have
\be 
t_m=\frac{1}{(m+1)!} \left[\left(\frac{25}{24}-k\right)^{m+1}-\left(\frac{1}{24}-k\right)^{m+1}\right]\ .
\ee
It follows that
\begin{align}
\omega_{0,1}(z)&=\sum_{m=1}^\infty \frac{\Gamma(\frac{1}{2})t_m}{(2m+1)\Gamma(m+\frac{1}{2})} z^{2m+2} \, \mathrm{d}z \\
&=\frac{1}{2} \left( \cos \left(\frac{\sqrt{24k-25} \, z}{\sqrt{6}}\right)-\cos \left(\frac{\sqrt{24k-1}\, z}{\sqrt{6}}\right)\right) \, \mathrm{d}z\ . 
\end{align}
One can rewrite this formula nicely as follows. Let us parametrize the central charge of the dual CFT in the following standard way
\be 
c=24k=1+6(b+b^{-1})^2\ .
\ee
Then we simply have
\vspace{.2cm}
\begin{tcolorbox}[ams align]
\omega_{0,1}(z)=\sin\left(b z\right) \sin (b^{-1} z )\,  \mathrm{d}z\ .\label{eq:omega01}
\end{tcolorbox}
\noindent
Finally, we want to compute the forms $\mathrm{d}\eta_\ell(z)$ from their definition in \eqref{eq:definition etal residue}. They are given by
\begin{align}
\mathrm{d}\eta_\ell(z)&=\frac{\Gamma(\ell+\frac{1}{2})}{\Gamma(\frac{1}{2})} \Res_{\zeta=0} \zeta^{-2\ell-1} B(z,\zeta) \\
&=\frac{(2\ell+1)!!}{2^\ell z^{2\ell+2}}-2 \sum_{s=0}^\infty \frac{4^{s} B_{s+\ell+1}(\frac{1}{24})}{\ell!\, (2s)!\, (s+\ell+1)} z^{2s}\ . 
\end{align}
This completes the determination of the quantities entering the topological recursion \eqref{eq:topological recursion}.

We now want to make the relation between the fake partition function \eqref{eq:multiboundary correlator} and the differentials $\omega_{g,n}(z_1,\dots,z_n)$ more explicit. We have from the definition \eqref{eq:topological recursion intersection number}
\begin{align}
\Zfake^\text{p}_g(q_1,\dots,q_n)&=\sum_{\ell_i \ge 0} [\psi_i^{\ell_i}] \frac{\mathrm{e}^{\frac{\psi_i}{24}}}{1-q_i \mathrm{e}^{\psi_i}} \int_{\bM_{g,n}} \mathrm{e}^{k \kappa_1} \td(\bM_{g,n})\,  \prod_{i=1}^n \psi_i^{\ell_i} \\
&=2^{3-3g-n} \sum_{\ell_i \ge 0} \prod_{i=1}^n [\psi_i^{\ell_i}] \frac{\mathrm{e}^{\frac{\psi_i}{24}}}{1-q_i\,  \mathrm{e}^{\psi_i}} [\mathrm{d} \eta_{\ell_i}(z_i)] \, \omega_{g,n}(z_1,\dots,z_n)\ ,
\end{align}
where we wrote $q_i=\mathrm{e}^{-\beta_i}$.
The bracket notation means that we are isolating the coefficient of the corresponding term in the expression.
So the partition function of primary states is obtained from the differentials by replacing $\mathrm{d}\eta_{\ell}(z_i)$ by the coefficient of $\psi^\ell$ in the series expansion of $\mathrm{e}^{\frac{\psi}{24}}(1-q\,  \mathrm{e}^\psi)^{-1}$ around $\psi=0$. This constitutes now an efficient algorithmic way to compute the partition function with an arbitrary number of boundaries. 
Of course, we checked that the result as computed from the topological recursion is consistent with the explicit computations \eqref{eq:index theorem 20} and \eqref{eq:index theorem 30}.

We can express this relation in a slightly nicer way as follows. Since the singular part of $\mathrm{d}\eta_\ell(z)$ is simple (see eq.~\eqref{eq:definition etal powerseries}), the operation of extracting the coefficient of $\mathrm{d}\eta_\ell(z)$ is represented by the contour integral
\be 
\frac{2^\ell}{2\pi i (2\ell+1)!!} \oint_0 \mathrm{d} z\ z^{2\ell +1} 
\ee
This means that
\be 
\frac{1}{2\pi i}\oint_0 \mathrm{d}z\ \sum_{\ell=0}^\infty \frac{2^\ell \, z^{2\ell+1}}{(2\ell+1)!!} \, [\psi^\ell] \frac{\mathrm{e}^{\frac{\psi}{24}}}{1-q \mathrm{e}^\psi}
\ee
does the operation of replacing the coefficient of $\mathrm{d}\eta(z)$ with the right coefficient of $\mathrm{e}^{\frac{\psi}{24}}(1-q \mathrm{e}^\psi)^{-1}$. We rewrite the appearing function as follows:
\begin{align} 
f_q(z)&\equiv\sum_{\ell=0}^\infty \frac{2^\ell \, z^{2\ell+1}}{(2\ell+1)!!} \, [\psi^\ell] \frac{\mathrm{e}^{\frac{\psi}{24}}}{1-q \mathrm{e}^\psi}\\
&= \sum_{\ell=0}^\infty \sum_{m=0}^\infty \frac{2^\ell \, z^{2\ell+1}}{(2\ell+1)!!} q^m [\psi^\ell] \mathrm{e}^{(m+\frac{1}{24})\psi} \\
&= \sum_{\ell=0}^\infty \sum_{m=0}^\infty \frac{2^\ell \, z^{2\ell+1}}{(2\ell+1)!! \, \ell!}  q^m \left(m+\frac{1}{24}\right)^\ell \\
&=\sum_{m=0}^\infty \sqrt{\frac{6}{24m+1}} \sinh \left(\sqrt{\frac{24m+1}{6}}\, z\right) q^m
\end{align}
This series is absolutely converging as long as $|q|<1$. To summarize, we have the relation
\be 
\Zfake^\text{p}_g(q_1,\dots,q_n)=2^{3-3g-n}  \Res_{z_1=z_2=\dots =0}\, \prod_i f_{q_i}(z_i)\, \omega_{g,n}(z_1,\dots,z_n)\ , \label{eq:relation Zfake and omega}
\ee
The inverse of the relation between $\Zfake^\text{p}(q_1,\dots,q_n)$ and $\omega_{g,n}(z_1,\dots,z_n)$ is actually simpler to state. We will work it out for the special case $n=1$. First notice that for $m \ge 0$
\be 
\Res_{q=1} q^m \Zfake_g(q)=-\Res_{q=\infty} q^m \Zfake_g(q) \ ,
\ee
since $\Zfake_q(q)$ is a rational function that clearly only has singularities at $q=1$ and $q=\infty$. We then compute
\begin{align}
\Res_{q=\infty} q^m f_q(z)&=\sum_{\ell \ge 0} \frac{z^{2\ell+1} 2^\ell}{(2\ell+1)!!\, \ell!}\left(-m-\frac{23}{24}\right)^\ell \\
&=\sqrt{\frac{6}{24m+23}} \sin \left(\sqrt{\frac{24m+23}{6}}\, z\right)\ .
\end{align}
Plugging this identity in eq.~\eqref{eq:relation Zfake and omega} leads to
\be 
\Res_{q=1} q^m \Zfake_q^\text{p}(q)=-2^{2-3g} \sum_{\ell \ge 0} \left(m+\frac{23}{24}\right)^\ell \frac{(-1)^\ell}{\ell!} \, \omega_{g,1}^{(\ell)}\ , \label{eq:residue q=1 q^m}
\ee
where we wrote
\be
\omega_{g,1}(z)=\sum_{\ell \ge 0} \omega_{g,1}^{(\ell)} \, \mathrm{d} \eta_\ell(z)\ .
\ee 
We can view \eqref{eq:residue q=1 q^m} as a meromorphic function in $m$. We can in particular multiply both sides by $(m+\frac{23}{24})^{-\ell-1}$ and take the residue at $m=-\frac{23}{24}$, which selects a single term on the right hand side. This gives
\be 
\omega_{g,1}^{(\ell)}=2^{3g-2} (-1)^{\ell+1} \Res_{q=1} q^{-\frac{23}{24}} \log^\ell q\, \Zfake_g^\text{p}(q)\ .
\ee
Summing this equation over $\ell$ can be done as a formal power series, but since the definition of $\mathrm{d}\eta_\ell(z)$ involves the factor $(2\ell+1)!!$, this series has zero convergence radius.
\subsection{The dilaton equation}
We now show that the topological recursion implies the following very simple equation
\be 
\Res_{q_{n+1}=\infty} q_{n+1}^{k-2}(1-q_{n+1}) \Zfake^\text{p}_g(q_1,\dots,q_n,q_{n+1})=(2g-2+n) \Zfake^\text{p}_g(q_1,\dots,q_n)\ . \label{eq:dilaton equation}
\ee
This is the analogue of the dilaton equation in topological gravity and we will hence refer to it by that name. Even though we derive this equation only for the fake partition function, we actually conjecture that it holds true for the full partition function, i.e.\ we can replace $\Zfake_g^\text{p}$ by $Z_g^\text{p}$ in \eqref{eq:dilaton equation}. While this extension is of course trivially true for $g=0$, one can also check that it is true for $g=1$ and $n=1$ by using the explicit formulas for the partition functions derived in Section~\ref{subsec:g=1 partition functions}.  In view of eq.~\eqref{eq:primary partition function sections}, this conjecture is quite non-trivial from the point of view of algebraic geometry on $\bM_{g,n}$.

\begin{figure}
\begin{center}
\begin{tikzpicture}
\node at (-4.7,0) {$\begin{aligned} \Res_{q=\infty}\end{aligned}$};
\draw[very thick,out=0, in=180] (-1.5,.5) to (0,1);
\draw[very thick,out=0, in=180] (0,1) to (1.5,.5);
\draw[very thick,in=180,out=-45] (-3,1.5) to (-1.5,.5);
\draw[very thick,out=0,in=225] (1.5,.5) to (3,1.5);
\draw[very thick,out=0, in=180] (-1.5,-.5) to (0,-1);
\draw[very thick,out=0, in=180] (0,-1) to (1.5,-.5);
\draw[very thick,in=180,out=45] (-3,-1.5) to (-1.5,-.5);
\draw[very thick,out=0,in=-225] (1.5,-.5) to (3,-1.5);
\draw[very thick, bend right=30] (-.6,.05) to (.6,.05); 
\draw[very thick, bend left=30] (-.5,0) to (.5,0); 
\draw[very thick] (3.2,0) circle (.6 and 1.6);
\draw[very thick] (-3.2,0) circle (.6 and 1.6);
\node at (4.3,0) {$\dfrac{1}{q}$};
\draw[very thick] (5.4,0) circle (.6 and 1.6);
\draw[very thick,out=-45,in=180] (5.6,1.5) to (7.1,.5);
\draw[very thick,out=45,in=180] (5.6,-1.5) to (7.1,-.5);
\draw[very thick,out=0, in=0] (7.1,.5) to (7.1,-.5);
\node at (3.2,0) {$q$};
\node at (5.4,0) {$q^{-1}$};
\node at (-2,-4) {$=(2-2g+n) \ \times$};
\begin{scope}[shift={(3.8,-4)}]
	\draw[very thick,out=0, in=180] (-1.5,.5) to (0,1);
	\draw[very thick,out=0, in=180] (0,1) to (1.5,.5);
	\draw[very thick,in=180,out=-45] (-3,1.5) to (-1.5,.5);
	\draw[very thick,out=0, in=180] (-1.5,-.5) to (0,-1);
	\draw[very thick,out=0, in=180] (0,-1) to (1.5,-.5);
	\draw[very thick,in=180,out=45] (-3,-1.5) to (-1.5,-.5);
	\draw[very thick, bend right=30] (-.6,.05) to (.6,.05); 
	\draw[very thick, bend left=30] (-.5,0) to (.5,0); 
	\draw[very thick,out=0, in=0] (1.5,.5) to (1.5,-.5);
	\draw[very thick] (-3.2,0) circle (.6 and 1.6);
\end{scope}
\end{tikzpicture}
\end{center}
\caption{A pictorial depiction of the dilaton equation for $g=1$ and $n=1$.} \label{fig:dilaton equation}
\end{figure}
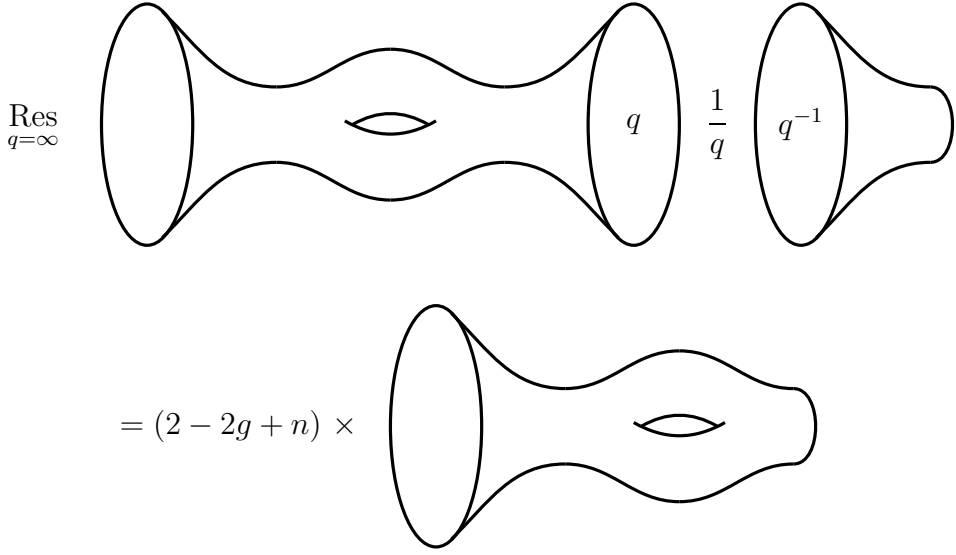

We should also note that there is a perhaps nicer way of expressing the dilaton equation. Recognizing that the primary disk partition function equals
\be 
\Zfake_0^\text{p}(q^{-1})=q^k (1-q^{-1})=-q^{k-1}(1-q)\ ,
\ee
we can write
\be 
\Res_{q_{n+1}=\infty } \frac{1}{q_{n+1}}\Zfake_0^\text{p}(q_{n+1}^{-1}) \Zfake^\text{p}_g(q_1,\dots,q_n,q_{n+1})=(2-2g-n) \Zfake^\text{p}_g(q_1,\dots,q_n)\
\ee
In pictures, we hence suggestively have Figure~\ref{fig:dilaton equation}.

We now prove the dilaton equation \eqref{eq:dilaton equation} as follows.
Note that the primary fake partition function has only poles for $q_{n+1}=1$ and $q_{n+1}=\infty$. This follows essentially from the definition as an integral \eqref{eq:multiboundary correlator}. Indeed, since the moduli space is finite-dimensional, it is a rational function and one easily sees that the expansion of the function
\be 
\frac{\mathrm{e}^{\frac{\psi}{24}}}{1-q\mathrm{e}^\psi}
\ee
in $\psi$ produces terms that have the claimed property. To make the following presentation simpler, we will just demonstrate the dilaton equation in the case of $n=0$. The case $n>0$ is not more difficult.
We have
\begin{align} 
\Res_{q=1} q^{k-2}(1-q) \Zfake^\text{p}_g(q)&=-\Res_{q=\infty} q^{k-2}(1-q) 2^{2-3g} \sum_{\ell \ge 0} \left([\psi^\ell] \frac{\mathrm{e}^{\frac{\psi}{24}}}{1-q\, \mathrm{e}^\psi} \right) \omega_{g,1}^{(\ell)} \\
&=-2^{2-3g} \sum_{\ell \ge 0} \frac{(-1)^\ell}{\ell!} \left[\left(k-\frac{25}{24}\right)^\ell-\left(k-\frac{1}{24}\right)^\ell\right] \omega_{g,1}^{(\ell)}\ , \label{eq:dilaton equation LHS}
\end{align}
where $\omega_{g,1}^{(\ell)}$ is the coefficient of $\mathrm{d}\eta_\ell(z)$ in the expansion of $\omega_{g,1}(z)$. On the other hand, the dilaton equation of the topological recursion gives
\begin{align}
(2g-2+n) \Zfake^\text{p}_g&=2^{3-3g} \Res_{z=0} \Phi(z) \omega_{g,1}(z)\ ,
\end{align}
where $\mathrm{d}\Phi(z)=\omega_{0,1}(z)$. We can choose
\begin{align} 
\Phi(z)&=\sqrt{\frac{3}{2(24k-25)}} \sin \left(\sqrt{\frac{24k-25}{6}}\, z\right)-\sqrt{\frac{3}{2(24k-1)}} \sin \left(\sqrt{\frac{24k-1}{6}}\, z\right)\\
&=\sum_{m \ge 0} \frac{(-1)^m z^{2m+1}}{2(2m+1)!} \left[\left(4k-\frac{25}{6}\right)^m-\left(4k-\frac{1}{6}\right)^m\right]\ .
\end{align}
Since only the singular parts of $\omega_{g,1}(z)$ are contributing to the dilaton equation, we can compute
\begin{align}
(2g-2+n) \Zfake^\text{p}_g&=2^{3-3g} \Res_{z=0} \Phi(z) \omega_{g,1}(z)\\
&=2^{3-3g} \sum_{\ell \ge 0} \omega_{g,1}^{(\ell)} \frac{ (2\ell+1)!! (-1)^\ell}{2^{\ell+1}(2\ell+1)!} \left[\left(4k-\frac{25}{6}\right)^\ell-\left(4k-\frac{1}{6}\right)^\ell\right] \ .
\end{align}
Simplifying the right hand side and comparing to \eqref{eq:dilaton equation LHS} then yields \eqref{eq:dilaton equation}.

\section{Discussion and future directions} \label{sec:discussion}

\subsection{Discussion}

\paragraph{Comparison to Cotler-Jensen.}
We should compare our result with the computation of Cotler and Jensen \cite{Cotler:2020ugk} for the wormhole partition function on the same topology for regular three-dimensional gravity. Their result is (before summing over relative mapping class group actions)
\be 
Z_{0,2}^\text{CJ}(\tau_1,\tau_2)=\frac{\sqrt{\Im (\tau_1)\Im (\tau_2)}}{2 \pi^2 |\eta(\tau_1) \eta(\tau_2)|^2 |\tau_1+\tau_2|^2}  \ . \label{eq:Z0,2 Cotler-Jensen}
\ee
This partition function is invariant under joint modular transformations of the form \cite{Collier:2021rsn}
\be 
(\tau_1,\tau_2) \mapsto (\gamma \cdot \tau_1,M \cdot \gamma \cdot M\cdot \tau_2)\ , \quad\gamma \in \PSL(2,\ZZ)\ , \quad M=\begin{pmatrix}
-1 & 0 \\ 0 & 1
\end{pmatrix}\ .
\ee
The full partition function is then obtained by summing over the relative modular transformations, 
\be 
\sum_{\gamma \in \PSL(2,\ZZ)} Z_{0,2}^\text{CJ}(\tau_1,\gamma \cdot \tau_2)\ . \label{eq:Z0,2 modular sum}
\ee
One can carry out the analysis that we discussed in this paper also in the case of ordinary 3d gravity, which allows for a more direct comparison with \cite{Cotler:2020ugk}. Let us explain this in the simple case of the two-sided genus 0 wormhole, the more general case is discussed below. One can immediately integrate out the boundary wiggles as described in Section~\ref{subsec:alternative derivation} for chiral gravity. This then gives the partition function of primary states. The remaining phase space (the analogue of $\bM_{g,n}^{\LL}$) is in this case $(\mathscr{T} \times \mathscr{T})/\ZZ$, where $\mathscr{T}$ is the Teichm\"uller space of a cylinder, i.e.\ describing the length and the Dehn twist of the wormhole. A useful way to think about $\mathscr{T}/\ZZ$ is $\CC^\times$, i.e.\ $\CC$ with the origin removed. Then the angular coordinate parametrizes the twist and the radial coordinate the length of the neck. Thus $\mathscr{T}$ is the universal cover and the mapping class group is isomorphic to $\ZZ$. As in Section~\ref{subsec:alternative derivation}, we are now instructed to count holomorphic sections of the cotangent bundle on this space. Denoting the coordinates on $\CC^\times \times \CC^\times$ as $(z,w)$, the relevant sections are thus of the form $z^{h} w^{\bar{h}}$ with $h,\,\bar{h} \in \RR_{\ge 0}$ and $h-\bar{h} \in \ZZ$. From the analogous formula to \eqref{eq:primary partition function sections}, we then get
\begin{align}
Z^\text{p}_{0,2}(\tau_1,\tau_2)&=M(\tau_1,\tau_2)\int_0^\infty \mathrm{d}h\int_0^\infty \mathrm{d}\bar{h}\ \delta_\ZZ(h-\bar{h}) (q_1q_2)^{h} (\bar{q}_1\bar{q}_2)^{\bar{h}} \\
&= \frac{M(\tau_1,\tau_2)(1-|q_1q_2|^2)}{2\pi \Im(\tau_1+\tau_2)(1-q_1q_2)(1-\bar{q}_1\bar{q}_2)}\ . \label{eq:Z0,2 canonical quantization}
\end{align}
Here 
\be 
\delta_\ZZ(h-\bar{h})=\sum_{n \in \ZZ} \delta(h-\bar{h}-n)\ .
\ee
$M(\tau_1,\tau_2)$ arises from the normalization of the integral over $h$ and $\bar{h}$. Cotler and Jensen argued that it is given by
\be 
M(\tau_1,\tau_2)=\sqrt{\Im(\tau_1)\Im(\tau_2)}\ , \label{eq:normalization measure integral}
\ee
which follows from a more careful discussion of the symplectic form than what we have given. 

To make contact with the formula of Cotler and Jensen one notices that one can rewrite this as
\begin{align}
Z^\text{p}_{0,2}(\tau_1,\tau_2)=\sqrt{\Im(\tau_1)\Im(\tau_2)}\ \sum_{n \in \ZZ} \frac{1}{|\tau_1+\tau_2+n|^2}\ ,
\end{align}
whose $n=0$ term agrees with the partition function of primaries in \eqref{eq:Z0,2 Cotler-Jensen}. As explained in their paper, \eqref{eq:Z0,2 modular sum} is then the correct way to include all the relative Dehn twists and in particular the sum over $n$ arises by choosing $\gamma=\begin{pmatrix}
1 & n \\ 0 & 1
\end{pmatrix}$.
Of course our result differ from the one of Cotler and Jensen by the  absence of the Casimir ground state energies. This is because we did not include a metaplectic correction in the canonical quantization scheme. Contrary to the case of chiral gravity, it is consistent to include it for ordinary gravity and would supply the correct ground state energies, similar to what we discussed in Section~\ref{subsec:vacuum Virasoro character}.

\paragraph{Non-chiral 3d gravity and divergence of partition functions.} The most obvious extension is the application of these methods to ordinary three-dimensional gravity. The canonical quantization procedure is in principle similar, but the non-compactness of phase space makes almost all partition functions diverge. One can easily see that an analogous formula to \eqref{eq:primary partition function sections} holds for ordinary 3d gravity:
\begin{multline}
Z^\text{p}_g(\tau_1,\dots,\tau_n)=\int_0^\infty  \prod_{i=1}^n \mathrm{d}h_i\, \mathrm{d}\bar{h}_i\, \sqrt{\Im(\tau_i)}\, \delta_\ZZ(h_i-\bar{h}_i)\, q_i^{h_i} \, \overline{q}_i^{\bar{h}_i} \\
\times \dim \Gamma\left(\bM_{g,n},|\mathscr{L}|^{2k} \otimes \bigotimes_{i=1}^n \LL_i^{h_i} \otimes \overline{\LL}_i^{\bar{h}_i}\right)\ .
\end{multline}
Here, $\Gamma(\bM_{g,n},L)$ denotes all non-holomorphic sections of the complex (but not holomorphic) line bundle $L$. The non-holomorphic sections implement precisely the invariance under the diagonal mapping class group and can be identified with holomorphic sections of the corresponding line bundle on $(\mathscr{T}_{g,n} \times \mathscr{T}_{g,n})/\Map_{g,n}$. Clearly as soon as $\dim \bM_{g,n}>0$, there are infinitely such non-holomorphic sections and the dimension is simply infinite. We also inserted the normalization of the integral over the conformal weights, in analogy to \eqref{eq:normalization measure integral}, but these are again not easily fixed from the formalism discussed in this paper and are thus essentially an educated guess.
We do not understand the physical relevance of this divergence in the partition functions. It seems to be a very serious sickness of the gravitational path integral of 3d gravity and is the main reason why we considered chiral gravity in this paper.

\paragraph{Choice of compactification of phase space.} Throughout the paper, we worked with the Deligne-Mumford compactification of phase space. The connection with $\PSL(2,\RR)$ Chern-Simons theory suggests that this is a natural compactification to consider. Indeed, flat $\PSL(2,\RR)$-bundles from the Teichm\"uller component equip the Riemann surface $\Sigma$ with a natural hyperbolic metric. For hyperbolic surfaces, one naturally gets the Deligne-Mumford compactification by allowing geodesic lengths to pinch. However, we want to stress that all of this corresponds to a particular \emph{choice}. We should first mention that had we not compactified the phase space then canonical quantization would always give an infinite-dimensional Hilbert space and we would have found divergent answers. Second, the choice of a compactification is in some sense arbitrary, provided that the symplectic form extends to the compactification. Our choice means that we are allowing nodal singularities for Cauchy slices. Since canonical quantization cannot be used in situations where the thermal circle contracts, this does not make a statement about the possible singularities of the thermal circle. One should however keep in mind that the two are potentially treated differently and our results depend on this choice.

\paragraph{The mapping class group in Chern-Simons theory and Teichm\"uller TQFT.} We want to stress again that gauging the mapping class group is not a natural operation in Chern-Simons theory. In particular, in a path integral treatment, we would integrate over \emph{all} $\PSL(2,\RR)$ connections and there is no natural mapping class action on this space. We only get a well-defined action on the constrained phase space of flat $\PSL(2,\RR)$ connections on the Riemann surface $\Sigma$. A related theory to $\SL(2,\RR)$ Chern-Simons theory is Teichm\"uller TQFT that is obtained by only quantizing the Teichm\"uller component of phase space \cite{Verlinde:1989ua, Teschner:2003at, Teschner:2005bz,Kashaev:1998fc,  EllegaardAndersen:2011vps, Mikhaylov:2017ngi}. It assigns a Hilbert space to Teichm\"uller space, whose states correspond to conformal blocks of Virasoro symmetry with central charge $c_\text{L}=1+6(b+b^{-1})^2$, where $b$ plays the role of $\hbar$.\footnote{This relation is a gravitational analogue  to the relation of $\SU(2)$ Chern-Simons theory and the conformal blocks of the $\mathfrak{su}(2)$ Kac-Moody algebra \cite{Witten:1988hf}.} The mapping class group defines a projective representation whose projective phase is again given by $\exp \left( 2\pi i \, c_\text{L} \right)$ \cite{Kashaev:1998xp}. Thus another definition of the chiral gravity would be Teichm\"uller TQFT with gauged mapping class group (which is non-anomalous for $k \in \ZZ$). 

\paragraph{Erratic behaviour in $k$.} One very interesting feature of our computations is the somewhat erratic of partition functions as a function of $k$ and by extension in Newton's constant, as is already visible in the genus 1 answer \eqref{eq:intro partition function Z1,1}. This is somewhat surprising from a path integral point of view, since a semiclassical evaluation of the gravity path integral around a saddle always leads to a expansion in $G_\text{N}$. It was commonly always assumed that semiclassical gravity can only reproduce certain smooth self-averaging quantities \cite{Cotler:2016fpe}. It is unclear to us whether the oscillatory behaviour is an accident of whether it has deeper physical meaning.

\paragraph{Summing over topologies.} In this paper we have computed gravity partition functions on three-manifolds with a fixed topology, but of course one should sum over all these topologies. In particular, this operation should reproduce the partition function of the putative dual chiral CFT. Let us however remark that such a sum seems badly divergent, even when only summing over manifolds of the form $\Sigma \times \S^1$. Indeed, the dimension of Hilbert space is expected to grow very quickly with $g$.\footnote{Our analysis gives easy access to the regime of fixed $g$ and large $k$, but not to fixed $k$ and large $g$. Thus we do not know how to quantify this intuition.} It is unclear what this divergence means for a holographic dual. Various arguments might indicate that a naive sum over topologies vastly overcounts the true states of quantum gravity \cite{Marolf:2020xie, Eberhardt:2021jvj}. This would ultimately mean that pure 3d gravity defined by a semiclassical path integral is not a consistent theory on its own.

\subsection{Future directions}

\paragraph{Extended topological recursion.} We found a topological recursion that computes the fake partition functions of three-dimensional chiral gravity on $\Sigma \times \S^1$. In principle, one can imagine that there is a modified version of topological recursion that computes the full partition function. The full partition function can be computed via an integral over the inertia stack of moduli space, as described in Appendix~\ref{app:Kawasaki index theorem}. The inertia stack consists of all loci of Riemann surfaces with prescribed automorphism groups. Since one can quotient by the orbifold groups, all the loci are themselves again isomorphic to moduli spaces of lower complexity. Thus one might hope to evaluate Kawasaki's index theorem completely by considering a type of extended topological recursion that also takes into account the twisted sectors of the inertia stack, which would then lead to a complete solution of the model.

\paragraph{Consequences of the dilaton equation.} We derived a dilaton equation \eqref{eq:dilaton equation} for our model. Our proof is valid for the fake partition function, but the equation holds conjecturally for the full partition function. Given that the primary partition functions count certain sections of line bundles on moduli space \eqref{eq:primary partition function sections}, the dilaton equation is a non-trivial statement in algebraic geometry. It would be very interesting to give mathematical proof of this relation.

\paragraph{Conical defects.} It is in possible to enrich the theory with conical defects, i.e.\ massive point particles. In this case, the relevant phase space is the moduli space of Riemann surfaces with conical defects. It can again be constructed from Teichm\"uller space which are the $\PSL(2,\RR)$ bundles with prescribed elliptic holonomy around marked points. This moduli space behaves very discontinuously in the conical defect angles as was emphasized in \cite{Turiaci:2020fjj}. There does not seem to be a fundamental obstacle to compute the partition functions in the presence of such conical defects. They were considered before in e.g. \cite{Krasnov:2000ia, Benjamin:2020mfz, Chandra:2022bqq}.

\paragraph{Seifert manifolds.} We restricted to three-manifolds of the form $\Sigma \times \S^1$ in this work. One possible generalization would be Seifert manifolds which played an important role in \cite{Maxfield:2020ale}. Seifert manifolds are defined to be the total spaces of circle bundles over (orbifold) Riemann surfaces. As such, $\Sigma \times \S^1$ is a special case of a Seifert manifold. For Seifert manifolds, it was demonstrated in \cite{Beasley:2005vf} for the case of $\SU(2)$ Chern-Simons theory that the partition function can still be written as a particular integral over phase space. Thus our analysis can perhaps be extended to cover this case as well.

\paragraph{Supergravity.} Another obvious generalization is to consider chiral or non-chiral supergravity. For $\mathcal{N}=1$ supersymmetry, it is related to $\OSp(1|2)/\ZZ_2$ Chern-Simons theory in the same was as ordinary gravity is related to $\PSL(2,\RR)$ Chern-Simons theory. In this case, it is natural to expect the phase space to be given by the compactified super moduli space $\overline{\mathfrak{M}}_g$ (possibly with asymptotic boundaries). In the case of JT gravity, this was discussed in \cite{Norbury:2017, Stanford:2019vob, Norbury:2020vyi}.

\section*{Acknowledgements}
I thank Edward Witten for many interesting conversations and initial collaboration. I also thank Nathan Benjamin, Alexander Maloney and Gustavo Joaquin Turiaci for useful discussions. I thank Bertrand Eynard for correspondence and Jordan Cotler and Kristan Jensen for comments on the first version of this paper. 
I gratefully acknowledge
support from the grant DE-SC0009988 from the U.S. Department of Energy.

\appendix

\section{The Kawasaki index theorem} \label{app:Kawasaki index theorem}
We describe here the modifications of the Riemann-Roch-Hirzebruch theorem to orbifolds due to Kawasaki \cite{kawasaki}. It is a direct consequence of the Lefschetz fixed point formula that is derived in \cite{AtiyahSegalII, AtiyahSingerIII}. In the following, $\mathcal{M}$ denotes a smooth complex orbifold.
\subsection{The inertia stack}
To formulate the index theorem, one needs to integrate over a larger orbifold called the inertia stack $I\mathcal{M}$ that is a disjoint union of orbifolds, one of which is $\mathcal{M}$. In language familiar to physicists, $\mathcal{M}$ consists of the untwisted sector of the orbifold, whereas $I\mathcal{M}$ consists of both the untwisted and twisted sectors. 

From the analytic viewpoint $I\mathcal{M}$ is constructed as follows. Let $\mathcal{U}=(U_i)_{i \in I}$ be an atlas for $\mathcal{M}$. Every open set $U_i$ comes equipped with a group action of a group $G_i$ and a covering space $\tilde{U}_i$ such that $U_i=\tilde{U}_i/G_i$. 
Now for every $g \in G_i$ consider the fixed point sets $\tilde{U}_i^g$ inside $\tilde{U}_i$. On $\tilde{U}_i^g$ we still have a well-defined group action of the centralizer $Z_{G_i}(g)$ of the group element $g$ in $G_i$. We then set $U^g_i=\tilde{U}_i^g/Z_{G_i}(g)$. Clearly $U_i^g$ only depends on the conjugacy class of $g\in G_i$. The $U_i^g$'s with $i \in I$ and $g \in \text{Conj}(G_i)$ provide then an atlas for a disjoint union of orbifolds of different dimensions, since they can be patched together in an obvious way.  This is the inertia stack $I\mathcal{M}$.
In particular, $I\mathcal{M}$ contains $\mathcal{M}$ as one of its components because we can always choose $g=e$ the identity for any open set and since $U_i^e=U_i$, patching these open sets together gives back the original orbifold. There is always a canonical map $I\mathcal{M} \to \mathcal{M}$ given by inclusion of the fixed point sets into the original orbifolds.

For a stack of the form $\mathcal{M}=X/G$ where $G$ is a finite group and $X$ a smooth manifold, one can define the inertia stack alternatively as follows. Define first the inertia manifold as
\be 
IX=\{ (g,x) \in G \times X\, |\, g \cdot x=x \}\ .
\ee
$IX$ has a natural $G$ action given by $g \cdot (h,x)=(g h g^{-1},g \cdot x)$. Then one can define
\be 
I \mathcal{M}=IX/G\ .
\ee
Thus, points in the inertia stack are in general labeled by a point $x\in \mathcal{M}$ together with a conjugacy class $[g] \subset G_x$, where $G_x$ is the local isotropy group of $x$.

Let's exemplify this construction on $\bM_{1,1}$. In this case, the stabilizer groups are all abelian ($\ZZ_2$ for a generic point, $\ZZ_4$ for $\tau=i$ and $\ZZ_6$ for $\tau=\mathrm{e}^{\frac{2\pi i}{3}}$). Hence the centralizers are all isomorphic to the group $G_i$ itself. $I\bM_{1,1}$ contains one component $\bM_{1,1}$ for the identity acting at every point. But since any point carries also a $\ZZ_2$ action, we get a further copy of $\bM_{1,1}$. For $\tau=i$, we get two more points that carry a $\ZZ_4$ action and for $\tau=\mathrm{e}^{\frac{\pi i}{3}}$, we get four more points that carry a $\ZZ_6$ action. Overall, we hence have
\be 
I\bM_{1,1}=\bM_{1,1} \sqcup \bM_{1,1} \sqcup i/\ZZ_4 \sqcup i/\ZZ_4 \sqcup \mathrm{e}^{\frac{\pi i}{3}}/\ZZ_6 \sqcup \mathrm{e}^{\frac{\pi i}{3}}/\ZZ_6 \sqcup \mathrm{e}^{\frac{\pi i}{3}}/\ZZ_6 \sqcup \mathrm{e}^{\frac{\pi i}{3}}/\ZZ_6 \ , \label{eq:IM11 decomposition}
\ee
in hopefully obvious notation.
\subsection{The index theorem}
Let now $E$ be a vector bundle on $\mathcal{M}$. We can lift $E$ to a vector bundle $I\mathcal{M}$ in an obvious way by pulling it back along the natural map $I \mathcal{M} \to \mathcal{M}$. On $I\mathcal{M}$, $g$ acts on the fiber of $E$ at $(x,[g])$ and we may decompose the bundle according to the eigenspaces under the action of $g$. One then introduces the virtual trace bundle\footnote{This virtual bundle formally defines an element of the K-theory group $K_0(I \mathcal{M}) \otimes_\ZZ \CC$.}
\be 
\tr(E)\equiv\sum_\lambda \lambda E_\lambda\ ,
\ee
i.e.\ an eigenspace is weighted by its eigenvalue. We also form the following formal linear combination of the conormal bundle $\mathcal{N}^*$ of the component of $I \mathcal{M}$ inside $\mathcal{M}$
\be 
\Lambda^\bullet \mathcal{N}^*=\mathscr{O}- \mathcal{N}^*+\Lambda^2 \mathcal{N}^* - \dots\ ,
\ee
i.e. the alternating sum of the exterior algebra. This bundle is also called the  K-theoretic Euler character of $\mathcal{N}^*$. 

The index theorem now states
\be 
\chi(\mathcal{M},E)=\int_{I\mathcal{M}} \td(I\mathcal{M}) \,  \frac{\ch(\tr(E))}{\ch(\tr(\Lambda^\bullet \mathcal{N}^*))}\ . \label{eq:Kawasaki index theorem}
\ee
The normal bundle appears essentially because of the localization formula in equivariant cohomology. 
Since the Chern character respects tensor products and direct sums as $\ch(E \otimes F)=\ch(E) \ch(F)$ and $\ch(E + F)=\ch(E) + \ch(F)$ (i.e.\ it is a ring homomorphism from the K-theory ring to the ordinary cohomology ring), we can uniquely extend its definition to virtual bundles. For example 
\be 
\ch( \tr(E))= \sum_\lambda \lambda \ch (E_\lambda)\ .
\ee
The action of $g$ on the normal bundle can never have eigenvalue 1. This means that the Chern character $\ch (\tr (\Lambda^\bullet \mathcal{N}^*))$ can always be formally inverted.

\subsection{Modular forms from the orbifold index theorem} \label{subapp:modular forms from the orbifold index theorem}
Let's exemplify the theorem with the help of $\bM_{1,1}$ and the bundle $\mathscr{L}^k$ of modular forms. Sections satisfy by definition
\be 
f\left(\frac{a \tau+b}{c \tau+d}\right)=(c \tau+d)^k f(\tau)
\ee
under the natural $\SL(2,\ZZ)$ action.
Let us also recall from Section~\ref{subsec:M11 modular forms} that the canonical bundle of $\bM_{1,1}$ was given by cusp forms of weight 2.
The generic point is stabilized by $-\mathds{1}$ with eigenvalue $(-1)^k$. The additional stabilizers of the point $\tau=i$ are the $\SL(2,\ZZ)$ group elements
\be 
\begin{pmatrix}
0 & 1 \\ -1 & 0
\end{pmatrix}\qquad \text{and} \qquad  \begin{pmatrix}
0 & -1 \\ 1 & 0
\end{pmatrix}
\ee
with eigenvalues $(-i)^k$ and $i^k$ respectively. The additional stabilizers of the point $\tau=\mathrm{e}^{\frac{2\pi i}{3}}$ are
\be 
\begin{pmatrix}
0 & -1 \\ 1 & 1
\end{pmatrix}\ , \qquad 
\begin{pmatrix}
-1 & -1 \\ 1 & 0
\end{pmatrix}\ ,\qquad
\begin{pmatrix}
0 & 1 \\ -1 & -1
\end{pmatrix}\ , \qquad 
\begin{pmatrix}
1 & 1 \\ -1 & 0
\end{pmatrix}\ ,
\ee
with eigenvalues $\mathrm{e}^{\frac{\pi ik}{3}}$, $\mathrm{e}^{\frac{2\pi ik}{3}}$, $\mathrm{e}^{\frac{4\pi ik}{3}}$ and $\mathrm{e}^{\frac{5\pi ik}{3}}$. 

The action on the normal bundles follows immediately from the fact that the cotangent bundle is given by the bundle $\mathscr{L}^{-2}$ on $\mathcal{M}_{1,1}$ and in particular on the fixed points. Thus in the cases in the cases with a non-trivial normal bundle we have
\be 
\Lambda^\bullet \mathcal{N}^*=1-\mathscr{L}^2\ .
\ee
We can then evaluate the index theorem. Higher cohomology $\mathrm{H}^1(\bM_{1,1},\mathscr{L}^k)$ vanishes for $k \ge 0$ and hence
\begin{multline}
\dim \H^0(\bM_{1,1},\mathscr{L}^k)=(1+(-1)^k)\int_{\bM_{1,1}} \td(\bM_{1,1}) \mathrm{e}^{k \kappa_1} +\sum_{\omega=i,\, -i} \int_{i/\ZZ_4} \frac{\ch(\omega^k \mathscr{L}^k)}{\ch(1-\omega^2 \mathscr{L}^2)} \\
+\sum_{\omega=\mathrm{e}^{\frac{\pi i}{3}},\, \mathrm{e}^{\frac{2\pi i}{3}},\, \mathrm{e}^{\frac{4\pi i}{3}},\, \mathrm{e}^{\frac{5\pi i}{3}}} \int_{\mathrm{e}^{\frac{\pi i}{3}}/\ZZ_6} \frac{\ch(\omega^k \mathscr{L}^k)}{\ch(1-\omega^2 \mathscr{L}^2)}\ .
\end{multline}
Here the first two terms come from the identity element and the generic $\ZZ_2$ automorphism. Since the integrals over $i/\ZZ_4$ and $\mathrm{e}^{\frac{\pi i}{3}}$ are 0-dimensional, we just have to extract the degree 0 of the Chern characters. We have
\be 
\frac{\ch_0 (\omega^k \mathscr{L}^k)}{\ch_0 (1-\omega^2 \mathscr{L}^2)}=\frac{\omega^k}{1-\omega^2}\ .
\ee
The integral $\int_{i/\ZZ_4} 1$ evaluates to $\frac{1}{4}$ because we have to divide it as usual by the order of the automorphism group and similarly for the $\ZZ_6$ fixed point. Using also the evaluation of the first integral explained in Section~\ref{subsec:M11 modular forms}, we obtain
\begin{multline}
\dim \H^0(\bM_{1,1},\mathscr{L}^k)=\frac{k+5}{24}\, (1+(-1)^k) + \sum_{\omega=i,\, -i} \frac{\omega^k}{8} \\
+ \sum_{\omega=\mathrm{e}^{\frac{\pi i}{3}},\, \mathrm{e}^{\frac{2\pi i}{3}},\, \mathrm{e}^{\frac{4\pi i}{3}},\, \mathrm{e}^{\frac{5\pi i}{3}}}  \frac{\omega^k}{6(1-\omega^2)}
\end{multline}
It is simple to check that the right-hand side is indeed an integer and predicts the correct dimensions consistent with the fact that the ring of modular forms is freely generated by the two Eisenstein series $E_4$ and $E_6$.

\subsection{Leading correction to fake partition function}
We can also use the Kawasaki index theorem to evaluate the leading correction to the fake partition function in the large $k$ limit. There are some low-genus cases where where the following is not the leading correction, namely $(g,n)=(1,1)$, $(2,0)$, in which case the surface carries generically a $\ZZ_2$-automorphism. For $(g,n)=(1,1)$, $(1,2)$, $(2,0)$, $(2,1)$, $(3,0)$ there are also other corrections that contribute at the same order in $k$, since there are other components of $I\bM_{g,n}$ with the same codimension. For example for $(g,n)=(3,0)$ to locus of hyperelliptic surfaces would lead to a correction of the same order.

As we mentioned in Section~\ref{subsec:corrections}, the leading correction comes generically from the locus where a genus 1 component splits off from the surface, i.e.\ $\mathscr{D}_{1,\emptyset} \cong \bM_{g-1,n+1} \times \bM_{1,1}$ which has codimension 1 and automorphism group $\ZZ_2$. The conormal bundle to such a divisor is
\be 
\mathcal{N}^*=\LLleft \otimes \LLright\ ,
\ee
as was explained for example in Section~\ref{subsec:general case}. Let's say that $\LLright$ lies on the $\bM_{1,1}$. We have to work out the action of the non-trivial automorphism on the normal bundle and the prequantum line bundle $\mathscr{L}^k$. For $\mathcal{N}^*$, this is easy. Since the $\ZZ_2$ only acts on $\bM_{1,1}$, sections of $\LLleft$ are invariant under the automorphisms. One the other hand, $\LLright$ is by definition the cotangent bundle at the node and  the $\ZZ_2$-involution of the torus acts as $-1$ on the cotangent bundle at every point on the torus. Thus the automorphism acts as $-1$ on the conormal bundle. This had to be so, since it could only act as $+1$ or $-1$ and a $+1$ action is not possible, since then the fixed point locus would be bigger. We can similarly work out the action of the $\ZZ_2$ on the prequantum line bundle $\mathscr{L}^k$. $\mathscr{L}^k$ splits into an outer product $\mathscr{L}^k =\mathscr{L}_{\bM_{g-1,n+1}}^k \boxtimes \mathscr{L}_{\bM_{1,1}}^k$ on the two components. Again $\ZZ_2$ does not act on the first factor. On the second factor, the line bundle corresponds to modular forms of weight $k$, on which the $\ZZ_2$ automorphism acts as $(-1)^k$.
Thus Kawasaki's index theorem predicts \eqref{eq:leading correction index theorem} as leading correction to the naive index theorem in a large $k$ limit.

\section{Some algebraic geometry on \texorpdfstring{$\bM_{g,n}$}{Mg,n}} \label{app:Mgn algebraic geometry}
In this appendix, we review some background about the cohomology of $\bM_{g,n}$. We also explain
the application of the Grothendieck-Riemann-Roch theorem to the universal curve $\bC_{g,n}$ over $\bM_{g,n}$ that leads to formulas  for the Chern characters of the Hodge bundle and tangent bundle of moduli space in terms of the more basic cohomology classes.  This computation is standard in algebraic geometry and far-reaching generalizations exist. Our presentation follows loosely \cite{Zvonkine_intro}. Since we are not aware of an explicit formula for the Chern characters of the tangent bundle in the literature, we will explain the computation in detail. We also explain strong consistency checks on our formula.\footnote{The preprint \cite{Bini_Chern} also explains a similar computation, but we do not agree with their formula. Their formula fails in particular the checks that we perform here.}

\subsection{The universal curve}
Let us first recall the definition of the universal curve. The universal curve $\bC_{g,n}$ is a fiber bundle over $\bM_{g,n}$ whose fiber at a given curve in $\bM_{g,n}$ is the (nodal stable) curve itself. Thus, the total space of $\bC_{g,n}$ has one more complex dimension than $\bM_{g,n}$ and consists of curves together with points on them, $(\Sigma_{g,n},z)$, $z \in \Sigma_{g,n}$. There is a canonical projection map
\be 
\pi: \bC_{g,n} \longrightarrow \bM_{g,n}
\ee
that takes the pair $(\Sigma_{g,n},z)$ to the curve $\Sigma_{g,n}$. We have the isomorphism $\bM_{g,n+1} \cong \bC_{g,n}$, where the last point describes the location in the fiber and $\pi$ becomes the usual forgetful morphism. The perspective of $\bC_{g,n}$ is however slightly different. Because the fiber over $\bM_{g,n}$ is described by any point on the curve, this point could also coincide with a marked point or a node. Thus the isomorphism with $\bM_{g,n+1}$ involves collapsing components of the nodal curve that contain the marked point $n+1$ and which become unstable when forgetting the $(n+1)$-st marked point. We will in the following switch back and forth between the two points of view.

There is a natural line bundle on $\bC_{g,n}$ that is called the relative dualizing sheaf $\omega_\pi$, whose fiber at a curve with marked point is the cotangent bundle at the marked point. 
 It is now very important that in the case of a nodal curve, $\omega_\pi$ is defined to allow poles of the one-forms at the boundary divisors, as long as the residues on the two branches of the nodes are opposite. $\omega_\pi$ is called the relative dualizing sheaf, because it is the sheaf that allows one to still write a natural Serre duality even in the presence of nodes.

The universal curve carries $n$ canonical divisors $\mathscr{D}_i$. These are the divisors where the the point in the fiber coincides with the $i$-th marked point. Alternatively, we can also say that there are $n$ sections of the universal curve 
\be 
\sigma_i: \bM_{g,n} \longrightarrow \bC_{g,n}
\ee
that map a curve to the curve together with the $i$-th marked point. Then $\mathscr{D}_i$ is the image of $\sigma_i$. Similarly to the notation used in in Section~\ref{subsec:bundles on Mgn}, we denote the corresponding Poincar\'e dual classes by $\delta_i \in \H^2(\bC_{g,n},\QQ)$. Under the identification of $\bC_{g,n}$ with $\bM_{g,n+1}$, we identify $\mathscr{D}_i \cong \mathscr{D}_{0,\{i,n+1\}}$ and $\delta_i \cong \delta_{0,\{i,n+1\}}$.

$\bC_{g,n}$ has also a codimension 2 locus consisting of the nodes in singular curves, which we call $\Sing$.\footnote{We use $\Sing$ instead of the commonly used $\Delta$ in the literature to avoid confusions with the cohomology classes $\Delta_\ell$ that we introduced in Section~\ref{subsec:bundles on Mgn}.} We will by abuse of notation also denote by $\Sing$ the corresponding Poincar\'e dual class $\Sing \in \H^4(\bC_{g,n},\QQ)$. $\Sing$ has natural double cover, whose fibers are a choice of branch of the node curve (i.e.\ on which of the two components of the curves it lies). Thus the double cover consists of the marked point of the normalization of the curve (which means that we separate the nodal curve at the nodes into several disconnected curves). 

\subsection{\texorpdfstring{$\psi$- and $\kappa$-classes}{psi- and kappa-classes}} \label{subapp:psi and kappa classes}
Let us collect some further background on $\psi$- and $\kappa$-classes. The precise definition of $\psi_i$ is 
\be 
\psi_i=c_1(\LL_i)\ , \qquad \LL_i=\sigma_i^*(\omega_\pi)\ ,
\ee
compare also with footnote~\ref{footnote:definition Li}. 

There is an important comparison result. Under the projection
\be 
\pi: \bM_{g,n+1} \longrightarrow \bM_{g,n}\ ,
\ee
we can compare the classes $\psi_i$ and $\pi^*(\psi_i)$, where the first is the $\psi$-class as defined in $\bM_{g,n+1}$ and the second is the pullback of the corresponding $\psi$-class in $\bM_{g,n}$. One might naively think that the two are the same, but this is incorrect. Essentially, $\psi_i$ is defined in terms of the line bundle $\omega_\pi$ on the universal curve $\bC_{g,n+1}$ while the pullback $\pi^*(\psi_i)$ was defined through the line bundle $\omega_\pi$ on the universal curve $\bC_{g,n}$. In the former, we allow the holomorphic differential to have poles at the boundary divisor where the $i$-th and $(n+1)$-st marked point coincide. In the latter case, there is no $(n+1)$-st marked point and hence no corresponding pole is allowed. This means that
\be 
\psi_i-\pi^*(\psi_i)=\delta_{0,\{i,n+1\}}\ , \label{eq:comparison lemma psi classes}
\ee
where we recall that $\mathscr{D}_{0,\{i,n+1\}}$ is the boundary divisor where the $i$-th and the $(n+1)$-th marked point coincide and bubble off into a 3-pointed sphere and $\delta_{0,\{i,n+1\}}$ the corresponding Poincar\'e dual class.

We furthermore also observe that the restriction of $\LL_i$ on $\bM_{g,n+1}$ to the divisor $\mathscr{D}_{0,\{i,n+1\}}$ is trivial, since the $i$-th point on $\mathscr{D}_{0,\{i,n+1\}}$ lies on a genus 0 component with two marked point and one node. This component does not have any non-trivial moduli and thus $\LL_i$ is trivial. This means the the intersection $\psi_i \, \delta_{0,\{i,n+1\}}=0$ vanishes.

One can derive the following immediate consequence from these results that we will need below. We will change our perspective and think about $\bM_{g,n+1}$ as $\bC_{g,n}$.
Let us compute $\pi_*(\delta_i^{m+1})$. For $m=0$, this is clearly equal to one since every $\mathscr{D}_i$ intersects the fiber of the universal curve exactly once. We then have for $m \ge 1$
\be 
\pi_*(\delta_i^{m+1})=\pi_*\left(\delta_i^m\left(\psi_i-\pi^*(\psi_i)\right)\right) \\
=-\psi_i \, \pi_*(\delta_i^m)=(-\psi_i)^m\ , \label{eq:pushforward delta}
\ee
where the first equality is \eqref{eq:comparison lemma psi classes}. The second follows from the vanishing intersection of $\psi_i$ and $\delta_i$ together with the general properties of the pushforward. Finally, 
 the last equality follows by induction.

Let us also recall that we defined the $\kappa$-classes on $\bM_{g,n}$ as
\be 
\kappa_m=\pi_*(\psi_{n+1}^{m+1})\ .
\ee
By definition, $\psi_{n+1}$ is the first Chern class of the line bundle $\LL_{n+1}$ on $\bM_{g,n+1}$. Let us work out the relation between $\LL_{n+1}$ and $\omega_\pi$. Both are line bundles on $\bM_{g,n+1} \cong \bC_{g,n}$. Away from singularities, the two line bundles are isomorphic, since the fiber is the cotangent space at the $(n+1)$-th marked point. When the $i$-th marked point collides with the $(n+1)$-th marked point, $\omega_\pi$ is completely regular since its definition makes no reference to marked points. $\LL_{n+1}$ is however allowed to have poles which follows from the definition of $\omega_\pi$ on $\bC_{g,n+1}$. Thus we have
\be 
\LL_{n+1} \cong \omega_\pi^\text{log}\equiv \omega_\pi\left(\sum_i \mathscr{D}_i\right)\ . \label{eq:definition omegalog}
\ee
Here, we defined $\omega_\pi^\text{log}$ as the relative dualizing sheaf that also allows poles at the punctures. It is the usual dualizing sheaf twisted by the divisors $\mathscr{D}_i$. In particular this means that we can also define the $\kappa$-classes as
\be 
\kappa_m \equiv \pi_*\left(c_1(\omega_\pi^\text{log})^{m+1}\right)\ . \label{eq:kappa omegalog}
\ee
This is the definition of Arbarello and Cornalba \cite{Arbarello:1994sda}. It is aesthetically perhaps nicer since it only uses the universal curve for its definition.

\subsection{The GRR formula}
Let's also recall the GRR formula. It states that for maps 
\be 
\pi: X \longrightarrow Y\ ,
\ee
the Chern characters of line bundles $\mathscr{L}$ (or more generally coherent sheaves) behave under pushforward as follows:
\begin{align}
\ch(\pi_*(\mathscr{L}))=\pi_*\left(\ch(\mathscr{L}) \td(\pi)\right) \ . \label{eq:GRR}
\end{align}
We explained the definition of $\pi_*(\mathscr{L})$ in Section~\ref{subsec:bundles on Mgn}.
Here $\ch$ is the Chern character and $\td(\pi)$ is the relative Todd class, that can be defined as
\be 
\td(\pi)\equiv \frac{\td\left(\pi^*TY\right)}{\td\left(T X\right)}\ ,
\ee
We want to apply this to the situation where $X=\bC_{g,n}$ and $Y=\bM_{g.n}$.
First, we should remark that GRR is applicable, since even though the space $\bM_{g,n}$ has orbifold singularities, its fibers are stable Riemann surfaces without orbifold singularities. The GRR theorem is only sensible to the fibers which makes its application valid. 

\subsection{Computing \texorpdfstring{$\td(\pi)$}{td(pi)}}
The most non-trivial part of the application of the theorem is the computation of the relative Todd class. Naively one could have thought that we have a sequence
\be 
0 \longrightarrow \pi^*\left(T^* \bM_{g,n}\right)\xrightarrow{(\mathrm{d}\pi)^*} T^* \bC_{g,n} \xrightarrow{r}  \omega_\pi\longrightarrow 0\ .
\ee 
Here the first map is just the adjoint map of the differential $\mathrm{d}\pi$ between the tangent spaces of $\bC_{g,n}$ and $\bM_{g,n}$. The second map $r$ is the restriction of a section of the cotangent bundle to the marked point of the universal curve, which gives by definition a cotangent vector of the marked point. We can hence view this as a holomorphic one-form on the fiber of the universal curve, which is hence a section of the relative dualizing sheaf.

However, the caveat is that this sequence is \emph{not} exact, because contrary to $\omega_\pi$, sections of  $T^* \bC_{g,n}$ are not allowed to have poles at the boundary divisors. Thus the last map $r$ is not surjective.
We now explain how to repair this problem and get an exact sequence. Let us use a local model of the singular locus. The universal curve close to the locus can be modeled as
\be 
xy=q\ .
\ee
Here $q$ should be thought of as parametrizing the base $\bM_{g,n}$ and $(x,y)$ parametrize the fiber of the universal curve. All other coordinates will not be relevant. For $q=0$ we obtain a nodal curve whose two components are described by $x=0$ and $y=0$. Thus the canonical projection $\pi:\bC_{g,n} \to \bM_{g,n}$ takes the form
\be 
\pi(x,y)=q=xy\ .
\ee
So the cotangent space $T^* \bC_{g,n}$ is spanned by $\mathrm{d}x$ and $\mathrm{d}y$.  The cotangent bundle of $\pi^*(T^* \bM_{g,n})$ is instead spanned by $\mathrm{d}q=x \, \mathrm{d}y+y\,  \mathrm{d} x$. Finally, the line bundle $\omega_\pi$ is generated by the forms $\frac{\mathrm{d}x}{x}$ and $\frac{\mathrm{d}y}{y}$ modulo the relation $\frac{\mathrm{d}x}{x}+\frac{\mathrm{d}y}{y}=0$. Thus the composition of the two maps is trivial, $r \circ (\mathrm{d}\pi)^*=0$, but we already mentioned that they are not exact because $r$ is not a surjective map.

One can repair this failure of exactness by considering the following sequence
\be 
0 \longrightarrow \pi^*\left(T^* \bM_{g,n}\right) \longrightarrow T^* \bC_{g,n} \longrightarrow \omega_\pi^\text{log}\longrightarrow \omega_\pi^\text{log} \otimes \left( \mathscr{O}_\Sing \oplus \bigoplus_i \mathscr{O}_{\mathscr{D}_i} \right) \longrightarrow 0\ . \label{eq:exact sequence T star Cgn}
\ee
Here we used $\omega_\pi^\text{log}$ instead of $\omega_\pi$, which is the canonical bundle on the universal curve twisted by the divisors $\mathscr{D}_i$, see \eqref{eq:definition omegalog}. In other words, sections are holomorphic differentials that are allowed to have simple poles at the marked points. This change is just convenient and will simplify later computations, because of the relation \eqref{eq:kappa omegalog}. We compensated for it by including $ \bigoplus_i \omega_\pi^\text{log} \otimes \mathscr{O}_{\mathscr{D}_i}$ in the last term. Here $\mathscr{O}_\Sing$ and $\mathscr{O}_{\mathscr{D}_i}$ mean the sheaf of holomorphic functions on the corresponding subvarieties.\footnote{These are not vector bundles anymore. Thus the exact sequence should be understood as a sequence in sheaf cohomology. Indeed, $\mathscr{O}_{\mathscr{D}_i}$ is a (coherent) sheaf that is entirely supported on the divisor $\mathscr{D}_i$. } The last map is given by taking the residue at the nodes and the marked points respectively. It is now simple to check that this sequence is indeed exact, even in the vicinity of a node. 

Using the multiplicative property of the Todd class, one finds the following formula for the relative Todd class (after dualizing the sequence)
\be 
\td(\pi)=\frac{\td^*(\omega_\pi^\text{log})}{\td^*(\omega_\pi^\text{log} \otimes \mathscr{O}_\Sing) \prod_i \td^* (\omega_\pi^\text{log} \otimes \mathscr{O}_{\mathscr{D}_i})}\ . \label{eq:relative todd class 1}
\ee
Here, $\td^*$ means the Todd class of the dual bundle (or sheaf). 

To continue, one makes the following simple observations (for $i \ne j$)
\be 
\delta_i \, \delta_j=\Sing\, \delta_i =c_1(\omega_\pi^\text{log}) \, \delta_i=c_1(\omega_\pi^\text{log})\, \Sing=0\ . \label{eq:vanishing intersections}
\ee
The first two formulas are obvious. By definition of $\bM_{g,n}$, two marked points are never allowed to coincide and consequently the two divisors $\mathscr{D}_i$ and $\mathscr{D}_j$ do not intersect. The same goes for the intersection of $\mathscr{D}_i$ with $\Sing$, since a marked point is never allowed to coincide with a node. The third identity holds because the restriction of $\omega^\text{log}_\pi$ to $\mathscr{D}_i$ is trivial. The same reasoning applies for the restriction of $\omega^\text{log}_\pi$ to $\Sing$, except that there can be 2-torsion associated with the choice of branch. So the line bundle is rationally trivial, which is still good enough for $c_1(\omega_\pi^\text{log}) \, \Sing=0$ to hold in rational cohomology.

The triviality of $\omega^\text{log}_\pi$ when restricted to $\mathscr{D}_i$ or $\Sing$ also implies that we can simplify \eqref{eq:relative todd class 1} to
\be 
\td(\pi)=\frac{\td^*(\omega_\pi^\text{log})}{ \prod_i \td^* ( \mathscr{O}_{\mathscr{D}_i})\td^*(\mathscr{O}_\Sing)}\ . \label{eq:relative todd class 2}
\ee
 So it remains to compute these three Todd classes. We then take their pushforward below.
\begin{enumerate}
\item  Let us start with $\td^*(\omega_\pi^\text{log})$, which by simply takes the form
 \be 
 \td^*(\omega_\pi^\text{log})=\frac{c_1(\omega_\pi^\text{log})}{\mathrm{e}^{c_1(\omega_\pi^\text{log})}-1}
 \ee
 on $\bC_{g,n} \cong \bM_{g,n+1}$. 
 
\item It is also straightforward to work out $\td^*(\mathscr{O}_{\mathscr{D}_i})$. We have the short exact sequence
 \be 
 0 \longrightarrow \mathscr{O}(-\mathscr{D}_i) \longrightarrow \mathscr{O} \longrightarrow \mathscr{O}_{\mathscr{D}_i} \longrightarrow 0\ ,
 \ee
 where $\mathscr{O}(-\mathscr{D}_i)$ denotes functions on $\bC_{g,n}$ with a simple zero on $\mathscr{D}_i$. Since $\mathscr{O}$ is the trivial line bundle, we have
 \be 
 \frac{1}{\td^*(\mathscr{O}_{\mathscr{D}_i})}=\td^*(\mathscr{O}(-\mathscr{D}_i))=\td(\mathscr{O}(\mathscr{D}_i))=\frac{\delta_i}{1-\mathrm{e}^{-\delta_i}}\ .
 \ee
The last equality follows from the definition of the Chern class by noting that the constant function is a section of $\mathscr{O}(\mathscr{D}_i)$ with a simple zero on the divisor $\mathscr{D}_i$.

 \item  One has to work a bit more for the Todd class of $\mathscr{O}_\Sing$. $\Sing$ has several components, namely one associated to every boundary divisor of $\bM_{g,n}$. Indeed, by definition $\Sing$ is the set of all nodal surfaces together with a marked point that coincides with the node. In the language of $\bM_{g,n+1}$, a component of $\Sing$ thus looks like Figure~\ref{fig:component Delta}.
 
\begin{figure}
\begin{center}
\begin{tikzpicture}
\draw[very thick] (0,0) circle (1);
\draw[very thick, looseness=5] (-3,1.5) .. controls (-.35,.9) and (-.35,-.9) .. (-3,-1.5);
\draw[very thick, looseness=5] (3,1.5) .. controls (.35,.9) and (.35,-.9) .. (3,-1.5);
\draw[very thick, bend right=30] (1.9,.05) to (3.1,.05); 
\draw[very thick, bend left=30] (2,0) to (3,0); 
\draw  (0,-.2) node[cross out, draw=black, very thick, minimum size=5pt, inner sep=0pt, outer sep=0pt] {};
\draw  (-2.5,.3) node[cross out, draw=black, very thick, minimum size=5pt, inner sep=0pt, outer sep=0pt] {};
\draw  (2.3,.7) node[cross out, draw=black, very thick, minimum size=5pt, inner sep=0pt, outer sep=0pt] {};
\node at (0,.2) {$n+1$};
\end{tikzpicture}
\end{center}
\caption{A component of the singular locus $\Sing$ from the point of view of $\bM_{g,n+1}$.} \label{fig:component Delta}
\end{figure}
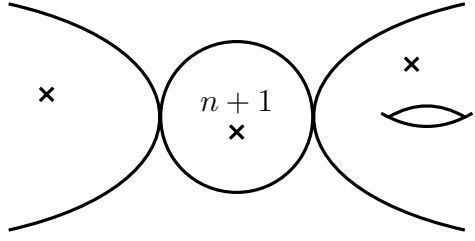
 
Thus let us write $\Sing=\sum_\Gamma \Sing_\Gamma$ as the sum of the different components and $\Sing$ runs over the stable graphs (defined in Section~\ref{subsec:review topological recursion}) that correspond to codimension 1 boundary classes in $\bM_{g,n}$. From the Figure~\ref{fig:component Delta} it should be clear that the intersection of any two distinct components vanishes and thus $\Sing_\Gamma \, \Sing_{\Gamma'}=0$ for $\Gamma \ne \Gamma'$. Thus we have 
\be 
\frac{1}{\td^*(\mathscr{O}_\Sing)}=\prod_\Gamma \frac{1}{\td^*(\mathscr{O}_{\Sing_\Gamma})}=1+\sum_\Gamma\left(\frac{1}{\td^*(\mathscr{O}_{\Sing_\Gamma})}-1\right) \label{eq:todd class product Gamma to sum}
\ee
and we can proceed to compute each $\td^*(\mathscr{O}_{\Sing_\Gamma})$ separately.
 
Since boundary divisors are normal crossing divisors in moduli space we can write locally $\Sing_\Gamma=\mathscr{D}^{(1)}_\Gamma \cap \mathscr{D}^{(2)}_\Gamma$ as the intersection of the two divisors. Let us denote by $\delta^{(1)}_\Gamma$ and $\delta^{(2)}_\Gamma$ the corresponding cohomology classes.
  Then we have the exact sequence
 \be 
 0 \longrightarrow \mathscr{O}(-\mathscr{D}_\Gamma^{(1)}-\mathscr{D}_\Gamma^{(2)}) \longrightarrow \mathscr{O}(-\mathscr{D}_\Gamma^{(1)}) \oplus \mathscr{O}(-\mathscr{D}_\Gamma^{(2)}) \longrightarrow \mathscr{O} \longrightarrow \mathscr{O}_{\Sing_\Gamma} \longrightarrow 0 \ . \label{eq:exact sequence Delta}
 \ee
The maps are almost obvious. The first map maps a function $f$ that vanishes on both divisors $\mathscr{D}_\Gamma^{(1)}$ and $\mathscr{D}_\Gamma^{(2)}$ to $(f,f)$. The next map maps a pair of functions $(f^{(1)},f^{(2)})$ that vanish on the respective divisors to the difference $f^{(1)}-f^{(2)}$. The final map is the restriction of a function $f$ to $\Sing_\Gamma$. This sequence is clearly exact. We thus have
 \begin{align} 
 \frac{1}{\td^*(\mathscr{O}_{\Sing_\Gamma})}&=\frac{\td^*(\mathscr{O}(-\mathscr{D}^{(1)}_\Gamma) \oplus \mathscr{O}(-\mathscr{D}^{(2)}_\Gamma))}{\td^*(\mathscr{O}(-\mathscr{D}^{(1)}_\Gamma-\mathscr{D}^{(2)}_\Gamma))} \\
 &=\frac{\delta_\Gamma^{(1)} \delta_\Gamma^{(2)} (1-\mathrm{e}^{-\delta_\Gamma^{(1)}-\delta_\Gamma^{(2)}})}{(\delta_\Gamma^{(1)}+\delta_\Gamma^{(2)})(1-\mathrm{e}^{-\delta_\Gamma^{(1)}})(1-\mathrm{e}^{-\delta_\Gamma^{(2)}})}\\
  &=\Sing_\Gamma\, \frac{ (1-\mathrm{e}^{-\delta_\Gamma^{(1)}-\delta_\Gamma^{(2)}})}{(\delta_\Gamma^{(1)}+\delta_\Gamma^{(2)})(1-\mathrm{e}^{-\delta_\Gamma^{(1)}})(1-\mathrm{e}^{-\delta_\Gamma^{(2)}})}\ . \label{eq:toddstar class ODeltaGamma}
\end{align}
\end{enumerate}
The GRR formula \eqref{eq:GRR} involves the pushforward of the relative Todd class under $\pi$ to $\bM_{g,n}$ and hence we should determine the various pushforwards.
Using eqs.~\eqref{eq:pushforward delta} and \eqref{eq:kappa omegalog}, this is immediate for the first two cases discussed above,
\begin{subequations} \label{eq:pushforwards todd star classes omega and Di}
\begin{align}
\pi_*\left(\td^*(\omega_\pi^\text{log})\right)&=-\frac{1}{2}(2g-2+n)+\sum_{m\ge 1} \frac{B_{2m}}{(2m)!} \kappa_{2m-1}\ , \\
\pi_*\left(\frac{1}{\td^*(\mathscr{O}_{\mathscr{D}_i})}\right)&=\frac{1}{2}-\sum_{m\ge 1} \frac{B_{2m}}{(2m)!} \psi_i^{2m-1}\ .
\end{align}
\end{subequations}
We used that $\kappa_0=\pi_*(\omega_\pi^\text{log})$ computes the degree of the line bundle $\omega_\pi^\text{log}$ on the surface $\Sigma$, which is $2g-2+n$.
Thus the main remaining work is to cast the pushforward of \eqref{eq:toddstar class ODeltaGamma} in a useful form. Let us first rephrase \eqref{eq:toddstar class ODeltaGamma} in terms of the Chern classes of the normal bundle to $\Sing$. For any smooth divisor $\mathscr{D}$, $\mathscr{O}(-\mathscr{D})$ is in fact the same as the conormal line bundle. Indeed, sections of $\mathscr{O}(-\mathscr{D})$ are functions with a simple zero on the divisor $\mathscr{D}$, while sections of the conormal bundle are by definition first order approximations to such functions. 
 Hence $\delta_\Gamma^{(1)}$ and $\delta_\Gamma^{(2)}$ are the two Chern roots of the normal bundle of $\Sing_\Gamma$ in $\bM_{g,n+1}$. In Section~\ref{subsec:general case}, we already discussed the normal bundles of divisors in moduli space and saw that the normal bundle to a divisor $\mathscr{D}$ is isomorphic to $\LLleft^{-1} \otimes \LLright^{-1}$, where $\LLleft$ and $\LLright$ are the usual two line bundles of the two nodes. In our case, one of the nodes is on the 3-punctured sphere component in Figure~\ref{fig:component Delta} and thus does not contribute to the first Chern class. Using the definition of the $\psi$-classes, we hence find
\be 
\delta_\Gamma^{(1)}=-\psileft\ , \qquad \delta_\Gamma^{(2)}=-\psiright
\ee
on the singular locus $\Sing_\Gamma$, where as usual $\psileft$ and $\psiright$ denote the two $\psi$-classes of the two nodes (that are indistinguishable). We then have
\begin{align}
\pi_*\left(\frac{1}{\td^*(\mathscr{O}_{\Sing_\Gamma})}\right)&= \frac{1}{|\Aut(\Gamma)|} (\xi_\Gamma)_* \left(\frac{ (1-\mathrm{e}^{\psileft+\psiright})}{(-\psileft-\psiright)(1-\mathrm{e}^{\psileft})(1-\mathrm{e}^{\psiright})}\right) \\
&=\frac{1}{|\Aut(\Gamma)|}\sum_{m \ge 1}  \frac{B_{2m}}{(2m)!} (\xi_\Gamma)_*  \left(\frac{\psileft^{2m-1}+\psiright^{2m-1}}{\psileft+\psiright} \right)\ , \label{eq:pushforward todd star class Delta Gamma}
\end{align}
where we recall that $\xi_\Gamma$ is the inclusion of the boundary divisor $\mathscr{D}_\Gamma$ in $\bM_{g,n}$.
 
\subsection{Evaluating the GRR theorem}
Now that we have explained how to compute $\td(\pi)$, we can apply the theorem \eqref{eq:GRR} to various instances of interest. For this, we can choose the line bundle $\mathscr{L}$ on $\bC_{g,n}$ freely. 

\paragraph{The Hodge bundle.} We start by choosing $\mathscr{L}=\mathscr{O}$, the trivial line bundle. Then the push-forward is
\be 
\pi_* \mathscr{O}=\H^0(\Sigma,\mathscr{O})-\H^1(\Sigma,\mathscr{O})=\mathscr{O}-\H^0(\Sigma,\omega_\pi)^*=\mathscr{O}-\EE^*\ ,
\ee
where we used Serre duality (and hence in $\EE$ we need to allow for simple poles at the nodes for this step to be valid). So
\begin{align}
\ch(\EE^*)&=1-\pi_* \td(\pi) \\
&= 1-\pi_*\left(\frac{\td^*(\omega_\pi^\text{log})}{\prod_i \td^*(\mathscr{O}_{\mathscr{D}_i}) \prod_\Gamma \td^*(\mathscr{O}_{\Sing_\Gamma})}\right) \label{eq:pushforward expression Chern character Hodge bundle} \\
&=1-\pi_*\left(\td^*(\omega_\pi^\text{log})\right)-\sum_i \pi_*\left(\frac{1}{\td^*(\mathscr{O}_{\mathscr{D}_i})}\right)-\sum_\Gamma \pi_*\left(\frac{1}{\td^*(\mathscr{O}_{\Sing_\Gamma})}\right) \\
&=g-\sum_{m \ge 1} \frac{B_{2m}}{(2m)!} \Bigg[\kappa_{2m-1}-\sum_i \psi_i^{2m-1}\nonumber\\
&\qquad\qquad\qquad\qquad+\sum_\Gamma \frac{1}{|\Aut(\Gamma)|} \, (\xi_\Gamma)_*\left(\frac{\psileft^{2m-1}+\psiright^{2m-1}}{\psileft+\psiright}\right) \Bigg]\ .
\end{align}
Here we first used that various intersections vanish to rewrite the product of classes as a sum, see eqs.~\eqref{eq:vanishing intersections} and \eqref{eq:todd class product Gamma to sum}. We then used eqs.~\eqref{eq:pushforwards todd star classes omega and Di} and \eqref{eq:pushforward todd star class Delta Gamma} to rewrite the result in terms of known quantities.
Finally upon taking the dual, we learn that
\begin{align} 
\ch_{2m-1}(\EE)&=\sum_{m \ge 1} \frac{B_{2m}}{(2m)!} \Bigg[\kappa_{2m-1}-\sum_i \psi_i^{2m-1}\nonumber\\
&\qquad\qquad\qquad\qquad+\sum_\Gamma \frac{1}{|\Aut(\Gamma)|} \, (\xi_\Gamma)_*\left(\frac{\psileft^{2m-1}+\psiright^{2m-1}}{\psileft+\psiright}\right) \Bigg]\ , \label{eq:Chern characters Hodge bundle}
\end{align}
while the even Chern characters vanish (except for $\ch_0(\EE)=g$). Some further exercises in symmetric function theory turn this into a famous formula for the Chern classes $\lambda_i=c_i(\EE)$, which was first found by Mumford \cite{Mumford1983}.
\paragraph{The bundle of quadratic differentials.} Let's repeat the computation for the bundle of quadratic differentials that we denote by $\EE^{(2)}$. Quadratic differentials are by definition sections of $\omega_\pi \otimes \omega_\pi^\text{log}$, i.e.\ they are allowed to have simple poles at the marked points. Near a node, the quadratic differential is allowed to have double poles, but there are conditions on the singular terms at the node. We stress already here that these conditions are different than for the cotangent bundle of $\bM_{g,n}$. In fact, quadratic differentials lead to the cotangent bundle of uncompactified moduli space, $\mathcal{M}_{g,n}$. It will be slightly more convenient to start with the line bundle $(\omega_\pi^\text{log})^{-1}$, which has the dual pushforward, since
\begin{align} 
\pi_*(\omega_\pi \otimes \omega_\pi^\text{log})&=\H^0(\Sigma,\omega_\pi \otimes \omega_\pi^\text{log})=\EE^{(2)}\ , \\
\pi_*  ((\omega_\pi^\text{log})^{-1})&=-\H^1(\Sigma,(\omega_\pi^\text{log})^{-1})=-\H^0(\Sigma,\omega_\pi \otimes \omega_\pi^\text{log})^*=-(\EE^{(2)})^*\ .
\end{align}
So we get
\begin{align}
\ch((\EE^{(2)})^*)&=-\pi_* \left(\ch((\omega_\pi^\text{log})^{-1})\td(\pi)\right) \\
&=-\pi_* \left(\frac{\mathrm{e}^{-c_1(\omega_\pi^\text{log})}\, \td^*(\omega_\pi^\text{log})}{\prod_i \td^*(\mathscr{O}_{\mathscr{D}_i}) \prod_\Gamma \td^*(\mathscr{O}_{\Sing_\Gamma})}\right)\ . \label{eq:pushforward expression for Chern character quadratic differentials}
\end{align}
Observe that
\be 
\frac{x\,  \mathrm{e}^{-x}}{\mathrm{e}^x-1}=\frac{x}{\mathrm{e}^x-1}-x\,  \mathrm{e}^{-x}\ ,
\ee
and hence
\be 
\mathrm{e}^{-c_1(\omega_\pi^\text{log})}\, \td^*(\omega_\pi^\text{log})=\td^*(\omega_\pi^\text{log})- c_1(\omega_\pi^\text{log})\, \mathrm{e}^{-c_1(\omega_\pi^\text{log})}
\ee
Te products in \eqref{eq:pushforward expression for Chern character quadratic differentials} can again be rewritten as a sum of the involved Todd classes because the intersections \eqref{eq:vanishing intersections} are trivial. We can then compare with \eqref{eq:pushforward expression Chern character Hodge bundle} and see that
\begin{align} 
\ch((\EE^{(2)})^*)&=\ch(\EE^*)-1+ \pi_* \left(c_1( \omega_\pi^\text{log}) \, \mathrm{e}^{-c_1(\omega_\pi^\text{log})} \right) \\
&=\ch(\EE^*)-1+\sum_{m=0}^\infty \frac{(-1)^m }{m!} \, \pi_*\left(c_1(\omega_\pi^\text{log})^{m+1}\right) \\
&=\ch(\EE^*)-1+\sum_{m=0}^\infty \frac{(-1)^m \, \kappa_m }{m!} \ ,
\end{align}
where we used the definition \eqref{eq:kappa omegalog}.
Finally, we can take the dual and use that the degree of $\omega_\pi^\text{log}$ is $2g-2+n$ and hence $\kappa_0=2g-2+n$ to get
\be 
\ch(\EE^{(2)})=\ch(\EE)+2g-3+n+\sum_{m=1}^\infty \frac{\kappa_m}{m!}\ , \label{eq:Chern character quadratic differentials}
\ee
where $\ch(\EE)$ is given by \eqref{eq:Chern characters Hodge bundle} above.
In particular,
\be 
\ch_0(\EE^{(2)})=3g-3+n
\ee
gives the correct dimension of the space of quadratic differentials.

\paragraph{Tangent bundle.} Finally, we want to repeat the same exercise for the cotangent bundle of $\bM_{g,n}$. Let us first recall the following standard result of deformation theory due to Kodaira and Spencer. The Kodaira Spencer map identifies
\be 
T \bM_{g,n} \big|_\Sigma=\H^1(\Sigma,T\Sigma)\ .
\ee
But since $\H^0(\Sigma,T\Sigma)=0$, the right hand side is the pushforward of the relative tangent bundle (or sheaf), i.e.
\be 
T \bM_{g,n}=-\pi_* (T_{\bC_{g,n}/\bM_{g,n}})\ .
\ee
$T_{\bC_{g,n}/\bM_{g,n}}$ fits into the short exact sequence
\be 
0 \longrightarrow T_{\bC_{g,n}/\bM_{g,n}}  \longrightarrow T \bC_{g,n} \longrightarrow T \bM_{g,n} \longrightarrow 0\ . \label{eq:short exact sequence relative tangent space}
\ee
After dualizing, we recognize this sequence essentially as \eqref{eq:exact sequence T star Cgn}. Let us denote
\be 
\Omega_\pi^\text{log}=T_{\bC_{g,n}/\bM_{g,n}}^*\ .
\ee
Then $\Omega_\pi^\text{log}$ is the sheaf of differentials on $\Sigma$ with possible simple poles at the divisors $\mathscr{D}_i$. However, contrary to $\omega_\pi^\text{log}$, sections of $\Omega_\pi^\text{log}$ are not allowed to have simple poles at the nodes. To summarize, we have
\be 
T \bM_{g,n}=-\pi_* \left((\Omega_\pi^\text{log})^{-1}\right)\ .
\ee
To relate $\Omega_\pi^\text{log}$ to $\omega_\pi^\text{log}$, we combine the sequences \eqref{eq:exact sequence T star Cgn} and \eqref{eq:short exact sequence relative tangent space}, which gives
\be 
0 \longrightarrow \Omega_\pi^\text{log} \longrightarrow \omega_\pi^\text{log} \longrightarrow \omega_\pi^\text{log} \otimes \mathscr{O}_\Sing \longrightarrow 0\ .
\ee
Thus we can dualize the sequence and conclude
\be 
\ch((\Omega_\pi^\text{log})^{-1})=\ch((\omega_\pi^\text{log})^{-1})-\ch^* (\omega_\pi^\text{log} \otimes \mathscr{O}_\Sing)= \mathrm{e}^{-c_1(\omega_\pi^\text{log})} -\ch^* (\mathscr{O}_\Sing)\ ,
\ee
where we again used the triviality of the line bundle $\omega_\pi^\text{log}$ when restricted to $\Sing$. $\ch^*$ denotes the Chern character of the dual bundle (or sheaf).
We again use the exact sequence \eqref{eq:exact sequence Delta} to compute $\ch^*(\mathscr{O}_\Sing)$:
\begin{align} 
\ch(\mathscr{O}_\Sing^*)&=\sum_\Gamma \ch(\mathscr{O}_{\Sing_\Gamma}^*) \\
&=\sum_\Gamma\bigg[1+\ch\left(\mathscr{O}\big(\mathscr{D}_\Gamma^{(1)}+\mathscr{D}_\Gamma^{(2)}\big)\right)-\ch\left(\mathscr{O}\big(\mathscr{D}_\Gamma^{(1)}\big)\right)-\ch\left(\mathscr{O}\big(\mathscr{D}_\Gamma^{(2)}\big)\right) \bigg]\\
&=\sum_\Gamma \left(1-\mathrm{e}^{\delta_\Gamma^{(1)}}\right)\left(1-\mathrm{e}^{\delta_\Gamma^{(2)}}\right)\ .
\end{align}
We then only have to put the pieces together. When comparing with the previous case for the quadratic differentials \eqref{eq:pushforward expression for Chern character quadratic differentials}, we get
\begin{align}
\ch(T \bM_{g,n})&=\ch((\EE^{(2)})^*)+\pi_*\left(\ch(\mathscr{O}_\Sing^*) \td(\pi)\right) \\
&=\ch((\EE^{(2)})^*)+\pi_*\left(\frac{\ch(\mathscr{O}_\Sing^*) \, \td^*(\omega_\pi^\text{log})}{\prod_i \td^*(\mathscr{O}_{\mathscr{D}_i}) \prod_\Gamma \td^*(\mathscr{O}_{\Sing_\Gamma})}\right)\ .
\end{align}
Since $\delta_i\,  \Sing=c_1(\omega_\pi^\text{log})\, \Sing=0$, this simplifies with the help of \eqref{eq:toddstar class ODeltaGamma} to
\begin{align}
\ch(T \bM_{g,n})&=\ch((\EE^{(2)})^*)+\sum_\Gamma \pi_*\left(\Sing\, \frac{\mathrm{e}^{\delta_\Gamma^{(1)}+\delta_\Gamma^{(2)}}-1}{\delta_\Gamma^{(1)}+\delta_\Gamma^{(2)}}\right) \\
&=\ch((\EE^{(2)})^*)+\sum_{m\ge 0} \frac{1}{m!} \, \pi_* \left(\Sing\, \big(\delta_\Gamma^{(1)}+\delta_\Gamma^{(2)}\big)^{m-1}\right) \\
&=\ch((\EE^{(2)})^*)+\sum_{m\ge 1} \frac{(-1)^{m+1}}{m!} \sum_\Gamma \frac{1}{|\Aut(\Gamma)|} (\xi_\Gamma)_* \left( \left(\psileft+\psiright\right)^{m-1}\right)\\
&=\ch((\EE^{(2)})^*)+\sum_{m \ge 1} \frac{(-1)^{m+1}\, \Delta_{m}}{m!}  \ , \label{eq:tangent bundle Chern characters}
\end{align}
where we used the definition \eqref{eq:definition Delta} for the boundary classes $\Delta_m$.
We can finally dualize to get the Chern character of the cotangent bundle,
\begin{align}
\ch(T^* \bM_{g,n})&=\ch(\EE^{(2)})-\sum_{m\ge 1} \frac{\Delta_m}{m!}  \ . \label{eq:Chern characters cotangent bundle}
\end{align}
So the full answer for the cotangent bundle is obtained by combining \eqref{eq:Chern characters Hodge bundle}, \eqref{eq:Chern character quadratic differentials} and \eqref{eq:Chern characters cotangent bundle}.
In particular, we have
\be 
c_1(\mathscr{K})=\ch_1(T^* \bM_{g,n})=\lambda_1+\kappa_1-\Delta_1=13\lambda_1+\sum_i \psi_i -2\Delta_1\ , \label{eq:c1 canonical bundle Mgnbar}
\ee
which is the well-known formula by Mumford and Harris derived in \cite{Harris_Kodaira}. In particular, $\Delta_1$ is the total boundary class of $\bM_{g,n}$, see eq.~\eqref{eq:definition Delta1}. We also have the following simple result for the even Chern characters
\be 
\ch_{2m}(T \bM_{g,n})=\frac{1}{(2m)!}\left(\kappa_{2m}-\Delta_{2m}\right)\ . \label{eq:even Chern characters tangent bundle Mgnbar}
\ee

\subsection{A consistency check}
Let us explain a strong consistency check that one can perform on this result. The Gauss-Bonnet theorem predicts
\be 
\int_{\bM_{g,n}} c_{3g-3+n}(T\bM_{g,n})  =\chi(\bM_{g,n})\ , \label{eq:Gauss-Bonnet theorem}
\ee
where $\chi(\bM_{g,n})$ is the orbifold Euler characteristic of compactified moduli space. For small values of $3g-3+n$ one can check this explicitly. Using the program \texttt{admcycles} \cite{admcycles}, we computed the left hand sides of \eqref{eq:Gauss-Bonnet theorem} for the following values of $(g,n)$. They agree with the known values of the Euler characteristics of $\bM_{g,n}$ \cite{Bini_Euler},\footnote{This differs from the famous calculation for the Euler characteristic of $\mathcal{M}_{g,n}$ by the inclusion of boundary divisors \cite{Harer_Euler}. It is simple combinatorics to deduce $\chi(\bM_{g,n})$ from $\chi(\mathcal{M}_{g,n})$.} which are
\begin{subequations}
\begin{align}
\chi(\bM_{0,4})&=2\ , \hspace{-.6cm} & \chi(\bM_{0,5})&=7\ ,\hspace{-.1cm} & \chi(\bM_{0,6})&=34\ , \hspace{-.1cm} & \chi(\bM_{0,7})&=213\ , \\
\chi(\bM_{1,1})&=\frac{5}{12}\ , \hspace{-.6cm} & \chi(\bM_{1,2})&=\frac{1}{2}\ ,\hspace{-.1cm} & \chi(\bM_{1,3})&=\frac{17}{12}\ , \hspace{-.1cm} & \chi(\bM_{1,4})&=\frac{35}{6}\ , \\
\chi(\bM_{1,5})&=\frac{389}{12}\ , \hspace{-.6cm} & \chi(\bM_{2,0})&=\frac{119}{1440}\ , \hspace{-.1cm} & \chi(\bM_{2,1})&=\frac{247}{1440}\ , \hspace{-.1cm} & \chi(\bM_{2,2})&=\frac{413}{720}\ , \\
 \chi(\bM_{3,0})&=\frac{8027}{181440} \ .  \hspace{-.6cm}  &&&&
\end{align}
\end{subequations}
This computation involves all the terms in the Chern characters and thus provides a simple check on our formula \eqref{eq:tangent bundle Chern characters}.

\section{Some details about genus 2}\label{app:genus 2}
In this appendix, we work out explicitly the sections of $\mathscr{L}^k$ in the case of a surface of genus 2. 
\subsection{Hyperelliptic surfaces}
We use an explicit parametrization of the genus 2 moduli space. First, we recall that every genus 2 surface can be written as a hyperelliptic surface of the form
\be 
y^2=\prod_{i=1}^6 (z-\lambda_i)\ . \label{eq:hyperelliptic surface}
\ee
Thus, we can view the $\lambda_i$'s as coordinates on $\mathcal{M}_2$. Of course permuting them leads to equivalent surfaces and there is an action of $\mathrm{PSL}(2,\CC)$ acting on them. So the moduli space $\mathcal{M}_{2}$ consists of unordered tuples $\lambda_i$ up to the action of $\mathrm{PSL}(2,\CC)$. Another way to think about unordered tuples $\lambda_i$ is to consider homogeneous polynomials or degree 6 in two variables, traditionally called binary sextics in the literature. Their zeros parametrize 6 values in $\mathbb{CP}^1$. 

Thus a section of a line bundle over $\bM_{2}$ is simply a symmetric function depending on the 6 variables $\lambda_i$ that satisfies certain boundary conditions at the boundary divisors and certain invariance conditions under the action of $\PSL(2,\CC)$.

\subsection{Invariance conditions}
We first explain the invariance condition under $\PSL(2,\CC)$ that a section of the prequantum line bundle $\mathscr{L}^k$ has to satisfy. On a hyperelliptic genus 2 surface, there are two holomorphic differentials, which we can take to be of the form
\be 
\omega_1=\frac{\mathrm{d}z}{y}\ , \qquad \omega_2=\frac{z \, \mathrm{d}z}{y}\ .
\ee
There are three quadratic differentials, namely
\be 
\omega_1^2\ , \qquad \omega_1\omega_2\ , \qquad \omega_2^2\ .
\ee
Recall from the discussion in Section~\ref{subsec:bundles on Mgn} that the prequantum line bundle can be realized as $\mathscr{L}^k=(\det \EE^{(2)})^k \otimes (\det \EE)^{-k}$. A section can thus be written as
\be 
f_k(\lambda_1,\lambda_2, \lambda_3,\lambda_4,\lambda_5,\lambda_6)\left(\frac{\omega_1^2 \wedge \omega_1 \omega_2 \wedge \omega_2^2}{\omega_1 \wedge \omega_2}\right)^k\ .
\ee
Let us work out the invariance conditions on $f_k(\lambda_1,\lambda_2\dots)$.
Clearly, $f_k(\lambda_1,\lambda_2,\dots)$ has to be a symmetric function, but it also has to satisfy a further condition due to $\mathrm{PSL}(2,\CC)$ invariance. A simple computation yields the transformation 
\be 
\begin{pmatrix}
\omega_2 \\ \omega_1
\end{pmatrix} \longmapsto \prod_{i=1}^6 \sqrt{c \lambda_i+d} \begin{pmatrix}
 a & b \\ c & d
\end{pmatrix}\begin{pmatrix}
\omega_2 \\ \omega_1
\end{pmatrix}\ ,
\ee
under $\lambda_i \to \frac{a \lambda_i+b}{c \lambda_i+d}$ with $ad-bc=1$. We also transformed $z$ in the same fashion on the right hand side in order to bring the differentials back into their standard form. Consequently,
\be 
\frac{\omega_1^2 \wedge \omega_1 \omega_2 \wedge \omega_2^2}{\omega_1 \wedge \omega_2} \longmapsto    \frac{\omega_1^2 \wedge \omega_1 \omega_2 \wedge \omega_2^2}{\omega_1 \wedge \omega_2} \ \prod_i (c \lambda_i+d)^2\ ,
\ee
and thus we should have
\be 
f_k(\lambda_1,\lambda_2,\dots)=f_k\left(\frac{a \lambda_1+b}{c \lambda_1+d},\frac{a \lambda_2+b}{c \lambda_2+d},\dots\right)\prod_i (c \lambda_i+d)^{2k}\ . \label{eq:invariance condition Lk genus 2}
\ee
for a section of $\mathscr{L}^k$. 

\subsection{Non-separating degeneration}
Let us next discuss what sort of behaviour we expect near the degenerations. There are two types of degenerations, namely the separating and the non-separating type. To keep formulas shorter, we will gauge fix $\lambda_4=0$, $\lambda_5=1$ and $\lambda_6=\infty$. The non-separating degeneration corresponds to the collision of two $\lambda_i$'s. Since any two collisions are equivalent, we can consider $\lambda_3 \to 0$. In this limit, the hyperelliptic equation becomes
\be 
y^2=z^2 (z-1)(z-\lambda_1)(z-\lambda_2)\ .
\ee
Upon defining $\tilde{y}=\frac{y}{z}$, this becomes the standard hyperelliptic surface of a genus 1 surface,
\be 
\tilde{y}^2=(z-1)(z-\lambda_1)(z-\lambda_2)\ .
\ee
The differentials become
\begin{align} 
\omega_1(z) \to  \frac{\mathrm{d}z}{z \tilde{y}}\ , \qquad \omega_2(z) \to  \frac{\mathrm{d}z}{\tilde{y}}\ .
\end{align}
$\omega_2(z)$ is the standard differential on the genus 1 surface, whereas $\omega_1(z)$ has two additional poles at $z=0$, $\tilde{y}=\pm \sqrt{-\lambda_1\lambda_2}$, which are the two nodes of the surface. Similarly, $\omega_1^2$, $\omega_1\omega_2$ and $\omega_2^2$ are quadratic differentials that have up to second order poles at the nodes. These are the standard differentials on the nodal surface and we hence conclude that $f_k(\lambda_1,\lambda_2,\dots)$ should be regular in the limit $\lambda_i \to \lambda_j$.

\subsection{Separating degeneration} \label{subapp:separating degeneration}
Next we discuss the separating degeneration. It can be identified with the limit where both $\lambda_2 \to 0$ and $\lambda_3 \to 0$ at the same rate, but their ratio is kept fixed. So let's set $\lambda_2=q^2$, $\lambda_3=q^2 \lambda$ and send $q \to 0$.\footnote{The reason to use $q^2$ here is that $\lambda_i-\lambda_j$ is only well-defined up to sign for small $\lambda_i-\lambda_j$ and thus a good local parameter is $q^2=\lambda_i-\lambda_j$.} Then the hyperelliptic equation becomes
\be 
y^2=z(z-1)(z-\lambda_1)(z-q^2)(z-q^2 \lambda)\ .
\ee
There are now two different ways to take the limit. We can either naively proceed and obtain
\be 
\tilde{y}^2=z_\text{L}(z_\text{L}-1)(z_\text{L}-\lambda_1)\ ,
\ee
where $\tilde{y}=\frac{y}{z_\text{L}}$. We wrote $z_\text{L}$ for $z$. This again a genus 1 surface. We can also first rescale $z \to q^2 z_\text{R}$ and $y \to q^3 \tilde{y}$, which presents the surface in the form
\be 
\tilde{y}^2=z_\text{R} (q^2z_\text{R}-1)(q^2z_\text{R}-\lambda_1)(z_\text{R}-1)(z_\text{R}-\lambda)\ .
\ee
The limit $q \to 0$ now gives again a genus 1 surface of the form
\be 
\tilde{y}^2=\lambda_1 z_\text{R}(z_\text{R}-1)(z_\text{R}-\lambda)\ .
\ee
We thus obtain two genus 1 surfaces that are connected at a single node. The node is located at $z_\text{L}=0$ and $z_\text{R}=\infty$. Let's work out what happens to the differentials. We start with the left surface.
\begin{align}
\omega_2(z) \to \frac{\mathrm{d}z_\text{L}}{\sqrt{z_\text{L}(z_\text{L}-1)(z_\text{L}-\lambda_1)}}\ ,
\end{align}
which is the standard differential on the surface. For $\omega_1(z)$, we find instead
\be 
\omega_1(z) \to \frac{\mathrm{d}z_\text{L}}{z_\text{L}\sqrt{z_\text{L}(z_\text{L}-1)(z_\text{L}-\lambda_1)}}\ .
\ee
A good local parameter of the surface near $z_\text{L}=0$ is given by $\tilde{y}$, since $z_\text{L}=\lambda_1^{-1} \tilde{y}^2$. The differential near $z_\text{L}$ behaves as
\be 
\frac{2\lambda_1^{-1} \tilde{y} \mathrm{d} \tilde{y}}{\lambda_1^{-1} \tilde{y}^3} =\frac{2\mathrm{d}\tilde{y}}{\tilde{y}^2}\ . 
\ee
Thus the differential has a double pole at the node and is thus not a valid differential on the left surface. Similarly we find
\begin{align} 
\omega_1(z) &\to \frac{q^2\mathrm{d}z_\text{R}}{ \sqrt{\lambda_1 q^6z_\text{R}(z_\text{R}-1)(z_\text{R}-\lambda)}}=\frac{\mathrm{d}z_\text{R}}{q \tilde{y}}\ , \\
\omega_2(z) &\to \frac{q^4z_\text{R}\mathrm{d}z_\text{R}}{ \sqrt{\lambda_1 q^6z_\text{R}(z_\text{R}-1)(z_\text{R}-\lambda)}}=\frac{q z_\text{R} \mathrm{d}z_\text{R}}{\tilde{y}}\ .
\end{align}
The limit of $\omega_2(z)$ has again a double pole at the node $z_\text{R}=\infty$. For the wedge products, $q \omega_1 \wedge \omega_2$ tends to a well-defined object in the limit, because the correct combination
\be 
q \omega_1 \wedge \omega_2 \longrightarrow \frac{\mathrm{d}z_\text{R}}{\tilde{y}_\text{R}} \wedge \frac{\mathrm{d}z_\text{L}}{\tilde{y}_\text{L}}
\ee
survives, whereas the terms with the double poles vanish in the limit. For the quadratic differentials, we note that $\omega_2^2$ tends to the quadratic differential that is only non-zero on the left part of the surface and $q^2 \omega_1^2$ tends to the quadratic differential that is only non-zero on the right part of the surface. Finally $\omega_1\omega_2$ tends to a quadratic differential that is non-zero on both sides of the surface and has double poles at the residues.
So we find that
\be 
q^2\omega_1^2 \wedge \omega_1 \omega_2 \wedge \omega_2^2 
\ee
tends to a well-defined product of the three quadratic differentials on the surface. Overall we find that
\be 
q\, \frac{\omega_1^2 \wedge \omega_1 \omega_2 \wedge \omega_2^2}{\omega_1 \wedge \omega_2} 
\ee
tends to a well-defined element of the fiber of $\mathscr{L}$
in the limit. Thus, we need to require that $f_k(\lambda_1,\lambda_2,\dots)$ has a zero of order $q^k$ at this degeneration in order to get a well-defined section of $\mathscr{L}^k$ that also extends to the boundary of $\bM_2$.
\subsection{A side note about a relation in cohomology}
Our discussion here can be used to derive a relation in $\H^2(\bM_2,\QQ)$, which we can compare with the literature in order to cross-check our analysis. Let's consider a section of the line bundle
\be 
\det \EE^{(2)} \otimes (\det \EE)^{-3}
\ee
In this case, the invariance conditions on such a section simply means that it is a function on moduli space (i.e.\ a section of the trivial line bundle on $\mathcal{M}_2$). However, writing the section as
\be 
f(\lambda_1,\lambda_2,\lambda_3,\lambda_4,\lambda_5,\lambda_6) \, \frac{\omega_1^2 \wedge \omega_1 \omega_2 \wedge \omega_2^2}{(\omega_1 \wedge \omega_2)^3}\ , 
\ee
$f$ can behave non-trivially near the boundary of moduli space. From our discussion, it follows that $f$ behaves regularly near the non-separating degeneration of moduli space. Near the separating degeneration, the discussion of \ref{subapp:separating degeneration} implies that $f$ is allowed to have a first order pole there. Let's denote as in Section~\ref{subsec:bundles on Mgn} by $\delta_1\equiv \delta_{1,\emptyset}$ the cohomology class of the separating divisor in $\H^2(\mathcal{M}_{2,0})$ and by $\delta_\text{irr}$ the cohomology class of the non-separating divisor. We thus learn that
\be 
c_1(\EE^{(2)})-3c_1(\EE)=\delta_1\ ,
\ee
since the first Chern class is by definition the zero divisor minus the pole divisor. 
Recalling the definition $c_1(\EE)=\lambda_1$ and from \eqref{eq:Chern character quadratic differentials} that  $c_1(\EE^{(2)})=\lambda_1+\kappa_1$, this reads
\be 
\kappa_1-2\lambda_1=\delta_1\ .
\ee
We can furthermore use Mumford's formula \eqref{eq:Mumford formula}
\be 
\lambda_1=\frac{1}{12}(\kappa_1+\Delta_1)=\frac{1}{12}(\kappa_1+\delta_\text{irr}+\delta_1)\ .
\ee
We can thus solve for $\kappa_1$ in terms of boundary divisors, which yields
\be 
5\kappa_1=\delta_\text{irr}+7\delta_1\ .
\ee
This relation in cohomology is well-known \cite{Arabello_Cohomology}. No similar relation exists for $g \ge 3$.
\subsection{The classical invariants}
Let us get back to our study of sections of $\mathscr{L}^k$.
We learned from our study of the degenerations that $f_k(\lambda_1,\lambda_2,\dots)$ does not have any poles, even at the degenerations. Thus, it has to be a polynomial. The invariance condition under $\PSL(2,\CC)$ then tells us that the order of the polynomial is $2k$ in every variable. 
The symmetric polynomials satisfying
\be 
f_k(\lambda_1,\lambda_2,\dots)=f_k\left(\frac{a \lambda_1+b}{c \lambda_2+d}\, ,\, \frac{a \lambda_1+b}{c \lambda_2+d}\, , \, \dots\right) \prod_i (c \lambda_i+d)^w
\ee
for some weight $w$ (but without the further condition at the degenerations) are well-known. A list of generators is (see e.g.\ \cite{Igusa})
\begin{align}
A&=\sum_{\text{fifteen}} (12)^2(34)^2(56)^2\ , \\
B&=\sum_{\text{ten}} (12)^2(34)^2(56)^2(45)^2(56)^2(46)^2\ , \\
C&=\sum_{\text{sixty}} (12)^2(23)^2(13)^2(45)^2(56)^2(46)^2 (14)^2(25)^2(36)^2\ , \\
D&=\prod_{i<j} (ij)^2\ , \\
E&=\prod_{\text{fifteen}} \det \begin{pmatrix}
1 & \lambda_1+\lambda_2 & \lambda_1 \lambda_2 \\
1 & \lambda_3+\lambda_4 & \lambda_3 \lambda_4 \\
1 & \lambda_5+\lambda_6 & \lambda_5 \lambda_6 
\end{pmatrix}\ .
\end{align}
Here the notation $(ij)$ means $\lambda_i-\lambda_j$. The sums and products run over all permutations of the labels $1,\dots,6$ that act non-trivially on the given expression. One readily checks that $A$ has weight 2, $B$ has weight 4, $C$ has weight 6, $D$ has weight 10 and $E$ has weight 15.
The generators satisfy a single relation of the form
\be 
E^2=F(A,B,C,D)\ ,
\ee
where $F$ is of graded polynomial of degree 30 in the other generators. 
\subsection{Sections of \texorpdfstring{$(\det \EE)^k$}{(det E)k} and Siegel modular forms}
In order to make contact with more classical results in the literature, we will first determine a set of generators for sections of $(\det \EE)^k$. These can be identified with Siegel modular forms of genus 2. On top of the invariance requirement that is solved by the invariants $A$, $B$, $C$, $D$ and $E$, we also need to impose that
\be 
f_k(q^2 \lambda_1,q^2 \lambda_2,q^2 \lambda_3,\lambda_4,\lambda_5,\lambda_6) \sim \mathcal{O}(q^k)\ . \label{eq:order of vanishing}
\ee
This comes from the fact that $q \omega_1 \wedge \omega_2$ is a well-defined section as $q \to 0$.
In fact, this requirement is already satisfied by $B$ and $D$ that vanish to orders $q^4$ and  $q^{12}$, respectively. So $D$ vanishes faster than required and is hence a cusp form. The complete list of generators that satisfy these requirements are
\begin{align} 
E_4&=B\ , \\
E_6&=AB-3C\ , \\
\chi_{10}&=D\ , \\
\chi_{12}&=AD\ , \\
\chi_{35}&=ED^2\ .
\end{align}
They satisfy a single relation of the form
\be 
\chi_{35}^2=F(E_4,E_6,\chi_{10},\chi_{12})\ ,
\ee
where $F$ is a graded homogeneous polynomial in the generators. In particular, the Hilbert series of the Siegel modular forms of genus 2 is
\be 
\sum_{k=0}^\infty \dim \H^0(\bM_{2},(\det \EE)^k) \, t^k=\frac{1+t^{35}}{(1-t^4)(1-t^6)(1-t^{10})(1-t^{12})}\ .
\ee
This is a classical result of Igusa \cite{Igusa}.
\subsection{Sections of \texorpdfstring{$\mathscr{L}^k$}{Lk}}
Finally, we can discuss the matter of interest and determine the ring of sections of the prequantum line bundle $\mathscr{L}^k$. By definition from \eqref{eq:invariance condition Lk genus 2}, the weight $w=2k$ is even and hence the ring in question will be a subring of the free polynomial algebra $\CC[A,B,C,D]$. It will be more convenient to use instead similar generators as the one that appeared for the Siegel modular forms and consider
\be 
\alpha=A\ , \qquad \beta=B\ , \qquad \gamma=AB-3C\ , \qquad \delta=D\ .
\ee
These generators still do not have any relation, but $\gamma$ has a higher order of vanishing at the separating degeneration than $C$. In fact, let $\ell$ denote the order of vanishing at the separating degeneration as in \eqref{eq:order of vanishing}. Then the weights and orders of vanishing of the generators $\alpha$, $\beta$, $\gamma$ and $\delta$ are
\begin{align} 
k[\alpha]&=1\ , & k[\beta]&=2\ , & k[\gamma]&=3\ , & k[\delta]&=5\ , \\
\ell[\alpha]&=0\ , & \ell[\beta]&=4\ , & \ell[\gamma]&=6\ , & \ell[\delta]&=12\ .
\end{align}
We need the order of vanishing to be at least as high as the order of vanishing. Thus, $\beta$, $\gamma$ and $\delta$ give directly generators of the ring $\bigoplus_{k \ge 0} \H^0(\bM_2,\mathscr{L}^k)$. Any element in the ring $\bigoplus_{k \ge 0} \H^0(\bM_2,\mathscr{L}^k)$ can thus be written as
\be 
\alpha^m \beta^n \gamma^p \delta^q
\ee
where
\be 
k=m+2n+3p+5q\ , \qquad m \le 2n+3p+7q\ .
\ee
Thus all that remains is to count the number of integer solutions to these constraints, which will give the Hilbert series. Let us first consider the refined Hilbert series that also keeps track of the order of vanishing. The Hilbert series that only accounts for the use of $\beta$, $\gamma$ and $\delta$ is
\be 
P_{\beta,\gamma,\delta}(x,t)=\frac{1}{(1-t^2x^4)(1-t^3x^6)(1-t^5x^{12})}\ ,
\ee
where $t$ keeps track of the weight of the generators and $x$ of the order of vanishing.
Now in the formal expansion around $t=0$, every term of the form $x^\ell t^k$ gives rise to several terms in the actual Hilbert series. In fact, we should replace every monomial
\be 
x^\ell t^k \longmapsto t^k + t^{k+1}+ \dots t^{\ell}=\frac{t^k-t^{\ell+1}}{1-t}
\ee
to get the actual Hilbert series for the number of sections of $\mathscr{L}^k$.
This is because every monomial in $\beta$, $\gamma$ and $\delta$ whose weight and order of vanishing are $k$ and $\ell$ can be multiplied by $\alpha^m$ with $0 \le m \le \ell-k$ to give a section of $\mathscr{L}^{k+m}$.
Thus it follows that the full Hilbert series takes the form
\begin{align} 
P(t) &\equiv \sum_{k\ge 0} \dim \H^0(\bM_2,\mathscr{L}^k)\, t^k \\
&=\frac{P_{\beta,\gamma,\delta}(1,t)-t P_{\beta,\gamma,\delta}(t,1)}{1-t} \\
&=\frac{1}{1-t}\left(\frac{1}{(1-t^2)(1-t^3)(1-t^{5})}-\frac{t}{(1-t^4)(1-t^6)(1-t^{12})}\right)\ . 
\end{align}
\bibliographystyle{JHEP}
\bibliography{bib}
\end{document}